\documentclass[12pt]{article}
\begin{document}
\input epsf
\def\be{\begin{equation}}
\def\bea{\begin{eqnarray}}
\def\ee{\end{equation}}
\def\eea{\end{eqnarray}}
\def\d{\partial}
\def\la{\lambda}
\def\eps{\epsilon}

\begin{titlepage}
\begin{center}
\hfill EFI-07-12\\

\vskip 1cm

{\LARGE {\bf 1/2--BPS states in M theory\\ \ \  \\
 and defects in the dual CFTs
  }}

\vskip 1.5cm

{\large  Oleg Lunin }

\vskip 1cm

{\it Enrico Fermi Institute, University of Chicago,
Chicago, IL 60637
}

\vskip 3.5cm

\vspace{5mm}

\noindent

{\bf Abstract}

\end{center}

We study supersymmetric branes in $AdS_7\times S^4$ and $AdS_4\times S^7$. We
show that in the former case the membranes should be viewed as M5 branes with fluxes and we identify two types of such fivebranes (they are analogous to giant gravitons and to dual giants). In $AdS_4\times S^7$ we find both M5 branes with fluxes and freestanding stacks of membranes. 
We also go beyond probe approximation and construct regular supergravity solutions describing  geometries produced by the branes. The metrics are completely specified by one function which satisfies either Laplace or Toda equation and we give a complete classification of boundary conditions leading 
to smooth geometries. The brane configurations discussed in this paper are dual to various defects in three-- and six--dimensional conformal field theories. 

\vskip 4.5cm

\end{titlepage}

\newpage

\setcounter{tocdepth}{1}

\tableofcontents

\setcounter{tocdepth}{0}

\newpage

\section{Introduction}
\renewcommand{\theequation}{1.\arabic{equation}}
\setcounter{equation}{0}

The last decade saw a significant progress in our understanding of string theory and field theories due in large part to the discovery of the AdS/CFT correspondence \cite{mald,gkpw}. Most of the work on the subject has been devoted to $AdS_5/$CFT$_4$ duality where one can carry out reliable computations on both sides of the correspondence and compare the results. For chiral primaries the calculations on the bulk side can be performed using supergravity approximation and they match the outcome of field theory computations \cite{Freedm}.
In the case of theory on $AdS_5\times S^5$ one can go further: in spite of the presence of a background RR flux, a string can be quantized in certain limits and 
one finds a perfect agreement with boundary results for various unprotected quantities 
\cite{ppWave,FrlTs}. These developments led to a remarkable progress in understanding of 
${\cal N}=4$ SYM in four dimensions and strings on 
$AdS_5\times S^5$: one sees an emergence of integrable structures on both sides of the 
correspondence \cite{IntgrBlt}.

Unfortunately the same techniques cannot be used to study the examples of AdS/CFT which come from 
M theory. In this case neither the bulk side nor the field theories are well--understood. On the boundary 
one has either six--dimensional $(2,0)$ theory \cite{Witt20,Seiberg,CFT20} or a fixed point of an RG flow in three dimension \cite{Seiberg,CFT3D}, and it is not 
clear how to compute the correlation functions in either one of these cases. On the bulk side the fundamental degrees of freedom are described by an M2 brane and it is not 
known how to quantize this object. It seems that the supergravity is the only available approximation in this case and certain correlation functions have been computed in this regime \cite{AdSCorrel}. 

While supergravity description has a limited scope (it captures only a small subset of stringy modes), it 
also has certain advantages over the full quantization of a string: some semiclassical objects carrying
very large charges have a good approximate description in terms of geometries, while representation 
of these objects in terms of stringy modes is very complicated. To be described by a classical geometry, 
a state should 
be  semiclassical and it should preserve some amount of supersymmetry. In the simplest case of 
$1/2$--BPS objects the gravity solutions describing local states have been constructed for all known examples of AdS/CFT correspondence \cite{lunMath,LLM}. While the construction of \cite{lunMath} 
exhausts all 1/2--BPS states in $AdS_3$/CFT$_2$, in higher dimensional cases one  
should also look for the bulk description of non--local states. For $AdS_5$/CFT$_4$ one encounters  
one--, two-- or three--dimensional defects on the boundary and their gravity description was found in 
\cite{yama,myWils,gomRom}. In this paper we will present an analogous construction for the M theory examples of AdS/CFT.  

Since the field theory side of the correspondence is not well--understood, one cannot write a clean expression for the gauge--invariant operator corresponding to a defect on the boundary in the same way as it is done for a local operator or for a Wilson line in CFT$_4$. However one can use the symmetry arguments to show an existence of certain defects in the field theory and to identify the corresponding objects in the bulk. Then classification of the defects in field theory reduces to a corresponding problem 
in M theory on $AdS_m\times S^n$ where one looks for brane configurations preserving certain symmetries. If the number of branes is small, the geometry remains unchanged, so one should study the 
dynamics of the probe objects. Famous examples of such branes are known as "giant gravitons", they
exist in all four cases of AdS/CFT \cite{giantGrav,dualGiant}\footnote{For $AdS_5$/CFT$_4$ one can 
also map this bulk description into specific operators on the boundary \cite{GiantAki,berens}, and it is 
this map which is missing for M theory cases.} and the geometries of \cite{LLM} describe the 
backreaction of these objects. 
Other examples of 1/2--BPS branes in $AdS_5\times S^5$ were introduced in \cite{ReyYee,PawRey} 
and they were used as a dual description of Wilson lines in \cite{DrukFiol,GomisYama}. Unlike the giant gravitons which carry only D3 brane charge, these D branes have a nontrivial coupling to the Kalb--Ramond B 
field, so when the brane is shrunk to zero size it goes over to a fundamental string rather than to the perturbative graviton. This result is expected since in such limit the representation of the gauge group 
becomes small\footnote{It turns out that the natural order parameter is not the dimension of the representation, but the number of boxes $\Delta$ 
in the Young tableau. For $\Delta\sim 1$ one has fundamental strings in the bulk \cite{ReyYee,maldWils},
at $\Delta\sim N$ the description in terms of D branes takes over \cite{DrukFiol,GomisYama} and for 
$\Delta\ge N^2$ one has to look at the modified geometries \cite{myWils}. The same scaling works in the case of the giant gravitons \cite{berens,LLM}.} and a dual description of a Wilson line 
is given by a string \cite{ReyYee,maldWils}. In the opposite limit of a very large representation, 
the D branes 
cannot be treated as probes and one has to find a modified geometry \cite{myWils}.  This picture has a very natural counterpart in M theory examples of AdS/CFT. As we will see, the light defects correspond to a probe membrane in the bulk, but as the charge of the defect grows, the description in terms of M5 branes with fluxes takes over. Finally as the amount of flux becomes very large, the branes modify the geometry and the main goal of this paper is to construct the resulting metrics. We will do this for various defects which preserve 16 supercharges. 

While our main motivation comes from AdS/CFT, a classification of supersymmetric branes on curved backgrounds is a very interesting problem on its own right. In flat space a brane preserving a half of supersymmetries should have flat worldvolume, but in other symmetric spaces the situation is more interesting. For $AdS_5\times S^5$ background and its pp wave limit the supersymmetric branes have been classified in \cite{SkendTayl} and a similar analysis for the M theory pp--wave was presented in 
\cite{KimYee}. Here we will study the branes on $AdS_7\times S^4$ and $AdS_4\times S^7$ both in the probe approximation (which is valid if the number of branes is small) and beyond it.

This paper has the following organization. In the first two sections we consider the probe branes on 
$AdS_m\times S^n$ solutions of M theory, in particular we will see that M5 branes carry a non--zero amount of a membrane charge. We will also show that the relation between this charge and the position of the M5 brane is similar to the one which exists for the giant gravitons. It turns out that the analogy persists even further: some M5 branes have bounded charges (just as the usual giants) and for the other class the number of induced membranes is unlimited (such M5s should be identified with "dual giants").  From the brane probe analysis we also infer the symmetries preserved by the branes and the remaining part of the paper is devoted to construction of the geometries which preserve these symmetries. In section 4 we summarize
 the general solution of eleven dimensional supergravity with $SO(2,2)\times SO(4)^2$ isometries (and details of the computation are presented in the appendix A). It turns out that the solution is uniquely specified by one harmonic function and one real number $q$. This number is determined by the asymptotic geometry and we concentrate on the most interesting cases of $AdS_m\times S^n$ 
asymptotics. Since these two branches correspond to different values of $q$, we consider them separately in sections 5 and 6. In both cases we demonstrate that any harmonic function satisfying a very simple 
set of boundary conditions leads to the unique regular geometry. We also find a clear interpretation of these boundary conditions in terms of the probe branes. Once the general solution is constructed, one can  try to look at various limits and in section 7 we show that sending the warp factors of AdS or of the spheres to infinity, one recovers interesting geometries which have been found in the past (similar limits 
for the solutions of \cite{myWils} are discussed in the appendix B). 
 
While the geometries with $SO(2,2)\times SO(4)^2$ isometry describe a majority of the branes discussed in sections 2 and 3, some interesting 1/2 BPS configurations of M2 branes in $AdS_4\times S^7$ are not 
covered by this ansatz. In this case the symmetry is $SO(2,1)\times SO(6)$ and the local structure of the corresponding solutions can be easily found by making an analytic continuation of the geometries constructed in \cite{LLM}. Such solutions are specified by one function $D$ which satisfies Toda equation and in section \ref{SecAnalLLM} we discuss the boundary conditions for $D$ which lead to regular geometries.

\section{Branes in $AdS_7\times S^4$}
\renewcommand{\theequation}{2.\arabic{equation}}
\setcounter{equation}{0}

\label{SectA7S4}

Before delving into the construction of supergravity solutions, we study an easier problem 
which will provide some useful information about symmetries. Our starting point is a duality between M theory on $AdS_7\times S^4$ and (2,0) superconformal theory in $5+1$ dimensions. 
Although this theory is poorly understood, we know that it contains local operators as well as gauge invariant nonlocal defects. These objects are analogous to Wilson lines in 
AdS$_5$/CFT$_4$ correspondence, but since the fundamental field in (2,0) theory is a 2--form gauge potential (rather than a 1--form), the most natural defects are two dimensional surfaces. 

In the AdS$_5$/CFT$_4$ correspondence the Wilson lines have been studied several years ago 
\cite{ReyYee,maldWils}, where the loops in ${\cal N}=4$ SYM were shown to correspond to strings 
ending on the boundary. 
Recently it was observed \cite{DrukFiol,GomisYama} that if the dimension of the representation is large, 
then a better bulk description of the Wilson lines is given by D3/D5 branes with fluxes on their worldvolume.  We expect a similar picture to hold in AdS$_7$/CFT$_6$ case as well: if dimension of representation is small, the "Wilson surface" should be dual to an M2 brane ending on the boundary, but for a higher--dimensional representation a better description should be given by M5 brane with fluxes.  In this section we will explore such configurations. 

We begin with analyzing an M2 branes ending on the boundary. There are two ways of constructing such 
system: the intersection can be either one-- or two--dimensional.  We will be interested in configurations 
which preserve $16$ supercharges and it is the two--dimensional intersection which can produce such states (this is in a nice agreement with the fact that the natural objects in field theory are two--dimensional defects). To see this it is useful to go away from the near horizon regime and consider 
a membrane intersecting a stack of M5 branes in asymptotically flat space. 

Suppose M5 branes are oriented along $012345$, then one--dimensional intersection corresponds to M2$_1$ filling $067$ and two dimensional intersection corresponds to M2$_2$ stretching along $056$. The supersymmetries preserved by M5 brane satisfy the projection $\Gamma_{012345}\eta=\eta$, while for two different M2's we find 
$\Gamma_{067}\eta_1=\eta_1$ and $\Gamma_{056}\eta_2=\eta_2$. Since matrices 
$\Gamma_{012345}$ and $\Gamma_{067}$ anticommute, they cannot be diagonalized simultaneously, 
this means that no supersymmetry is preserved by both M5 and M2$_1$. 
Similar argument shows that a $1+1$--dimensional intersection of M5 and M2$_2$ 
preserves a quarter of supersymmetries, 
moreover, it is clear that this configuration preserves 
$SO(4)\times SO(4)\times U(1)_t$ bosonic symmetries as well. 

If we had a stack of M5 branes alone, 
it would preserve $16$ supercharges and $ISO(5,1)\times SO(5)$ bosonic symmetry, but as one takes the near horizon limit, the number of supersymmetries is doubled and bosonic symmetry is increased to 
$SO(6,2)\times SO(5)$ \cite{mald}. One may anticipate that such enhancement happens for the M2, 
M5 intersection as well. To check this we analyze bosonic symmetries from the point of view of the theory on the boundary. The vacuum of $(2,0)$ theory has an $SO(6,2)$ conformal group and the intersection 
with M2 brane is seen as a $1+1$ dimensional defect in such theory. Let us assume that the defect is flat 
(i.e it is an analog of a straight Wilson line discussed in \cite{DrukFiol,yama,myWils}). Then using the general analysis of the conformal groups performed in 
\cite{cardy}, we conclude that such a defect breaks $SO(6,2)$ to $SO(4)\times SO(2,2)$. Some of these symmetries were manifest even in the asymptotically--flat configuration, but the enhancement 
$U(1)\rightarrow SO(2,2)$ happened in the near--horizon limit. It turns out that in this limit the number of supersymmetries is also increased from $8$ to $16$, but we will postpone the explicit demonstration of this fact until section \ref{SectGenSolut}.

To summarize, we expect the $1+1$--dimensional defect in $(2,0)$ theory to preserve $16$ supercharges and $SO(4)\times SO(4)\times SO(2,2)$ bosonic symmetry. To analyze the bulk objects which are dual to such defects, it is convenient to write the metric of $AdS_7\times S^4$ in a way which makes this symmetry more explicit:
\bea\label{AdS7S4}
ds^2&=&4L^2(\cosh^2\rho ds_{AdS}^2+\sinh^2\rho d{\tilde\Omega}_3^2+d\rho^2)+
L^2(d\theta^2+\sin^2\theta d{\Omega}_3^2)\\
F_4&=&3L^3 \sin^3\theta d\theta\wedge d^3{\Omega},\qquad L^3=\pi Nl_P^3\nonumber
\eea
In the bulk one has three kinds of branes which preserve these symmetries, and it turns out that all of them are relevant for the dual description of the defects. In the next three subsections we will describe these objects in more detail. 

\subsection{Probe M2 brane}

\label{SectA7S4m2}

Let us begin with considering an M2 brane which ends on the boundary: this is a counterpart of the analysis \cite{ReyYee,maldWils} for Wilson line. Since the worldvolume of M2 should contain a 
timelike direction, 
to be consistent with symmetries this membrane should extend along $AdS_3$.
This makes an action especially simple in the static gauge where the worldvolume of the brane is parameterized by the AdS coordinates:
\bea
S_{DBI}=-T_{2}\int dV_{AdS}\cosh^3\rho
\eea
The equation of motion for $\rho$ following from this action sets $\rho=0$, so the profile of M2 brane is fixed uniquely. In particular this implies that at the location of the brane ${\tilde S}^3$ goes to zero size, 
so the brane preserves one of the $SO(4)$ symmetries. To preserve another $SO(4)$
the brane should be located at $\theta=0$ (this can always be accomplished by an appropriate $SO(5)$ rotation). 

The fact that M2 has no moduli is expected: by analogy with Wilson line, the $1+1$ dimensional defect should be characterized by a surface (which we choose to be a flat plane) and a 
representation of the gauge group. The analysis of \cite{ReyYee,maldWils} which we are 
mimicking here, 
established a correspondence between a Wilson line in the fundamental representation and a string, so we expect that a single M2 brane also corresponds to an object in the fundamental representation in 
$(2,0)$ theory. This implies that both a shape and a representation are fixed for the defect, which agrees 
with the fact that a profile of M2 brane has no parameters. 

To describe the defects corresponding to higher dimensional representations, one should consider multiple M2 branes which are placed on top of each other, but, as the number of such membranes becomes large, a better description emerges, and it involves polarized M5 branes \cite{CalMald}. 
This effect is 
similar to a description of Wilson lines in terms of D branes which was proposed in \cite{DrukFiol}.
There are two types of M5 branes which preserve the symmetries of the solution (\ref{AdS7S4}): 
the worldvolume can either be $AdS_3\times S^3$ or $AdS_3\times {\tilde S}^3$. Let us consider these two cases separately. 

\subsection{M5 brane wrapping ${S}^3$.}

\label{SubsA7M5S}

We begin with looking at M5 branes stretched along $AdS_3\times {S}^3$. To mimic the membrane charge, such brane should also have a self--dual three--form switched on and in general it is hard to describe self--dual fields using the action principle. However, for the case 
of M5 branes there have been several proposals in the literature \cite{AganSchw,PST} and we will 
follow the method of \cite{PST} which is based on introduction of a scalar auxiliary field $a$. 
The counterpart of the DBI action in this formalism is
\bea\label{PSTAct}
S_{PST}&=&
-T_5\int d^6\xi \left[\sqrt{-\mbox{det}(g_{mn}+i{\tilde F}_{mn})}+
\frac{\sqrt{-g}}{4(\nabla a)^2}\d_m a F^{*mnl}F_{nlp}\d^p a\right]
\eea
The dynamical variable is a two--form $B_{mn}$ and following \cite{PST}
we introduced
\bea\label{PSTDual}
F=2dB-C^{(3)},\quad F^{*mnl}=\frac{1}{6\sqrt{-g}}\eps^{mnlabc}F_{abc},\quad
{\tilde F}_{mn}=\frac{1}{\sqrt{(\nabla a)^2}}F^*_{mnl}\d^l a
\eea
We can fix the invariance under diffeomorphisms by choosing the 
static gauge where $\xi^0,\xi_1,\xi_2$ are identified with coordinates on $AdS_3$ and 
$\xi_4,\xi_5,\xi_6$ are identified with coordinates on ${S}^3$. For the gauge field 
we will choose a magnetic description
by setting $dB=L^3 b~dV_{{\Omega}}$ with constant $b$. 
The action (\ref{PSTAct}) has an additional gauge invariance which allows one to set $a$ to be an arbitrary function with non--vanishing gradient (see \cite{PST} for further discussion) and we will use this freedom to identify $a$ with radial direction on $AdS_3$. To be more specific,
we write
\bea\label{AdS3Metr}
ds_{AdS}^2=-\cosh^2\zeta d\tau^2+d\zeta^2+\sinh^2\zeta d\phi^2
\eea 
and set $a=\zeta$. With these conventions we find
\bea
&&F=L^3(2b-h)dV_{S},\quad
F^*=\frac{8\cosh^3\rho}{\sin^3\theta}L^3(2b-h)dV_{AdS},\quad
h\equiv\frac{1}{4}(\cos 3\theta-9\cos\theta+8)\nonumber\\
&&{\tilde F}=-\frac{2L^2\cosh^2\rho}{\sin^3\theta}(2b-h)\sinh 2\zeta~ d\tau\wedge d\phi,\quad
F^{*mnl}F_{nlp}=0\nonumber
\eea
We observe that unless $\rho=0$, M5 brane is located at a point on ${\tilde S}^3$ which implies that 
one of the $SO(4)$ symmetries is broken. Since we are interested in the symmetric case, from now on 
$\rho$ will be set to zero. 
 
Using all this information, one can simplify the action (\ref{PSTAct}):
\bea\label{PST1}
S_{PST}=-T_5\int d^3{\Omega}d^3 H~8L^6
\sqrt{\sin^6\theta+{\cal F}^2},\qquad
{\cal F}\equiv 2b-h
\eea
In general an action describing M5 brane contains two pieces: the contribution of the tension 
(\ref{PSTAct}) and a  Chern--Simons term:
\bea\label{Chern1}
S_{CS}=\frac{T_5}{2}\int F\wedge C^{(3)},
\eea
however in the present case this contribution vanishes, so the action (\ref{PST1}) is complete. 

To find the location of the brane in $\theta$ coordinate, one should minimize 
(\ref{PST1}) with respect to this variable. To do this it is convenient to rewrite the expression appearing under the square root in (\ref{PST1}) in terms of $x=\cos\theta$:
\bea
V=(1-x^2)^3+(x^3-3x+2-2b)^2
\eea
Taking a derivative of this expression with respect to $x$, we find a simple relation
\bea
\frac{dV}{dx}=12(1-x^2)(b+x-1)
\eea
This implies that $V$ reaches a minimum if $x=1-b$ provided that this expression lies in the 
interval $(-1,1)$. In other words, we found the relation between the location of the M5 brane and the value of the magnetic field:
\bea
\cos\theta=1-b
\eea
Notice that if $b=0$, then the M5 brane is located at 
$\theta=0$ which means that its worldvolume becomes degenerate (${S}^3$ shrinks to 
zero size). This would look like a $2+1$ dimensional object which should be identified with 
the M2 brane discussed in the previous subsection. 

Since M5 brane has a worldvolume flux, it carries an induced membrane charge. This effect is familiar from the physics of D branes \cite{Douglas}, and to compute the appropriate charge one should find 
the source term for the $C^{(3)}$:
\bea
\delta S=\frac{1}{6}\int d^6\xi \frac{\d{\cal L}}{\d C^{(3)}_{abc}}\delta C^{(3)}_{abc}
\eea
In the present case we have one unit of M5 brane charge which comes from a coupling with magnetic components of $\delta C^{(3)}_{abc}$, but due to the presence of last term in (\ref{PSTAct}) and Chern--Simons contribution (\ref{Chern1}), the M5 brane couples to the 
electric three--form potential as well:
\bea\label{TmpApr2}
\delta S^{PST}_{el}&=&-\frac{T_5}{2}\int d^6\xi\frac{\sqrt{-g}}{(2L)^6}
\frac{8L^3}{\sin^3\theta}(2b-h){\delta C}^{(3)}_{AdS}
=-\frac{T_5L^3}{2}\cos\theta\sin^2\theta~\Omega_3\int_{AdS}\delta C^{(3)}\nonumber\\
\delta S^{CS}_{el}&=&-\frac{T_5}{2}\int_{AdS}{\delta C}^{(3)}\ \cdot\ \int F
=-\frac{T_5L^3}{2}\cos\theta\sin^2\theta~\Omega_3\int_{AdS}\delta C^{(3)}
\eea
The same potential is sourced by a membrane which extends in the AdS space and in that case the 
coupling is given by
\bea\label{M2Coupl}
\delta S=T_2\int_{AdS}\delta C^{(3)}
\eea
Comparing this with (\ref{TmpApr2}), we find the membrane charge carried by the 
M5\footnote{We recall that the tensions are expressed in terms of the Planck scale as
$T_2=\frac{1}{(2\pi)^2l_P^3}$,  $T_5=\frac{1}{(2\pi)^5l_P^6}$.}:
\bea\label{IndGiant}
n_2=-\frac{T_5}{T_2}L^3\cos\theta\sin^2\theta~\Omega_3=-\frac{N}{4}\cos\theta\sin^2\theta
\eea
Notice that this expression is bounded\footnote{A similar quantization condition for the membrane charge 
has been previously derived in \cite{ramal1}. See also \cite{ramal2} for further discussion of branes with 
$AdS_3\times S^3$ worldvolume.}. The brane configurations with bounded charges have 
been encountered in the past: the simplest example is an original giant graviton of 
\cite{giantGrav} which has a bound on angular momentum. Another example which is closely related 
with M5/M2 system arose from studying branes in $AdS_5\times S^5$: a D5 brane placed on this background acquires a charge $n_f\le N$ under a Kalb--Ramond field \cite{PawRey}.  The fact that this charge is bounded by the amount of flux has a very nice field theoretic interpretation 
\cite{GomisYama}: the D5 brane corresponds to a Wilson line in an antisymmetric representation of the gauge group and the number of boxes in Young tableau which corresponds to such representation is bounded by $N$. It would be nice to find a similar explanation for (\ref{IndGiant}). 

The angle $\theta$ varies from zero to $\pi$, so the right hand side of equation (\ref{IndGiant}) 
changes sign at $\theta=\frac{\pi}{2}$. It appears that one deals with M2 branes\footnote{In our notation M2 corresponds to negative values of $n_2$: this can be reversed by redefining the orientation of M5 brane: we assumed that $\int dV_{AdS}\wedge dV_S>0$.} if 
$0<\theta<\frac{\pi}{2}$ and with anti--membranes if $\frac{\pi}{2}<\theta<\pi$, however both types
of branes preserve the same supercharges, so they can be superposed freely. Recently similar configurations of branes and antibranes were used to establish a relation between the partition functions of black holes and topological strings \cite{strom}. 

Finally we observe that to recover a single M2 from (\ref{IndGiant}) one should take 
$\theta\sim\frac{2}{N}\ll 1$, this means that M5 brane collapses to a $2+1$ dimensional object located at 
$\theta=0$. This agrees with consideration of section \ref{SectA7S4m2}. We also notice that quantization 
of charge in (\ref{IndGiant}) implies that M5 branes cannot be placed at arbitrary values of $\theta$, but rather one has a discrete sequence $\theta_{n}$. 

\subsection{M5 brane wrapping ${\tilde S}^3$.}

\label{SubsA7M5A}

Next we look at M5 brane with worldvolume $AdS_3\times {\tilde S}^3$. We again switch on the magnetic
field and use the sum of (\ref{PSTAct}) and Chern--Simons actions to describe the system. In the present 
case the static gauge involves identifying of $\xi^0,\xi_1,\xi_2$ with $AdS_3$ and 
$\xi_4,\xi_5,\xi_6$ with ${\tilde S}^3$, and we also identify $a$ with radial coordinate of AdS. 
As before, the field strength is taken to be proportional to the 
volume of the sphere: $dB=L^3 b~dV_{{\tilde S}}$. Since this M5 brane occupies a point of 
$S^4$, we can perform a $SO(5)$ rotation to place it at $\theta=0$ in the new coordinate system. 
With these conventions we find
\bea
&&F=2L^3b~dV_{\tilde S},\quad
F^*=2\frac{\cosh^3\rho}{\sinh^3\rho}L^3b~dV_{AdS},\nonumber\\
&&{\tilde F}=-\frac{L^2\cosh^2\rho}{\sinh^3\rho}b\sinh\zeta\cosh\zeta~ d\tau\wedge d\phi,\quad
F^{*mnl}F_{nlp}=0\nonumber
\eea
Substituting this into the PST action (\ref{PSTAct}), we find
\bea\label{PST2}
S_{PST}=-T_5\int d^3{\tilde\Omega} d^3 H~(2L)^6
\cosh^3\rho\sqrt{\sinh^6\rho+\frac{b^2}{16}}
\eea
Next we evaluate the Chern--Simons coupling between the background field and the brane. In the present setup, the relevant contribution comes from integrating the pullback of the {\it dual} gauge potential over the worldvolume of M5:
\bea
S_{CS}=T_5\int P[C^{(6)}]
\eea
We begin with evaluating the dual field strength from (\ref{AdS7S4}):
\bea
F_7=6\cdot (2L)^6\cosh^3\rho\sinh^3\rho d^3 H\wedge d^3{\tilde\Omega}\wedge d\rho
\eea
Integrating this expression with respect to $\rho$, we find a gauge potential which is invariant under $SO(2,2)\times SO(4)$:
\bea
C^{(6)}=2L^6(\cosh 6\rho-9\cosh 2\rho+8)~d^3 H\wedge d^3{\tilde\Omega}
\eea
and the Chern--Simons action becomes
\bea
S_{CS}=2T_5L^6\int (\cosh 6\rho-9\cosh 2\rho+8)~d^3 H\wedge d^3{\tilde\Omega}
\eea
To find the location of the M5 brane we should combine this with (\ref{PST2}) and minimize the resulting action with respect to $\rho$. As before, it is convenient to introduce a new variable 
$x=\cosh\rho$ and rewrite the Lagrangian in terms of it:
\bea
{\cal L}&=&-64x^3\sqrt{(x^2-1)^3+\frac{b^2}{16}}~+8(y^3-3y+2),\nonumber\\
&=&-8\sqrt{(y^2-1)^3+\frac{b^2}{2}(y+1)^3}~+8(y^3-3y+2),\quad y\equiv 2 x^2-1
\eea
Notice that the new variable $y=\cosh 2\rho$ is bounded from below. Extremizing the Lagrangian with respect to this variable, we find an equation
\bea
(y^2-1)^2y+\frac{b^2}{4}(y+1)^2=(y^2-1)\sqrt{(y^2-1)^3+\frac{b^2}{2}(y+1)^3}
\eea
In the region $y>1$, this equation has only one solution:
\bea\label{RhoEM2}
y=\frac{b}{2}+1:\qquad \sinh\rho=\frac{\sqrt{b}}{2}
\eea
Unlike the solution described in the previous subsection, this branch has an unbounded 
magnetic field, so it is analogous to the "dual giant graviton" \cite{dualGiant} (or to the probe D3 brane 
in the context of \cite{DrukFiol,myWils}). To see this more clearly, we again compute the coupling to the two form potential and extract the M2 brane charge:
\bea
\delta S_{el}=\frac{T_5}{4}\int d^6\xi \sqrt{-g}F^{*\zeta nl}\delta C^{(3)}_{nl\zeta}+
\frac{T_5}{2}\int F\wedge \delta C_{el}^{(3)}
=-2T_5bL^3\Omega_3\int_{AdS} \delta C^{(3)}
\eea
Comparing this with membrane coupling (\ref{M2Coupl}), we find the number of M2 branes:
\bea
n_2=-2\frac{T_5}{T_2}~4\sinh^2\rho L^3\Omega_3=-2N\sinh^2\rho
\eea
As before, the quantization of $n_2$ leads to a sequence of the allowed values of $\rho_n$ and 
for $n_2=-1$ the M5 collapses and we go back to the probe membrane discussed in section 
\ref{SectA7S4m2}. However unlike (\ref{IndGiant}) the present branch allows the charges to be arbitrarily large, so we have an analog of dual giant gravitons \cite{dualGiant} and D3 branes with 
$AdS_2\times S^2$ worldvolume on $AdS_5\times S^5$ background \cite{DrukFiol}. 

\bigskip

If many M5 branes are placed on $AdS_4\times S^7$ background, the probe approximation would break down and one would have to look for the modified geometry produced by the branes. A similar problem 
for giant gravitons was solved in \cite{LLM} and the main goal of this paper is to describe the corresponding construction for the branes discussed in this section. It turns out that the resulting supergravity solutions also describe branes on $AdS_7\times S^4$, so we first discuss the properties of these objects and we will come back to gravity solutions in section \ref{SectGenSolut}.

\section{Branes in $AdS_4\times S^7$}
\renewcommand{\theequation}{3.\arabic{equation}}
\setcounter{equation}{0}

\label{SectA4S7}

Let us now turn to another example of AdS/CFT correspondence and discuss a 
duality between M theory on $AdS_4\times S^7$ and ${\cal N}=8$ superconformal theory in $2+1$ dimension. This theory is defined as an infrared limit of three dimensional super Yang--Mills 
\cite{Seiberg,CFT3D}, and in particular one can try to trace various 
operators in the field theory under the 
RG flow. Due to operator mixings, it is very hard to establish a direct map between the quantities 
in the field theory (as defined in the UV) and gravity (which corresponds to the IR fixed point). However since the symmetries of the UV theory are preserved along the flow (and they are enhanced in the fixed point), we expect that any operator preserving certain symmetry would map into a supergravity configuration which preserves (at least) the same symmetry. One can apply this argument to learn some useful information about a map for local operators \cite{AdSCorrel}, but here we will be mostly concerned with Wilson lines and "Wilson surfaces" in the gauge theory. 

\subsection{One--dimensional intersection.}

\label{SubsM2Intrs}

It is interesting to look at Wilson lines which preserve certain amount of supersymmetry. 
The analysis of the four dimensional case is summarized in \cite{zarembo} and it can be easily extended to the three dimensional theory. One finds that to preserve $1/2$ of supersymmetries 
the Wilson line has to be straight and it is specified only by the representation of the gauge group. 
For the line in fundamental representation the dual description should be given by M2 brane ending on the boundary.  The fact that the line is one--dimensional suggests that the intersection of M2 brane with boundary should be one--dimensional as well, and it turns out that such intersection is also natural from the point of view of asymptotically--flat space. 

Indeed, suppose that the $AdS_4\times S^7$ emerged as a near horizon limit of a stack of M2 branes spanning directions $012$. Then in {\it flat space} they preserve Killing spinors satisfying a projection $\Gamma_{012}\eta=\eta$. One has two ways of introducing a probe brane which intersects this stack: M2$_1$ stretching along $034$ leads to one--dimensional intersection and 
M2$_2$ brane stretching along $013$ intersects a stack on $1+1$ dimensional surface. Looking at supersymmetries preserved by M2$_1$ and M2$_2$, we find that eight common supercharges are 
preserved by the original stack and M2$_1$, while combination of M2$_2$ and original M2's breaks all supersymmetries. 

We conclude that to preserve some supersymmetries in {\it asymptotically flat space}, one should 
look at one--dimensional intersection. In addition to eight supercharges, such intersection preserves $U(1)_t\times U(1)_1\times U(1)_2\times SO(6)$ bosonic symmetries. Here $U(1)_1$ 
and $U(1)_2$ correspond to the rotations in the spacial directions of M2 branes. 
As usual, we expect an enhancement of symmetry in the near horizon limit leading to 
$16$ supercharges and $SO(2,2)\times U(1)_1\times U(1)_2\times SO(6)$ bosonic symmetries.
To make these isometries more explicit, we write the metric of $AdS_4\times S^7$ as
\bea\label{AdSMetrU1}
ds^2&=&L^2(\cosh^2\rho dH_2^2+d\rho^2+\sinh^2\rho d\psi^2)+
4L^2(\cos^2\theta d\phi^2+d\theta^2+\sin^2\theta d\Omega_5^2)\nonumber\\
F_4&=&3L^3\cosh^2\rho\sinh\rho d\rho\wedge d\psi\wedge d^2 H,\quad 
L^6=\frac{\pi^2}{2}Nl_P^6
\eea
The object dual to a Wilson line is a probe M2 brane ending on the boundary and the symmetries
dictate that it has a worldvolume extending in $AdS_2$ and $\phi$. It is clear that the probe membrane should be located at $\rho=\theta=0$. As the dimension of representation becomes large, the M2 brane moves to a non--zero value of $\rho$ and breaks one of the $U(1)$ symmetries. Namely one can still parameterize the worldvolume by $AdS_2$ and $\phi$, but in addition a nontrivial dependence $\psi=\psi(\phi)$ should be introduced. Then one finds a metric induced on the M2 brane:
\bea
ds_3^2=L^2(\cosh^2\rho dH_2^2+
(\sinh^2\rho {\dot\psi}^2+4) d\phi^2),\quad {\dot\psi}\equiv \frac{d\psi}{d\phi}\nonumber
\eea
The action for M2 brane consists on the volume term and a Chern--Simons piece:
\bea
S_{M2}=-T_2\int d^3\xi~L^3\cosh^2\rho\sqrt{\sinh^2\rho {\dot\psi}^2+4}+
3 L^3 T_2\int d^3\xi {\dot\psi}\int_0^\rho \cosh^2\xi\sinh\xi d\xi\nonumber
\eea
and a variation with respect to $\rho$ gives an equation:
\bea
{\dot\psi}\cosh\rho=\sqrt{4+{\dot\psi}^2\sinh^2\rho}
\eea
This relation becomes an identity for any value of $\rho$ once we take 
\bea\label{PsiAndPhi}
\psi=2\phi.
\eea 
It is also interesting to compute an "angular momentum", i.e. a variable which is canonically conjugate to 
$\psi$:
\bea
J=\frac{\d {\cal L}}{\d {\dot\psi}}=L^3 T_2(\cosh\rho-1)
\eea
To summarize, we found that in $AdS_4\times S^7$ there is a set of supersymmetric M2 branes
which preserve $SO(2,2)\times SO(6)\times U(1)$ symmetry, and these membranes are parameterized  by a quantum number $J$. As one puts many such membranes together, the geometry will be modified and we will discuss the relevant supergravity solutions in section 
\ref{SecAnalLLM}.

\subsection{Two--dimensional intersections.}

\label{SubsA4M5S}

Another BPS configuration in flat space is given by a probe M5 brane which intersects the original stack of the membranes along $1+1$ dimensional manifold. The supersymmetries for this case were discussed in detail in the beginning of section \ref{SectA7S4} and we will not repeat that analysis here\footnote{If the stack of membranes 
is placed on a singularity of a Calabi--Yau 4--fold (rather than in 
flat space), then the 
intersection has fewer supersymmetries. An example of such configuration is discussed in \cite{yamaDef}.}. 
Let us discuss the bosonic symmetries preserved by the probe M5 branes in the geometry created by 
the membranes. From the asymptotically flat region one 
can read off $SO(4)\times SO(4)$ isometries (if M2 is stretched along $012$ and M5 is oriented in $013456$, then one has rotations in $3456$ and $789(10)$) as well as time translations. 
These symmetries are preserved by the intersecting branes, so they would be present in the 
$AdS_4\times S^7$ 
region as well. Since in the dual field theory one has a two dimensional defect, we expect that time translations are enhanced to $SO(2,2)$ conformal symmetry in the near horizon limit. Thus we conclude that to preserve 16 supercharges, a probe M5 brane placed on $AdS_4\times S^7$ should also preserve 
$SO(2,2)\times SO(4)\times SO(4)$ bosonic symmetries. To make this more explicit, we rewrite the $AdS_4\times S^7$ geometry as
\bea
ds^2&=&L^2(\cosh^2\rho dH_3^2+d\rho^2)+
4L^2(\cos^2\theta d\Omega_3^2+d\theta^2+\sin^2\theta d{\tilde\Omega}_3^2)\nonumber\\
F_4&=&3L^3\cosh^3\rho d\rho\wedge  d^3 H,\quad L^6=\frac{\pi^2}{2}Nl_P^6
\eea
A supersymmetric M5 brane can wrap either $AdS_3\times S^3$ or 
$AdS_3\times {\tilde S}^3$. Since these two types of branes are related to each other by a simple $Z_2$ reflection, it is sufficient to look at M5 wrapping 
$AdS_3\times {S}^3$. To preserve the symmetries, such brane should be placed at $\theta=0$ and a fixed value of $\rho$, and it can also have a worldvolume flux (see (\ref{PSTDual})):
\bea
F_{abc}dx^{abc}=2L^3e~d^3 H+2L^3b~d^3 {\Omega}-3L^3d^3 H\int_0^\rho\cosh^3\xi d\xi
\eea
Let us find the equations of motion for the brane. 
As in section \ref{SectA7S4}, we introduce an explicit parameterization  (\ref{AdS3Metr}) for $AdS_3$ 
and choose the gauge $a=\zeta$. Then equations (\ref{PSTDual}) become:
\bea
&&F^*=\frac{8}{\cosh^3\rho}L^3(2e-h)d^3{\Omega}+
\frac{\cosh^3\rho}{8}2L^3b~d^3 H,\quad h=3\int_0^\rho\cosh^3\xi d\xi\nonumber\\
&&{\tilde F}=-\frac{\cosh^2\rho}{4}L^2b\sinh\zeta\cosh\zeta~d\tau\wedge d\phi,
\nonumber\\
&&\frac{1}{4(\nabla a)^2}\d_m aF^{*mnl}F_{nlp}\d^p a=-\frac{2}{8\cosh^3\rho}b(e-\frac{h}{2})\nonumber
\eea
Substituting this into (\ref{PSTAct}), we find
\bea
S_{PST}=-T_5\int d^3\Omega d^3 H~ 8L^6\left[\cosh^3\rho\sqrt{1+\frac{b^2}{16}}-
\frac{1}{4}b(e-\frac{h}{2})\right]
\eea
This action should be supplemented by the Chern--Simons term (\ref{Chern1}):
\bea
S_{CS}=-\frac{T_5L^6}{2}\int d^3\Omega d^3 H~2bh
\eea
Minimizing the action with respect to $\rho$, we find an equation
\bea\label{RhoEfund}
\tanh\rho=\frac{b}{\sqrt{16+b^2}}:\qquad \sinh\rho=\frac{b}{4}
\eea
We observe that the value of $b$ is not bounded from above, so we are dealing with an analog of a "dual giant graviton". Notice that the "electric field" $e$ did not play any role in this analysis, it just led to an additional constant term in the action. In appears that in our gauge $e$ is an auxiliary field which does not affect dynamics. 

Notice that despite some similarity between two relations (\ref{RhoEM2}), (\ref{RhoEfund}) connecting 
$\rho$ and $b$, there is an important physical difference between the underlying M5 branes: as $\rho$ goes to zero, the worldvolume of M5 brane discussed in section \ref{SectA7S4} becomes degenerate, 
and the system effectively describes a $2+1$ dimensional membrane. This never happens for the M5 which led to (\ref{RhoEfund}): the volume element on this brane is
\bea
dV=8L^6\cosh^3\rho~d^3\Omega\wedge d^3 H
\eea
and it never degenerates. This is consistent with our earlier analysis which showed that $1+1$ dimensional intersection of M2 branes cannot be supersymmetric. 

Even though the M5 branes discussed in this subsection never degenerate into the membranes, they do carry an induced M2 charge:
\bea
\delta S_{el}=\frac{T_5}{4}\int d^6\xi \sqrt{-g}F^{*\zeta nl}\delta C^{(3)}_{nl\zeta}+
\frac{T_5}{2}\int F\wedge \delta C_{el}^{(3)}
=-2T_5bL^3\Omega_3\int_{AdS} \delta C^{(3)}
\eea
Comparing this with membrane coupling (\ref{M2Coupl}), we find the number of membranes:
\bea\label{A4S7nmb2}
n_2=-2\frac{T_5}{T_2}~16\sinh^2\rho L^3\Omega_3=-4\sqrt{2N}\sinh^2\rho
\eea
If we put many such M5 branes together, they are expected to modify the geometry, and in the next section we will describe the supergravity solutions produced by such configurations. 

\section{Geometries with $SO(2,2)\times SO(4)\times SO(4)$ symmetry}
\renewcommand{\theequation}{4.\arabic{equation}}
\setcounter{equation}{0}

\label{SectGenSolut}

In the previous sections we looked at various probes on the $AdS_4\times S^7$ and 
$AdS_7\times S^4$  and we saw that M5/M2 branes follow interesting profiles in these geometries. 
Once many branes are put together, one expects the probe approximation to break down, and the 
metric to be modified. Finding such modified geometries is the main goal of this paper and the construction will be described in this section.

In general it is very hard to find exact solutions of supergravity equations, but for configurations which have large amount of (super)symmetry one sometimes can succeed in constructing such 
solutions. In particular, it appears that the symmetries of M2/M5 configurations which were discussed in 
the previous sections are sufficient for finding the local geometry produced by them. We begin with 
describing the configurations involving M5 branes. The brane probe analysis suggests the bosonic symmetry $SO(2,2)\times SO(4)\times SO(4)$, and to enforce it in the supergravity solution, we choose an ansatz:
\bea\label{InAns}
ds^2&=&e^{2A}ds_H^2+e^{2B}ds_S^2+e^{2C}ds_{\tilde S}^2+
h_{ij}dx^i dx^j\nonumber\\
F_4&=&df_1\wedge dH_3+df_2\wedge d\Omega_3+
df_3\wedge d{\tilde\Omega}_3
\eea
Here $ds^2_S$ and $ds^2_{\tilde S}$ represent metrics on unit spheres $S^3$, ${\tilde S}^3$, and 
$ds_H^2$ is a metric on $AdS_3$ with unit radius. We also have an undetermined metric in two dimensions $h_{ij}dx^i dx^i$, and all scalars are functions of $x_1,x_2$. Starting with this ansatz, 
one should solve the equations for the Killing spinors:
\bea\label{InSUSY}
\nabla_m\eta+\frac{1}{288}\left[{\gamma_m}^{npqr}-
8\delta^n_m\gamma^{pqr}\right]G_{npqr}\eta=0
\eea
and the details of the computations are presented in the Appendix \ref{SectAppA}\footnote{A partial analysis of the system appeared earlier in \cite{yama}.}. Here we just mention that the equations 
for the spinor are simplified if one introduces a coordinate 
$x_2\equiv y=e^{A+B+C}$ and chooses $x_1$ to be orthogonal to it. This fixes the residual diffeomorphism invariance in (\ref{InAns}) and further manipulations lead to the unique {\it local} solution of the SUSY variations (\ref{InSUSY}). 

In the Appendix \ref{SectAppA} we show that equations (\ref{InSUSY}) guarantee that the system 
(\ref{InAns}) preserves (at least) 16 supercharges, and we derive the expressions for the metric and 
fluxes in terms of the warp factors 
$e^A$, $e^B$, $e^C$:
\bea\label{EqnMetr}
&&y=e^{A+B+C},\qquad h_{ij}dx^idx^j=g_y^2(dx^2+dy^2),\qquad dx=*dy\\
&&g_y^{-1}=y\sqrt{-e^{-2A}+e^{-2B}+e^{-2C}},\quad 
e^F\equiv \sqrt{e^{2A}-c_1^2 e^{2C}-c_2^2 e^{2B}}\nonumber
\eea
\bea\label{Eqnf1}
&&df_1=2c_1c_2ye^{2A-2F}*d(B-C)+de^{4A-F},\quad f_0\equiv 
f_1-e^{2A+F}\\
\label{Eqnf2}
&&e^{A-3B}df_2=2e^{-A-B}df_0+2c_1e^C*d(2B+C)+c_2e^{B-3A}df_1\\
\label{Eqnf3}
&&e^{A-3C}df_3=2e^{-A-C}df_0-2c_2e^B*d(2C+B)+c_1e^{C-3A}df_1\\
\label{Eqnf2*}
&&e^{F-3B}*df_2=c_1e^{C-3A}df_1+e^{-A-C}df_0-2c_2e^B*d(2B+C)\\
\label{Eqnf3*}
&&e^{F-3C}*df_3=-c_2e^{B-3A}df_1-
e^{-A-B}df_0-2c_1e^C*d(B+2C)
\eea
The constants $c_1$ and $c_2$ are not fixed completely, however they can be expressed in terms of one number $q$ through the relation (\ref{SumABCq})\footnote{Without a loss of generality we
 take $a=b=c=e_1=1$ in all relations appearing in the Appendix \ref{SectAppA}.}:
\bea\label{UnivABC}
c_1=q,\qquad c_2=-(q+1)
\eea
Finally, we have the expressions for the derivatives of $f_0$:
\bea\label{Eqnf0}
df_0=-2e^Ag_y^2\left[-e^{F+B+C}dy+e^A(c_1e^{2C}-c_2e^{2B})dx\right]
\eea

Notice that equations (\ref{Eqnf1})--(\ref{Eqnf3}) can be used to write the fluxes in terms of the 
warp factors, then (\ref{Eqnf2*}), (\ref{Eqnf3*}), (\ref{Eqnf0}) can be viewed as restrictions on two 
independent warp factors. 
In practice, in order to construct a solution it is convenient to combine (\ref{Eqnf2*}), 
(\ref{Eqnf3*}) into a relation (\ref{11DMaster}):
\bea\label{MasterEqn16}
d\log\frac{e^A-e^F}{e^A+e^F}+4*d\arctan\frac{c_2 e^{B-C}}{c_1}=
\frac{e^{F+B+C}}{e^{2A}-e^{2F}}\left[c_2e^{-4C}*df_3-c_1e^{-4B}*df_2\right]
\eea
The right hand side of the last equation is a one--form in two dimensions and it can be decomposed 
into exact and co--exact forms:
\bea\label{DefPsi1}
\frac{e^{F+B+C}}{e^{2A}-e^{2F}}\left[c_2e^{-4C}*df_3-c_1e^{-4B}*df_2\right]\equiv
-(d\Psi_2+*d\Psi_1)
\eea
While such decomposition always exists, it is not unique since the last equation is invariant under 
the "gauge transformation" which is parameterized by a harmonic function $\Psi$:
\bea
\Psi_2\rightarrow \Psi_2+\Psi,\quad \Psi_1\rightarrow \Psi_1+{\tilde\Psi}:\qquad
d\Psi+*d{\tilde\Psi}=0,\quad \Delta\Psi=0\nonumber
\eea
This freedom can be used to impose a convenient boundary condition on $\Psi_1$:
\bea\label{DefPsi2}
\Psi_1|_{y=0}=0,\qquad \Psi_1|_{x^2+y^2\rightarrow\infty}=0
\eea
We conclude that for any solution with $SO(2,2)\times SO(4)^2$ isometries, 
there exists a unique pair of functions $\Psi_1$, 
$\Psi_2$ defined by (\ref{DefPsi1}), (\ref{DefPsi2}) and equation (\ref{MasterEqn16}) can be rewritten in terms of them:
\bea
d\left[\log\frac{e^A-e^F}{e^A+e^F}+\Psi_2\right]+*d\left[4\arctan\frac{c_2 e^{B-C}}{c_1}+
\Psi_1\right]=0
\eea
The last equation implies that for any solution there exists a unique harmonic function $\Phi$:
such that
\bea\label{SolnThrPhi}
\log\frac{e^A-e^F}{e^A+e^F}+\Psi_2=\d_x\Phi,\quad
4\arctan\frac{c_2 e^{B-C}}{c_1}+\Psi_1=\d_y\Phi
\eea
There is also an inverse map:
once a harmonic function $\Phi$ is specified, one can use (\ref{SolnThrPhi}), (\ref{Eqnf0}), 
(\ref{Eqnf1})--(\ref{Eqnf3*}) to recover the solution. Although one has to solve nonlinear PDEs to 
find the solution in the closed form, later we will show that any harmonic function $\Phi$ leads to the unique solution. Thus locally we have a one--to--one correspondence between supersymmetric solutions (\ref{InAns}) and harmonic functions of two variables $\Phi(x,y)$:
\bea
(\d_x^2+\d_y^2)\Phi=0
\eea
The resulting local solution is smooth as long as derivatives of $\Phi$ remain finite and $y$ is not equal to zero. A generic function $\Phi$ leads to a geometry which becomes singular at $y=0$ (since 
at least one of the spheres collapses to zero size) and we need to impose some special boundary conditions to avoid a singularity.

\begin{figure}
\begin{center}
\epsfxsize=5.5in \epsffile{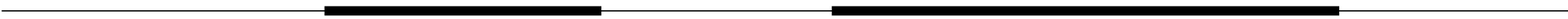}
\end{center}
\caption{Boundary condition for the harmonic function describing a typical regular geometry. The dark regions correspond to $\d_y\Phi|_{y=0}=2\pi~\mbox{sign}(c_1c_2)$ and light regions correspond to 
$\d_y\Phi|_{y=0}=0$ (see equation (\ref{y0Regul})). This is the most general boundary condition unless 
$c_1=c_2=-1/2$.
} \label{FigXbc}
\end{figure}

Let us consider a point on $x$ axis. Due to relation $y=e^{A+B+C}$, at least one of the spheres should 
collapse to zero size there (the remaining option of setting $e^A$ to zero leads to singularity), let us 
assume that it is $e^B$ that goes to zero while $e^C$ remains finite. Then assuming that $c_1,c_2\ne 0$ 
(which is true for the solutions with interesting asymptotics), we 
find that $\d_y\Phi$ must go to zero as we approach the $y$ axis (see (\ref{SolnThrPhi}), 
(\ref{DefPsi2})). Then $e^A$, $e^C$, $y^{-1}e^B$ and $g_y$ remain finite and non--zero, so to check the 
regularity we only need to look at $(S^3,y)$ part of the metric:
\bea
e^{2B}ds_S^2+g_y^2 dy^2=y^{-2}e^{2B}(y^2 ds_S^2+dy^2)
\eea
We conclude that the geometry remain smooth at $y=0$ if $\d_y\Phi=0$ there. 
Similarly, if $e^C$ goes to zero, then $\d_y\Phi$ goes to $2\pi~\mbox{sign}(c_1c_2)$ and the metric 
stays regular. Thus one has two types of Neumann boundary conditions at $y=0$:
\bea\label{y0Regul}
\d_y\Phi|_{y=0}=0\quad \mbox{or}\quad  \d_y\Phi|_{y=0}=2\pi~\mbox{sign}(c_1c_2)
\eea 
and we distinguish "light" points where $\d_y\Phi=0$ and "dark" points where 
$\d_y\Phi=2\pi~\mbox{sign}(c_1c_2)$. Then the $x$ axis splits into various regions and the 
boundary condition for a typical geometry is shown in figure \ref{FigXbc}. This coloring scheme is 
identical to the boundary conditions for the geometric duals of the Wilson lines in ${\cal N}=4$ SYM 
\cite{myWils}. 


\begin{figure}
\begin{center}
\epsfxsize=5in \epsffile{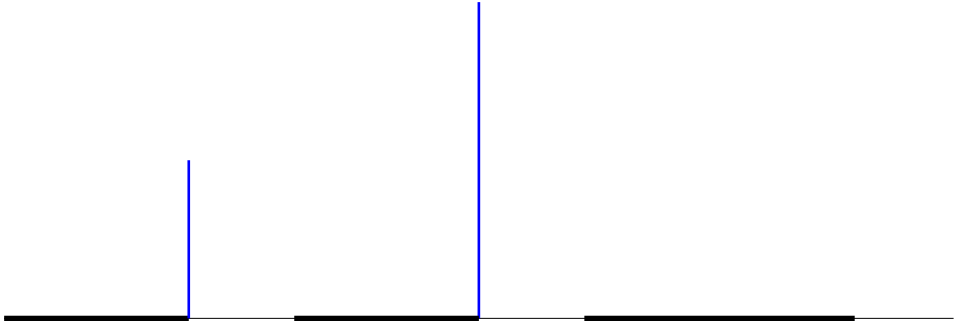}
\end{center}
\caption{Boundary condition for the harmonic function on $c_1=c_2=-1/2$ branch. The normal derivatives 
of $\Phi$ are fixed on the $x$ axis and along the branch cuts (see equations (\ref{y0Regul}), 
(\ref{xCutRegul})).
} \label{FigXyBc}
\end{figure}

While at generic values of $q$ one can only have the boundary conditions (\ref{y0Regul}), more general solutions are possible if $c_1=c_2=-1/2$. We will discuss the special nature of this case and derive a more general set of conditions in section \ref{SubsM2BndrCnd}, and here we just summarize the results. 
On the branch with $c_1=c_2=-1/2$, the harmonic function is allowed to have discontinuities in the upper half plane, but to yield regular geometries such cuts should be vertical and a boundary condition
\bea\label{xCutRegul}
\d_x\Phi|_{x=x_i,y<y_i}=0
\eea
should be satisfied along them. A typical boundary condition for this branch is depicted in figure 
\ref{FigXyBc}. 
In a presence of the cuts the relations (\ref{DefPsi1}), (\ref{DefPsi2}) do not fix $\Psi_1$ and $\Psi_2$ uniquely, and to eliminate extra degrees of freedom one should replace (\ref{DefPsi2}) by
\bea
\Psi_1|_{y=0}=0,\quad \Psi_2|_{x=x_i,y<y_i}=0,\quad \Psi_1|_{x^2+y^2\rightarrow\infty}=0
\eea
It turns out that a harmonic function corresponding to $AdS_4\times S^7$ has one branch cut and more complicated cut structures will be discussed in section \ref{SubsM2BndrCnd}.


The asymptotic behavior of the solution
is determined by boundary conditions at large values of $r=\sqrt{x^2+y^2}$. Such asymptotics also
fix the values of $c_1$, $c_2$, and in the next two sections we consider the two most interesting cases.
Some comments about solutions with general values of $c_1$ and $c_2$ will be given in section 
\ref{SubsCmnts}.

\section{$AdS_4\times S^7$ branch.}
\renewcommand{\theequation}{5.\arabic{equation}}
\setcounter{equation}{0}

\label{SectSugraA4S7}

In the previous section we summarized a general supersymmetric geometry which is invariant under $SO(2,2)\times SO(4)\times SO(4)$. We saw that the solution can be specified in terms of one harmonic function $\Phi$ and two constant $c_1$, $c_2$ which are subject to a constraint (\ref{UnivABC}). 
It turns out that there are two different values of $c_1$ which lead to solutions with interesting asymptotics, 
and here we discuss the case that leads to excitations of $AdS_4\times S^7$. But before we do this, let us explain how to recover $AdS_4\times S^7$ itself.

\subsection{Recovering $AdS_4\times S^7$.}
Since $AdS_4\times S^7$ arises as a near--horizon limit of M2 branes, one expects that to arrive at this geometry one should set $f_2=f_3=0$. We should start with determining the constants 
$c_1$ and $c_2$ for this solution. To do so one should look at various combinations of 
the equations (\ref{Eqnf1})--(\ref{Eqnf3*}),
and we encountered some of them while deriving the solution in the Appendix \ref{SectAppA}. 
In particular, one can see that the system (\ref{Eqnf2})--(\ref{Eqnf3*}) is equivalent to four equations (\ref{OldFluxes1}), (\ref{OldFluxes2}), (\ref{NewFluxes1}), (\ref{NewFluxes2}).
Setting $f_2=f_3=0$ in (\ref{OldFluxes1}), (\ref{OldFluxes2}), we arrive at two relations:
\bea\label{SpclEqn47}
&&e^{-A-B}df_0+2c_1e^C*d(B-C)=0\\
&&e^{-A-C}df_0+2c_2e^B*d(B-C)=0\nonumber
\eea
If we assume that $B-C$ is not a constant, then the consistency of these equations requires $c_2=c_1$, and combining this fact with relation (\ref{UnivABC}), we find that  $AdS_4\times S^7$ has
\bea\label{AdS4S7c}
c_1=-\frac{1}{2},\quad c_2=-\frac{1}{2},\quad q=-\frac{1}{2}
\eea
Since $c_1$ and $c_2$ are constants, they should take the above values for any solution which asymptotes to $AdS_4\times S^7$ and from now on the solutions with (\ref{AdS4S7c}) will be called 
"$AdS_4\times S^7$ branch". Notice that if we assume the relations (\ref{AdS4S7c}), then equations 
(\ref{OldFluxes1}), (\ref{OldFluxes2}) would imply that $f_2$ vanishes if and only if $f_3=0$.

\begin{figure}
\begin{center}
\begin{tabular}{ccc}
\epsfysize=1in \epsffile{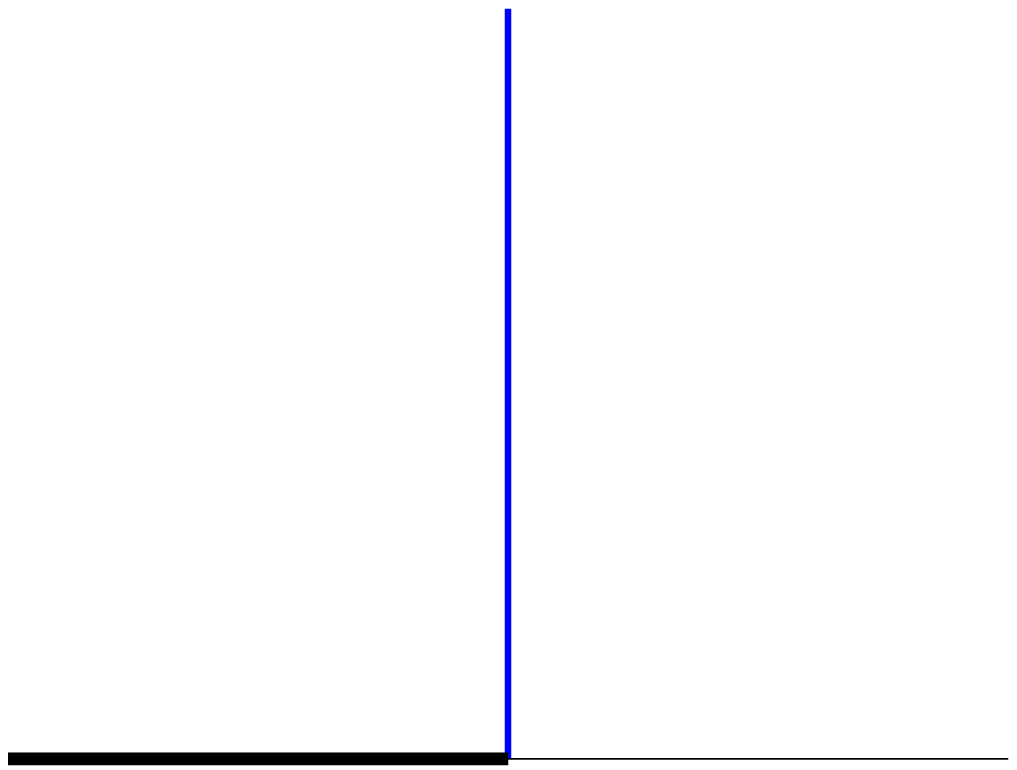}&{\hskip 1in}&
\epsfysize=1in \epsffile{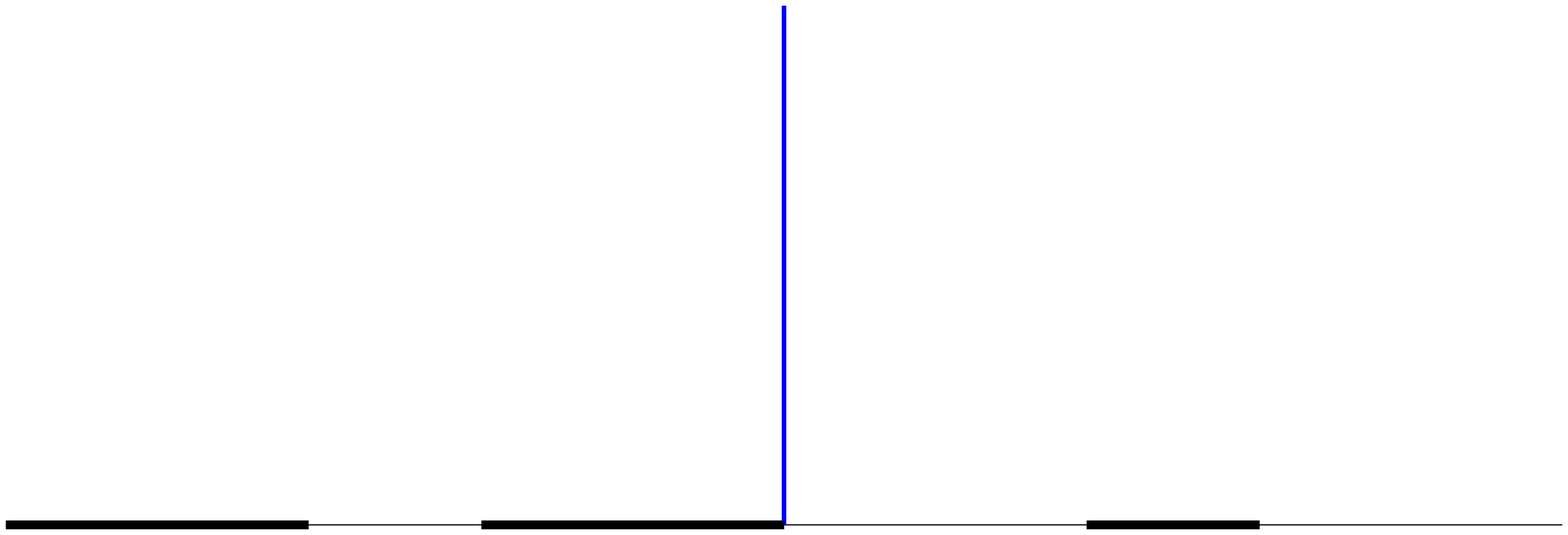}\\
\ &\ &\\
(a)&&(b)
\end{tabular}
\end{center}
\caption{
(a) Boundary conditions for the harmonic function (\ref{HarmA4S7}) describing $AdS_4\times S^7$.\newline
(b) Boundary conditions corresponding to a typical excitation (\ref{A4S7GenHarm}) of 
$AdS_4\times S^7$.
} \label{FigA4S7}
\end{figure}

Next we look at the relations (\ref{NewFluxes1}) and (\ref{NewFluxes2}). In the present case they become
\bea\label{SpclEqn47a}
e^{C-3A}*df_1=-6de^B,\qquad e^{B-3A}*df_1=6de^C
\eea
In particular we conclude that $e^{2B}+e^{2C}\equiv 4L^2$ must be a constant. Let us introduce a coordinate $\theta$:
\bea\label{Theta47}
e^B=2L\cos\theta,\qquad e^C=2L\sin\theta
\eea
One can use the equation (\ref{Eqnf1}) to eliminate the flux $f_1$ from (\ref{SpclEqn47a}):
\bea
de^B=\frac{1}{12}e^{B+2C-2F}d(B-C)-\frac{1}{6}e^{C-3A}*de^{4A-F}
\eea
If one substitutes the expressions for the $e^B$ and $e^C$ in terms of $\theta$, the last equation 
becomes
\bea\label{Temp3M16}
\left[1-\frac{L^2}{3}e^{-2F}\right]d\theta=\frac{1}{6}e^{-3A}*de^{4A-F}
\eea
We should also recall that $A$ and $F$ are not independent: they are related by (\ref{EqnMetr}), 
which in the present context means that $e^{2A}-e^{2F}=L^2$. Then it seems natural to define a 
coordinate $\rho$: $e^A=L\cosh\rho$, $e^F=L\sinh\rho$ and rewrite equation (\ref{Temp3M16}) 
in terms of it:
\bea\label{M16Dual}
2d\theta=*d\rho
\eea
At this point we know the warp factors and fluxes in terms of $\rho$, $\theta$, but the metric in the 
$(\rho,\theta)$ subspace is still undetermined. The simplest way to find it is to use the coordinates $(x,y)$. By definition,
\bea
y=4L^3 \cosh\rho\sin\theta\cos\theta,\quad dx=*dy \nonumber
\eea
then using the duality relation (\ref{M16Dual}), we find $x=-2L^3\sinh\rho\cos 2\theta$. 
Substituting this into (\ref{EqnMetr}), (\ref{InAns}), we recover the $AdS_4\times S^7$ geometry. Moreover, the relations (\ref{SolnThrPhi}) allow us to extract the harmonic function which corresponds to this space:
\bea\label{HarmA4S7}
\Phi_0&=&L^3\left[4(\rho\sinh\rho-\cosh\rho)\cos 2\theta+
8\cosh\rho\sin{2\theta}(\frac{\pi}{2}-\theta)\right]
\nonumber\\
&\sim&
-2\left[x\left(\log\frac{\sqrt{x^2+y^2}}{2L^3}-1\right)+y\arctan\frac{x}{y}\right]_{r\rightarrow\infty}\\
\d_y\Phi_0&=&4\left(\frac{\pi}{2}-\theta\right),\quad 
\d_x\Phi_0=-2\rho\nonumber
\eea
The boundary condition for this function is depicted in figure \ref{FigA4S7}a. One can see that the warp factors $e^B$ and $e^C$ jump at $(x,y)=0$ ($e^B=2L,e^C=0$ to the left of this point and 
$e^B=0,e^C=2L$ to 
the right) and along a rod $x=0,y<2L^3$. This branch cut is shown in figure \ref{FigA4S7}a. 


To describe a solution which asymptotes to $AdS_4\times S^7$, a harmonic function should approach 
$\Phi_0$ at large values of $\sqrt{x^2+y^2}$. 
In particular this implies that for the geometries on the $AdS_4\times S^7$ branch the $y=0$ line should be dark for large negative values of $x$ and it should be light for large positive values, 
a typical coloring is shown in figure 
\ref{FigA4S7}b. One can easily write the harmonic function corresponding to such 
picture\footnote{To simplify this expression we assumed that the regions change from dark to light 
at $x=0$ and 
$x_m^+x_m^->0$, $x_m^+>x_m^-$. We also assumed that the warp factors $e^B$, $e^C$ do not jump 
at nonzero values of $x$ and that the set of transition points is symmetric under $x\rightarrow -x$ (otherwise $\d_x\Phi\ne 0$  along the cut). A broader set of solutions will be discussed in section 
\ref{SubsM2BndrCnd}, in particular the most general solution with one cut is given by 
(\ref{HarmOneCut}).}: 
\bea\label{A4S7GenHarm}
\Phi&=&\Phi_0-\sum\theta(x^+_m)\left[-2(x-\xi)+2y\arctan\frac{x-\xi}{y}+(x-\xi)
\log[(x-\xi)^2+y^2]
\right]_{x^-_{m}}^{x^+_{m}}\nonumber\\
\d_y\Phi&=&\d_y\Phi_0-2\sum \theta(x^+_m)
\left(\arctan\frac{x - x^+_{m}}{y} -\arctan\frac{x - x^-_{m}}{y}\right)\\
\d_x\Phi&=&\d_x\Phi_0-\sum \theta(x^+_m)\log\frac{(x-x^+_{m})^2+y^2}{(x-x^-_{m})^2+y^2}
\nonumber
\eea
Unfortunately to extract a geometry corresponding to this data, one still has to solve some differential equations and we were not able to find an explicit map from $\Phi$ to the metric. However one can use a perturbative construction to show an existence and uniqueness of such map.

\subsection{Perturbation theory.}

\label{SubsA4S7Pert}

Let us now look at small perturbations around $AdS_4\times S^7$. We begin with writing the underlying harmonic function $\Phi$ as
\bea
\Phi\equiv \Phi_0+\Phi_1
\eea
where $\Phi_0$ corresponds to the $AdS_4\times S^7$ space and $\Phi_1$ is viewed as a perturbation. To make the perturbative expansion slightly more explicit, we introduce a small parameter $\eps$ (and we will take $\eps=1$ in the end\footnote{
Notice that since the solution approaches $AdS_4\times S^7$ at large values of $x$ and $y$, the real expansion parameter is $\frac{\eps L}{\sqrt{x^2+y^2}}$, so even for $\eps=1$ we expect to have a convergent series at least at large values of $r=\sqrt{x^2+y^2}$.}):
\bea
\Phi= \Phi_0+\eps\Phi_1
\eea
Since the fluxes are completely specified in terms of the warp factors by 
(\ref{Eqnf1})--(\ref{Eqnf3}), it seems sufficient to introduce the perturbative expansions
\bea
e^{B-C}=\cot\theta+\sum_1^\infty \eps^n G^{(n)},\quad e^{F-A}=\tanh\rho+
\sum_1^\infty \eps^n H^{(n)}
\eea
and try to use the equations (\ref{Eqnf1})--(\ref{SolnThrPhi}) to determine $G^{(n)}$, $H^{(n)}$ 
in terms of $\Phi^{(1)}$. To make the computations a little more transparent, we introduce one more expansion 
\bea\label{Expf2}
f_2=\sum_{n=1}^{\infty}\eps^n f_2^{(n)}
\eea
even though its coefficients are completely determined by $G^{(n)}$ and $H^{(n)}$ through equations 
(\ref{Eqnf1}), (\ref{Eqnf2}). Once the expansion for $f_2$ is known, the equation (\ref{DefPsi1}) allows 
one to determine $\Psi_1$ and $\Psi_2$. To see this we notice that on the $AdS_4\times S^7$ branch 
(i.e. for $c_1$ and $c_2$ given by (\ref{AdS4S7c})), one can combine 
(\ref{Eqnf2})--(\ref{Eqnf3*}) to obtain the equation relating $f_2$ and $f_3$:
\bea
e^{F-3B+C}*df_2+e^{A-2B}df_2=
-e^{F+B-3C}*df_3+e^{A-2C}df_3\nonumber
\eea
Combining this equation with its dual and using the relation
\bea
\frac{e^{2F+2B}+e^{2A+2C}}{e^{2B}+e^{2C}}=-\frac{1}{4}e^{2B}+e^{2A},
\eea
we express the differential of $f_3$ in terms of $f_2$:
\bea
e^{-4C}*df_3=\frac{1}{e^{2A}-\frac{e^{2B}}{4}}\left[
-e^{F-3B-C+A}df_2+\frac{1}{4}e^{-2B}*df_2\right]
\eea
This allows one to eliminate $f_3$ from (\ref{DefPsi1}):
\bea\label{Inter20Mr}
d\Psi_2+*d\Psi_1=\frac{2e^{F+B+C-4B}}{(e^{2C}+e^{2B})(e^{2A}-\frac{e^{2B}}{4})}
\left[-e^{F+B-C+A}df_2-(e^{2A}-\frac{e^{2B}}{2})*df_2\right]
\eea
Let us consider perturbative expansion of the last equation and look at the coefficient in front of 
$\eps^n$. Since the expansion (\ref{Expf2}) does not have contribution at zeroes order in $\eps$,
the right hand side of the last relation contains $G^{(k)}$ and $H^{(k)}$ for $k<n$ and since we are building the solution by induction, we assume that they are known (to begin the induction one also needs 
$G^{(0)}$ and $H^{(0)}$ which come from $AdS_4\times S^7$). The functions $f_2^{(k)}$ with $k<n$ are 
known as well, so the only undetermined terms in the right hand side of the last 
equation are the ones containing $f_2^{(n)}$. Differentiating (\ref{Inter20Mr}), we obtain the Poisson equation for $\Psi^{(n)}_1$ and it has a unique solution satisfying the boundary conditions (\ref{DefPsi2}). Plugging this solution back into 
(\ref{Inter20Mr}), one finds a unique expression for $\Psi^{(n)}_2$. We should stress that at this stage 
both $\Psi^{(n)}_1$ and $\Psi^{(n)}_2$ contain some integrals of the unknown $f_2^{(n)}$, however 
there is a unique linear map
\bea
f_2^{(n)}\quad\rightarrow \quad \Psi^{(n)}_1,\Psi^{(n)}_2
\eea
Substituting this result into (\ref{SolnThrPhi}), we find the unique expressions for $G^{(n)}$, 
$H^{(n)}$ in terms of $f_2^{(n)}$ and the solution in the previous orders. Then the relations
(\ref{Eqnf1}), (\ref{Eqnf0}) lead to the linear equations for $f_2^{(n)}$ which have a unique solution. At this point one completely determines the solution at the $n$--th order, and the entire series in constructed by induction.

We have outlined the procedure which allows one to start with any function $\Phi$ which asymptotes to $\Phi_0$ of the form (\ref{HarmA4S7}) and to construct the unique 1/2--BPS 
solution as a perturbative series in $\Phi-\Phi_0$. While the explicit realization of this construction might not be practical, the above construction guarantees that any harmonic function $\Phi$ with correct asymptotics leads to the unique solution. Combining this with the argument for regularity given in section \ref{SectGenSolut}, we conclude that any harmonic function $\Phi$ satisfying the boundary conditions 
(\ref{y0Regul}) and 
approaching (\ref{HarmA4S7}) at large values of $r$, leads to the unique regular supersymmetric 
solution of eleven dimensional supergravity. This statement is an M theory counterpart of the 
type IIB result derived in \cite{myWils}.


\subsection{Topology, fluxes and brane probes.}

\label{SubsTopolA4S7}

In the previous subsection we demonstrated that any harmonic function $\Phi$ with boundary conditions depicted in figure \ref{FigA4S7}b leads to the unique geometry with $AdS_4\times S^7$ asymptotics. By construction, the supergravity solutions described in this paper have no sources, so the space does not contain branes. However when the size of a dark or a light region becomes small, the supergravity description breaks down in the vicinity of such defect and a better semiclassical description is given in terms of brane probes. This situation has been encountered in the other examples of source--free BPS geometries as well 
\cite{LLM,myWils}. Here we will identify the relevant branes and relate the supergravity data with probe analysis of section \ref{SectA4S7}. 

If the sources become weak and a brane probe approximation takes over, the geometry is well--approximated by $AdS_4\times S^7$ everywhere except for the location of the branes where the space becomes singular. However the branes can still be detected by looking at the excited fluxes: $F_4$ of 
$AdS_4\times S^7$ gets small corrections (but their backreaction into the metric can be neglected away from the branes). To be able to support such fluxes, the geometry has to have non--contractible cycles, so we begin with analyzing topology of the solutions.


\begin{figure}
\begin{center}
\epsfxsize=5in \epsffile{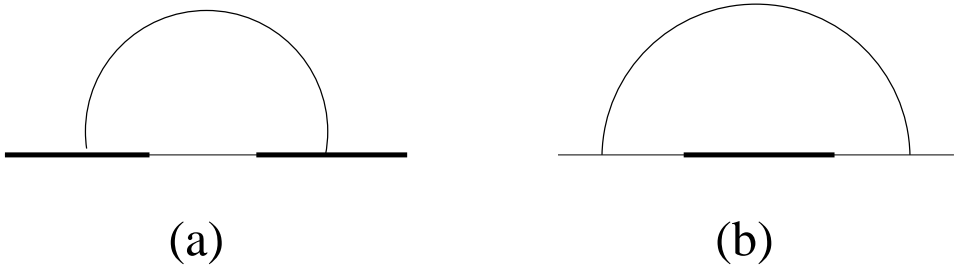}
\end{center}
\caption{
Two types of non--contractible four--spheres.
} \label{FigTop3}
\end{figure}


Let us start with a regular solution corresponding to a generic boundary condition depicted in figure 
\ref{FigA4S7}b. In the $(x,y)$ plane one can take an open contour which begins and ends in the dark 
regions of $y=0$ line and which goes through positive values of $y$ in the middle (see figure 
\ref{FigTop3}a). Restricting the metric (\ref{InAns}) to a four dimensional space composed of this contour and ${\tilde S}^3$, one finds a closed four dimensional manifold with topology of $S^4$ (similar "bubbles" 
have been encountered in 
\cite{LLM,myWils}). If there is a light region between the ends of the contour (as in figure \ref{FigTop3}a), 
then the resulting four--manifold is non--contractible. One can also show that the integral of $F_4$ over this manifold is non--zero, so a light region of finite extent should be identified with a stack of M5 branes 
wrapping $AdS_3\times{S}^3$. Similar non--contractible sphere can be constructed by combining a contour presented in figure \ref{FigTop3}b and $S^3$ and one concludes that a dark region of finite 
extent should be identified with M5 branes wrapping $AdS_3\times{\tilde S}^3$. 
We conclude that a generic geometry has a set of non--contractible four--manifolds with topology of 
$S^4$ and the distribution of these manifolds can be easily identified by looking at the coloring of 
$y=0$ line.

\begin{figure}
\begin{center}
\epsfxsize=4in \epsffile{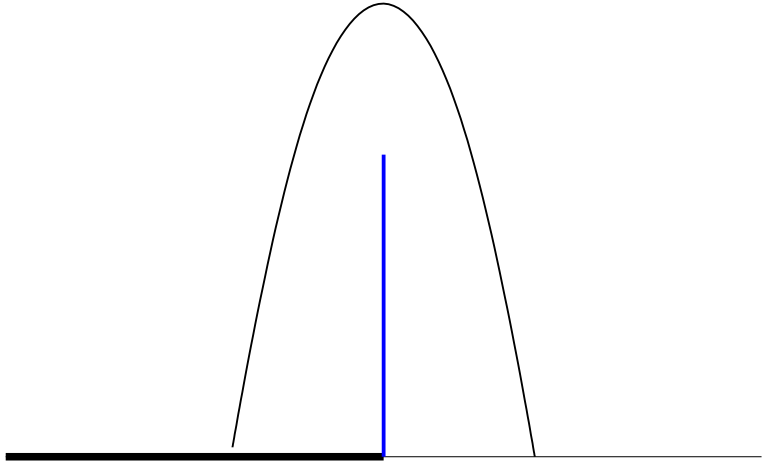}
\end{center}
\caption{
Construction of a non--contractible seven--manifold: one has to fiber both three--spheres over the 
one--dimensional contour.
} \label{FigTop7}
\end{figure}

This still leaves a question: how do we see the flux produced by the membranes? To extract an M2 brane charge one needs a non--contractible seven--manifold and it turns out that all geometries described in this section contain only one such manifold. This fact is consistent with analysis of section \ref{SectA4S7} 
where we saw that the probe M2 branes can always be viewed as fluxes on the worldvolume of M5 branes. There is only one exception from this rule: the branes which were used to produce the original 
$AdS_4\times S^7$. We will now see that these branes are very different from the rest and even in the generic distribution depicted in figure \ref{FigA4S7}b one can identify such "seed branes".  To read off a 
membrane charge one needs a noncontractible seven--manifold and for the geometries with 
$SO(2,2)\times SO(4)^2$ symmetry a natural candidate is a product of two 3--spheres and a contour in 
$(x,y)$ plane. To produce a closed manifold the contour should begin in the dark region and end in the light one (an example of such contour is depicted in figure \ref{FigTop7}). This procedure would produce 
a closed seven--manifold for every point on the $y=0$ line where the coloring changes from dark to light or vice versa. Let us show that for the solutions described by the harmonic function (\ref{A4S7GenHarm}) most of these 7--manifolds have trivial topology. 


Starting from a contour in figure \ref{FigTop7}, one can move its ends close to the transition point. As one approaches this point from the light region, the warp factor of ${\tilde S}^3$ can behave in two 
different ways: it either goes to zero or saturates to a finite value. In the first case both $e^B$ and $e^C$ go to zero at the transition point, so $e^F\rightarrow e^A$, then from (\ref{SolnThrPhi}) one concludes that 
$\d_x\Phi$ goes to infinity. Since both spheres collapse at the transition point, the ends of the contour can be moved from light to dark region and the seven manifold is contractible. The other possible scenario involves jumps in {\it both} warp factors:
\bea\label{BndrJump}
e^B|_{dark}\rightarrow R>0,\quad e^C|_{light}\rightarrow {\tilde R}>0, \quad
e^C|_{dark}=e^B|_{light}=0
\eea
In this situation the contour cannot be moved through the transition point, so one finds a non--contractible 
7--manifold. Notice that the jump described by (\ref{BndrJump}) leads to a finite value of $\d_x\Phi$.
We conclude that there are two different types of the transition points: the ones with finite $\d_x\Phi$ lead 
to a non--contractible seven--sphere associated with each point, and the ones with infinite $\d_x\Phi$ do not. If warp factors do jump at $(x,y)=(x^{(0)},0)$, by continuity they should also jump somewhere in the vicinity of this point. It turns out that the jumps always happen on a vertical rod of finite length (we already saw this in $AdS_4\times S^7$ example, and general case will be analyzed in the next subsection). If such branch cut is present, it should not be crossed by the contour which was used to construct the seven--manifold (otherwise this manifold will become singular at the intersection point), so to build a 
non--contractible sphere one should use a contour depicted in figure \ref{FigTop7}. 

To summarize, we showed that any finite dark (light) region leads to a non--contractible 4--manifold by taking a contour from figure \ref{FigTop3}a (\ref{FigTop3}b) and fibering ${\tilde S}^3$ ($S^3$) over it. 
We also showed that for every transition point with branch cut there exists a non--contractible 
7--manifold which is composed of a contour in figure \ref{FigTop7} and both three--spheres. No 
nontrivial topology is associated with transition points without cuts. Moreover, it is easy to see that 
$F_4$ has a non--vanishing flux over any nontrivial 4--manifold and the same is true about $*F_4$ 
and seven--manifolds. 
Thus the topology of a solution and distribution of fluxes are completely encoded in the boundary conditions for the harmonic function. 

After presenting this general analysis, we now specialize to the solutions parameterized by the harmonic function (\ref{A4S7GenHarm}). This function has $2n+1$ transitions points and only one of them ($x=0$)
has a finite value of $\d_x\Phi$. The topology of this solution can be read off from the diagram in figure 
\ref{FigA4S7}b: the geometry has $2n$ different non--contractible four--manifolds and one 
non--contractible seven--manifold. As we already mentioned, the four--manifolds carry non--zero values 
of magnetic flux, and the 7--manifold carries a membrane charge (which is measured by 
the integral of $*F_4$).


\begin{figure}
\begin{center}
\epsfxsize=5in \epsffile{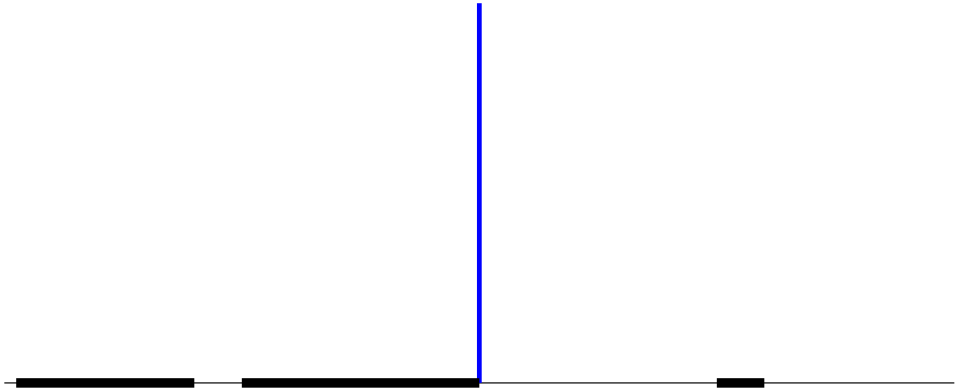}
\end{center}
\caption{
Small perturbations of $AdS_4\times S^7$: the defects should be interpreted as M5 branes.
} \label{FigA4S7Pert}
\end{figure}


Once we understood the general procedure for the extraction of fluxes, it is useful to look at small perturbations of $AdS_4\times S^7$. A typical boundary condition for this case is presented in figure 
\ref{FigA4S7Pert}: the coloring is almost identical to the one for the unperturbed solution, but there are small "defects" in the light and in the dark regions. As we already discussed, a dark defect carries magnetic charge on $S^3$, so it should be identified with M5 brane discussed in section 
\ref{SubsA4M5S}, while the light defect should be identified with its $Z_2$ image. In section 
\ref{SubsA4M5S} the brane was parameterized by its position in AdS space (or by the membrane charge, 
see (\ref{A4S7nmb2})) and in the present case this translates into the position of the defect on $y=0$ line:
\bea
x=-2L^3\sinh\rho
\eea
The map for the $Z_2$ image can be found by flipping the sign of $x$.


\subsection{Generalization: multiple membrane seeds and\\ Schwarz--Christoffel map}


\label{SubsM2BndrCnd}

So far we have been working with solutions described by the harmonic function of the form 
(\ref{A4S7GenHarm}) and typical boundary conditions for this function are shown in figure 
\ref{FigA4S7}b. In particular, one notices that there is only one vertical cut which is produced by 
the $\Phi_0$ piece in 
(\ref{A4S7GenHarm}). In this subsection we will discuss solutions with a more general cut structure (a typical example is presented in figure \ref{FigXyBc}): we will demonstrate that to produce a 
regular solution, 
all cuts should be vertical and a harmonic function should obey Neumann boundary conditions along the cuts. We will also outline the procedure for finding this function. Unfortunately in general case one has to 
invert a complicated holomorphic map, so the expression for the harmonic function will not be very 
explicit. However this situation is typical for solutions of Laplace equation with sophisticated boundary conditions: 
once the appropriate map is found the inversion problem is considered to be "trivial". 

While solutions described by (\ref{A4S7GenHarm}) have only one "seed M2 brane" and the remaining membrane charge is dissolved in M5s, one can naturally interpret a solution with $k$ cuts as a geometry with $k$ membrane seeds. 
The warp factors $e^B$, $e^C$ jump as one crosses a branch cut, and now we will show that despite this discontinuity, the geometry remains regular if the cuts are vertical and $\d_x\Phi=0$ along the cuts. 
In this subsection the analysis will be performed for an arbitrary value of $q$ which appeared in 
(\ref{UnivABC}), but in the end we will see that a nontrivial cut structure is only possible for $q=-1/2$.

Let us go back to the definitions of $\Psi_1$ and $\Psi_2$. If one assumes that these two functions are smooth in the upper half--plane, then relations (\ref{DefPsi1}) and (\ref{DefPsi2}) fix them uniquely.  However this is no longer true in the presence of branch cuts, to remove extra "gauge degrees of 
freedom" we add boundary conditions on the cuts:
\bea
\Psi_2|_{cut}=0
\eea 
This relation along with (\ref{DefPsi1}) and (\ref{DefPsi2}) leads to the unique expressions for 
$\Psi_1$, $\Psi_2$. Then restricting (\ref{SolnThrPhi}) to the branch cut, we find 
\bea\label{PhiCut}
\left[\log\frac{e^A-e^F}{e^A+e^F}-\d_x\Phi\right]_{cut}=0
\eea
Using $AdS_4\times S^7$ solution as a guide, we require both terms in this expression to be continuous in the vicinity of the cut, while the values of 
\bea
-\frac{\pi}{4}{\mbox{sign}}(c_1c_2)+\arctan\frac{c_2e^B}{c_1e^C}\nonumber
\eea
should differ by sign on the opposite sides of the cut. The combination of these two requirements leads to 
the prescription for crossing the cut:
\bea\label{FlipCut}
c_2e^B\leftrightarrow c_1e^C
\eea
Let us look at the first equation in (\ref{Eqnf1}): the terms without star remain invariant under the flip 
(\ref{FlipCut}), while the differential  $d(B-C)$ changes sign. We conclude that for regularity, 
the vector $d(B-C)$ must be pointing along the cut, while both $df_1$ and $de^{4A-F}$ must be transverse to it\footnote{To rule out the opposite arrangement, we notice that (\ref{Eqnf1}) implies a discontinuity of $*d(B-C)$.}. The discontinuity of both terms in the right hand side of the second 
equation in 
(\ref{Eqnf1}) implies that $df_0$ should also be transverse to the cut. To find the transformations of 
$df_2$ and $df_3$, 
it is convenient to construct combinations of (\ref{Eqnf2})--(\ref{Eqnf3*}) which do not contain $df_0$ 
(see (\ref{NewFluxes1}), (\ref{NewFluxes2})):
\bea
&&-6c_2de^B-2e^{F-3B}df_2-c_1e^{C-3A}*df_1-e^{A-3C}*df_3=0\nonumber\\
&&-6c_1de^C-2e^{F-3C}df_3+c_2e^{B-3A}*df_1+e^{A-3B}*df_2=0\nonumber
\eea
These relations should be interchanged by (\ref{FlipCut}), then one finds the following behavior under the flip:
\bea\label{FlipCut1}
c_1^3df_3\leftrightarrow c_2^3df_2
\eea
Then application of the flip to (\ref{Eqnf2}) gives a relation similar to (\ref{Eqnf3}), but the coefficient in front of $df_0$ has an extra factor of $c_2/c_1$. Since the crossing conditions (\ref{FlipCut}), 
(\ref{FlipCut1}) should arise from a symmetry of equations, we conclude that this type of branch cuts is only possible if $c_1=c_2$.

Once we established that $c_1=c_2=-1/2$ (see (\ref{UnivABC})), it is convenient to rewrite (\ref{Eqnf0}):
\bea
df_0=-2e^Ag_y^2\left[-e^{F+B+C}dy-\frac{1}{2}e^A(e^{2C}-e^{2B})dx\right]
\eea
As already mentioned, the left hand side of this relation is transverse to the cut, so it should flip sign under 
(\ref{FlipCut}). This leads to the conclusion that $e^F$ should vanish along the cut and $dx$ should point in the transverse direction. In other words, we showed that the cuts must be vertical and $\d_x\Phi$ 
should vanish along them (see equation (\ref{PhiCut})).

Let us summarize what we learned so far. We started with an assumption that $(x,y)$ plane has some 
cuts where the warp factors $e^B$ and $e^C$ are allowed to jump, but the geometry remains regular. We showed that this can only happen if $c_1=c_2=-1/2$ and 
\bea
\d_x\Phi|_{cut}=0
\eea
Let us now demonstrate that these conditions are also sufficient for ensuring the regularity of the solution. 

Since $e^F$ vanishes along the cut, we can parameterize the warp factors in terms of $e^A$ and an angle $\theta$:
\bea
e^B=2e^A\cos\theta,\quad e^C=2e^A\sin\theta:\quad y=2e^{3A}\sin 2\theta,\quad 
g_y^{-1}=2e^{2A}\cos 2\theta\nonumber
\eea
We observe that the reflection (\ref{FlipCut}) translates into $\theta\rightarrow\frac{\pi}{2}-\theta$, so $g_y^2$ remains invariant under it. To prove regularity we only need to analyze the metric in 
$(x,y,S^3,{\tilde S}^3)$ subspace:
\bea\label{JumpRegul}
ds_{x,y,S,{\tilde S}}^2=4e^{2A}\left[\cos^2\theta d\Omega_3^2+\sin^2\theta d{\tilde\Omega}_3^2+
d\theta^2\right]+9\tan^2 2\theta (de^A)^2+\frac{(e^{-2A}dx)^2}{4\cos^2\theta}
\eea
The term is the square brackets is a metric of seven dimensional sphere and it is smooth in the vicinity of the cut. The prefactors in the last two term ($\tan^2 2\theta$ and $\cos^{-2}\theta$) are invariant under $\theta\rightarrow\frac{\pi}{2}-\theta$, so they remain finite and continuous in the vicinity of the cut\footnote{We are considering a generic point where $y\ne 0$. A vicinity of the transition point 
where cut is glued to the $y=0$ axis requires a separate discussion, and one can demonstrate that there are no singularities there as well.}. Our previous analysis indicates that $de^A$ points in the transverse direction, i.e. it is proportional to $dx$, so we conclude that the last two terms in (\ref{JumpRegul}) should be combined together and the metric (\ref{JumpRegul}) is regular in the vicinity of the cut. 

We proved that for $c_1=c_2=-1/2$ there is a one--to--one map between regular geometries and 
harmonic functions satisfying Neumann boundary conditions:
\bea\label{ShwCrstBC}
&&\d_y\Phi|_{y=0}=2\pi,\quad x\in (-\infty,x_1)\cup (x_2,x_3)\cup\dots (x_{2p},x_{2p+1}),\nonumber\\
&&\d_y\Phi|_{y=0}=0,\quad x\in (x_1,x_2)\cup (x_3,x_4)\cup\dots (x_{2p+1},\infty),\\
&&\d_x\Phi=0 \ \mbox{at}\ x=x_p,y<y_p,\nonumber\\
&&\Phi\sim -2(x\log r+y\arctan\frac{x}{y}),\quad r=\sqrt{x^2+y^2}\rightarrow\infty\nonumber
\eea
It is convenient to introduce a graphical representation for these boundary conditions and an example of such coloring is depicted in figure \ref{FigXyBc}. If only one of $y_p$ is non--zero and distribution of 
transition points is symmetric, then $\Phi$ is given by (\ref{A4S7GenHarm}). Let us now discuss a construction 
of solutions with multiple cuts.

\begin{figure}
\begin{center}
\epsfxsize=5in \epsffile{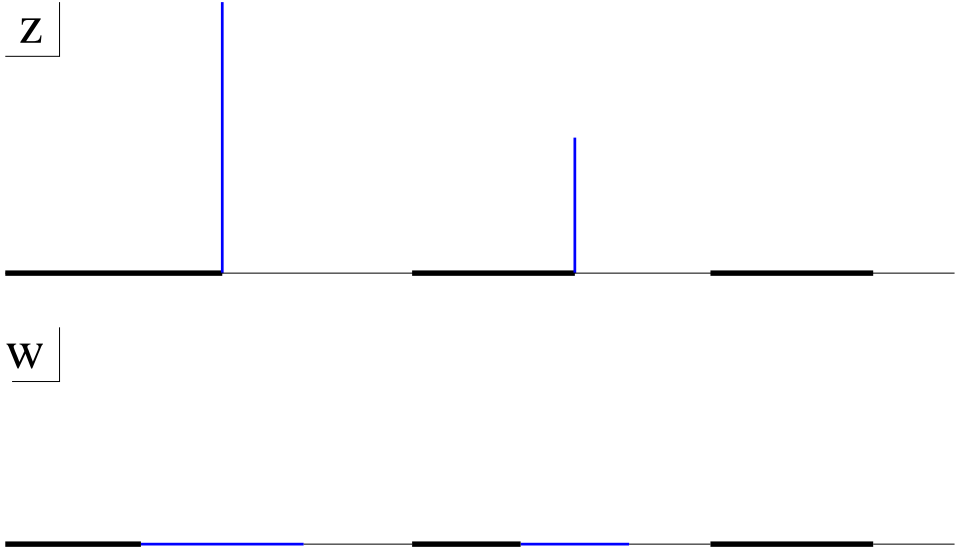}
\end{center}
\caption{
An illustration of the Schwarz--Christoffel map (\ref{SchwCrstMap}).
} \label{FigSCMap}
\end{figure}

The standard arguments allow us to write $\Phi$ in terms of Green's function:
\bea
\Phi(x,y)=\oint d^2\sigma\frac{\d\Phi}{\d n}G_N=2\pi\sum_p
\int_{x_{2p}}^{x_{2p+1}} d\xi G_N(x,y|\xi,0),\quad
x_0\equiv -\infty
\eea
and the challenge is to find a function $G_N$ which has vanishing normal derivatives on all components of the boundary:
\bea
(\d_x^2+\d_y^2)G_N(x,y|x',y')=-\delta(x-x')\delta(y-y'),\qquad \left.\frac{\d G_N}{\d n'}\right|_{bndry}=0
\eea
This problem has a very simple solution in the upper half plane (i.e. when there are no cuts), and this 
result was used to write down (\ref{A4S7GenHarm}), but there is no algorithmic method for writing a 
Green's function corresponding to a generic boundary condition. Fortunately, we are working in two dimensions, so conformal transformation can be used to map any region into an upper half plane. Moreover, for the configurations described by (\ref{ShwCrstBC}), such transformation is a particular case of a well--known 
Schwarz--Christoffel map: starting with $z=x+iy$, we go to a new variable $w$ by inverting the following function
\bea\label{SchwCrstMap}
z=f(w)=\int^w d\zeta\prod_p\frac{\zeta-w^{(2)}_p}{(\zeta-w^{(1)}_p)^{1/2}(\zeta-w^{(3)}_p)^{1/2}}
\eea 
The values $w^{(a)}_p$ should be chosen in such a way that function $f$ has right turning points in the 
$z$ plane (see figure \ref{FigSCMap} for the illustration of the map). Notice that if the is no cut at a 
transition point 
$x=x_p$, then one should take a limit $w^{(1)}_p=w^{(2)}_p=w^{(3)}_p$. It is easy to write the appropriate 
Green's function in the $w$ plane:
\bea\label{GreenNeum}
G_N(w|w')=\frac{1}{2\pi}\log|(w-w')^2|
\eea
and to translate this into $(x,y)$ variables one needs an inverse of the map (\ref{SchwCrstMap}). 
While this problem is solvable in principle, the computations for multiple cuts are quite involved 
so we will not present the explicit harmonic functions $\Phi(x,y)$ here. 


To summarize, in this subsection we looked for regular solutions with discontinuous warp factors $e^B$, $e^C$. We showed that such solutions are only possible in the $AdS_4\times S^7$ branch (where  $c_1=c_2=-1/2$) and to produce regular geometry the harmonic function should satisfy Neumann boundary conditions (\ref{ShwCrstBC}). We also gave a formal solution for such harmonic function in terms of the inverse of the Schwarz--Christoffel map (\ref{SchwCrstMap}). Once the harmonic function is fixed, we can use the arguments of section \ref{SubsA4S7Pert} to prove that a unique regular solution 
can be recovered from it\footnote{The key point in that construction was an asymptotic behavior of the solution, and even harmonic functions with multiple cuts lead to $AdS_4\times S^7$ asymptotics.}.
In the next subsection we will go back to the solutions with single cut and analyze them in the vicinity of the discontinuity.

\subsection{Structure of solutions with one cut}

Equation (\ref{A4S7GenHarm}) gives a harmonic function with one cut and a very special distribution of
transition points: we assumed that the picture was "antisymmetric" under the reflection of $x$ axis (see figure \ref{FigA4S7}b). We begin this subsection with analyzing the structure of the branch cut in 
(\ref{A4S7GenHarm}), and in contrast to the previous subsection we will work in $(x,y)$ coordinates to show the global picture of the cut. We will conclude by relaxing the symmetry requirement and writing the most general solution with one cut.

We begin with analyzing the differences between (\ref{HarmA4S7}) and behavior of (\ref{A4S7GenHarm}) near 
$x=x_m^+$. We begin with extracting a basic building block $\Phi_{x_m}$ from (\ref{A4S7GenHarm}):
\bea\label{DrvtNoRod}
\d_y\Phi_{x_m}=-2\arctan\frac{x-x_m}{y},\quad \d_x \Phi_{x_m}=-\log\left[(x-x_m)^2+y^2\right]
\eea
Unfortunately the expression (\ref{HarmA4S7}) for $\Phi_0$ is not as explicit: it is written in the parametric form\footnote{We consider the $x$ and $y$ derivatives of $\Phi$ since they look simpler than the entire function. Of course $\Phi$ is uniquely recovered from this data (up to irrelevant constant).}:
\bea
y=2L^3\cosh\rho\sin \zeta,&& x=-2L^3\sinh\rho\cos\zeta\nonumber\\
\d_y\Phi_0=2(\pi-\zeta),&& \d_x\Phi_0=-2\rho
\eea
Straightforward algebraic manipulations lead to the following expressions:
\bea
&&\tan\zeta=-\frac{1}{x}\left[\frac{s-(x^2+a^2-y^2)}{2}\right]^{1/2},\
e^{2\rho}=\frac{y^2+x^2+s}{a^2}+\sqrt{\frac{(y^2+x^2+s)^2}{a^4}-1}\nonumber\\
&&s\equiv\sqrt{(y^2+x^2-a^2)^2+4a^2x^2},\quad a\equiv 2L^3,
\eea
then we can write the derivatives of $\Phi_0$ in terms of $(x,y)$ coordinates:
\bea\label{JumpBlock}
&&\d_y\Phi_0=2\pi+2\arctan\left[\frac{\sqrt{s-(x^2+a^2-y^2)}}{\sqrt{2}x}\right]\nonumber\\
&&\d_x\Phi_0=-\log\left[\frac{y^2+x^2+s}{a^2}+\sqrt{\frac{(y^2+x^2+s)^2}{a^4}-1}\right]
\eea
Sending $a$ to zero, we recover (\ref{DrvtNoRod}) with $x_m=0$ up to constant shifts in $\d_x\Phi$ and 
$\d_y\Phi$. The value of $a$ gives the length of the cut at the transition point $x=0$. 

It is interesting to study the structure of the cut introduced by (\ref{JumpBlock}). One can see that if 
$a\ne 0$, then $\d_x \Phi_0$ is a continuous function, while $\d_y\Phi_0$ can jump at
$x=0$. To see this jump in more detail, we write the expression for $\d_y\Phi_0$ at small values of $x$:
\bea
&&s\sim |y^2-a^2|+\frac{x^2(y^2+a^2)}{|y^2-a^2|}\nonumber\\
y<a:&&\d_y\Phi_{0}\sim \pi-2\arctan\left[x\sqrt{\frac{a^2-y^2}{y^2x^2}}
\right],\nonumber\\
y>a:&&\d_y\Phi_{0}\sim\pi-2\arctan\left[\frac{x}{\sqrt{y^2-a^2}}\right]\nonumber
\eea
We see that as $x$ changes sign, $\d_y\Phi_{0}$ jumps if $y<a$ and it continuously goes through zero if $y>a$. This we have a vertical branch cut which extends from $(x,y)=(0,0)$ to 
$(x,y)=(x,a)$. Since in the vicinity of $x=0$ it is only $\d_y\Phi_{0}$ that can jump, we conclude that the same cut is present in the $y$ derivative of the complete harmonic function $\Phi$. This discontinuity translates into the jumps in the warp factors $e^B$ and $e^C$.

While (\ref{A4S7GenHarm}) gives a simple expression for a harmonic function with one cut, it is not completely general: to provide the condition $\d_x\Phi$ along the cut one has to assume that the distribution of the transition points is symmetric\footnote{Notice that this symmetry implies that dark regions are interchanged with light ones.} under $x\rightarrow -x$.  Let us now use the 
Schwarz--Christoffel map described in the previous subsection to write a completely general expression for the harmonic function with one cut.

The upper half plane with a cut can be mapped into the upper half by a very simple transformation:
\bea
w=\sqrt{z^2+a^2},\quad z=\sqrt{w^2-a^2},
\eea
which is parameterized by a positive real number $a$. The branch cut ($0<\mbox{Im}~z<a$) is mapped into the segment $-a<w<a$. Starting with Green's function 
(\ref{GreenNeum}) in the $w$ plane, we can find one in $z$ coordinates:
\bea
G_N(z|z')=\frac{1}{2\pi}\log|(z-z')^2+a^2|,
\eea
Then assuming that the regions $x\in (x_{2p},x_{2p+1})$ are dark\footnote{To have the right asymptotics we should set $x_0=-\infty$, while all other $x_k$ should remain finite. This leads to divergent integrals in $\Phi(x,y)$ and a divergent constant in $\d_x\Phi$, while $\d_y\Phi$ is finite. We regularize $\d_x\Phi$ by subtracting an infinite constant which is accomplished by dropping the boundary term for $\xi=x_0$ 
(this is indicated by a prime near the summation sign).}, we arrive at the expression for 
the harmonic function
\bea\label{HarmOneCut}
\Phi(x,y)&=&\sum_p \int_{x_{2p}}^{x_{2p+1}}d\xi \log|(x-\xi)^2-y^2+a^2+2i(x-\xi)y|\nonumber\\
&=&\frac{1}{2}\sum_p \int_{x_{2p}}^{x_{2p+1}}d\xi \log\left[\{(x-\xi)^2-y^2+a^2\}^2+4(x-\xi)^2y^2\right]\\
\d_y\Phi&=&\sum_p\left[\arctan\frac{x-\xi}{a-y}-\arctan\frac{x-\xi}{a+y}\right]_{x_{2p}}^{x_{2p+1}}
\nonumber\\
\d_x\Phi&=&\left.-\frac{1}{2}{\sum_p}'
\log\left[\{(x-\xi)^2-y^2+a^2\}^2+4(x-\xi)^2y^2\right]\right|_{x_{2p}}^{x_{2p+1}}\nonumber
\eea
For a symmetric distribution of transition points these expressions reduce to (\ref{A4S7GenHarm}).


\section{$AdS_7\times S^4$ branch}
\renewcommand{\theequation}{6.\arabic{equation}}
\setcounter{equation}{0}

\label{SectA7S4Brc}


In the previous section we set the value of $c_1$ in such a way that the solution approached $AdS_4\times S^7$ at the large distances. We saw that for such solutions the line $y=0$ was dark 
on the far left, it was light on the far right and all "defects" were located at finite values of $x$. Here we will discuss the other interesting class of solutions which asymptote to 
$AdS_7\times S^4$: as we will see it would correspond to the boundary conditions with $y=0$ line being light everywhere with an exception of a finite region. A possibility of more general boundary conditions will be discussed in the subsection \ref{SubsCmnts}. 

\subsection{Recovering $AdS_7\times S^4$.}

We begin with recovering $AdS_7\times S^4$ space. 
Since it arises as a near--horizon limit of M5 branes, one should set\footnote{The alternative solution with $f_1=f_2=0$ can be found by interchanging the spheres.} $f_1=f_3=0$. Let us compare (\ref{Eqnf3}) and (\ref{Eqnf3*})
\bea\label{Temp21M}
&&2e^{-A-C}df_0-2c_2e^B*d(2C+B)=0\\
&&-e^{-A-B}df_0-2c_1e^C*d(B+2C)=0\nonumber
\eea
Assuming that $f_0=-e^{2A+F}$ is not a constant, we arrive at the relation between coefficients $c_1$ and $c_2$:
\bea
c_2+2c_1=0,
\eea
and combining this with (\ref{UnivABC}), we determine $c_1$, $c_2$ for the $AdS_7\times S^4$ 
branch:
\bea\label{c12A7S4}
c_1=1,\quad c_2=-2,\quad q=1
\eea
Notice that this relation along with assumption $f_1=0$ uniquely determines 
$AdS_7\times S^4$ geometry. 

Let us now recall the relation (\ref{Eqnf1}):
\bea
e^{-2A+2F}de^{4A-F}=4e^{A+B+C}*d(B-C)\nonumber
\eea
and combine it with (\ref{Temp21M}) to eliminate a differential of $C$:
\bea\label{Temp21M2}
e^{-4A+F}de^{6A}=6e^{A+B+C}*dB
\eea
In the $AdS_4\times S^7$ case we found that a certain combination of the warp factors was constant and this led to a convenient parameterization (\ref{Theta47}). In the present case to introduce a similar set of coordinates it is convenient to eliminate $*dB$ from the last equation by trading it for some exact form. To do so we again compare the equations (\ref{NewFluxes1}), 
(\ref{NewFluxes2}):
\bea
6de^B=e^{F-3B}df_2,\quad 
6de^C=e^{A-3B}*df_2:\qquad *d e^B=e^{F-A}de^C\nonumber
\eea
Substituting this into (\ref{Temp21M2}), we find that $e^{2A}-e^{2C}\equiv 4L^2$ is a constant, so 
in an analogy with (\ref{Theta47}), it is convenient to introduce a new coordinate $\rho$:
\bea
e^{A}=2L\cosh\rho,\qquad e^{C}=2L\sinh\rho
\eea
Looking at the definition of $e^F$ (\ref{EqnMetr}), we conclude that in the present case
\bea
e^{2F}+4e^{2B}=4L^2,\nonumber
\eea
this suggests a natural parameterization
\bea
e^F=2L\cos\theta,\quad e^B=L\sin\theta
\eea
To establish the relation between $\rho$ and $\theta$ we can use (\ref{Temp21M2}):
\bea\label{M21Dual}
d\rho=\frac{1}{2}*d\theta
\eea
Notice that in comparison with (\ref{M16Dual}) $\rho$ and $\theta$ switched places. At this point we determined all warp factors as functions of $\rho$ and $\theta$, and to complete the construction of the solution we need to rewrite them in terms of $x$ and $y$. The expression for 
the coordinate $y$ follows from the definition (\ref{EqnMetr}): $y=2L^3\sinh 2\rho\sin\theta$ and to determine $x$ we use the relation (\ref{M21Dual}). The result is
\bea\label{A7S4Coord}
x=2L^3\cosh 2\rho\cos\theta,\qquad y=2L^3\sinh 2\rho\sin\theta
\eea
For some applications it might be useful to work in $(\rho,\theta)$ coordinates, we already know the warp factors, so one needs to evaluate the metric (\ref{EqnMetr}):
\bea
h_{ij}dx^idx^j=L^2\left[4d\rho^2+d\theta^2\right]
\eea
This completes the demonstration that the geometry is indeed $AdS_7\times S^4$. 


\begin{figure}
\begin{center}
\epsfxsize=5in \epsffile{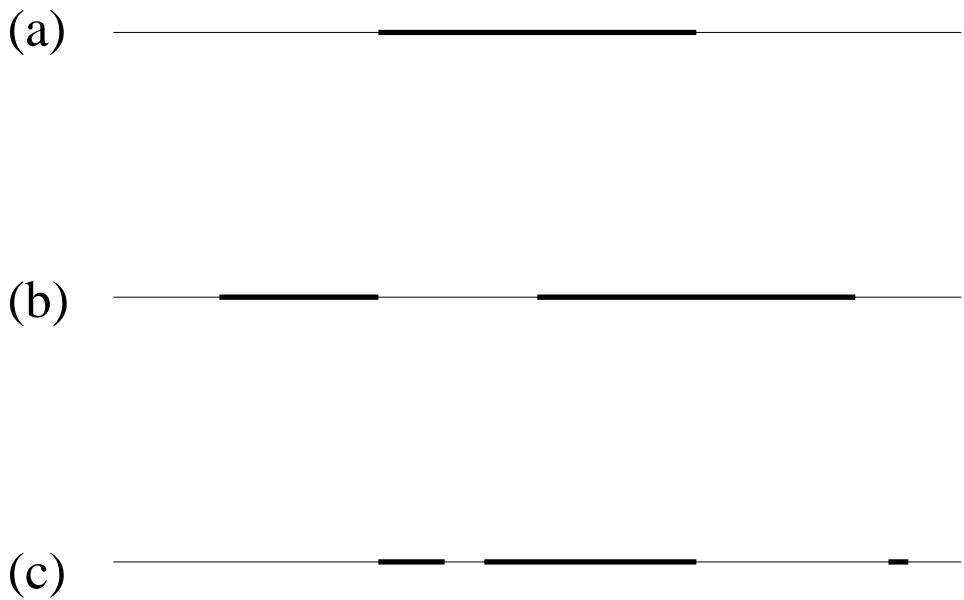}
\end{center}
\caption{
Boundary conditions for the $AdS_7\times S^4$ branch: (a) ground state; (b) typical excitation; 
(c) small perturbation corresponding to probe branes.
} \label{FigA7S4}
\end{figure}


To connect with the general picture we can also extract the harmonic function $\Phi$ which was defined in (\ref{SolnThrPhi}):
\bea\label{HarmA7S4}
\Phi_0&=&\sum\left[2y\arctan\frac{x-\xi}{y}+(x-\xi)
\log[(x-\xi)^2+y^2]
\right]_{\xi=-2L^3}^{\xi=2L^3}
\eea
In particular, it is clear that at $y=0$ one has a light line with some finite dark region and the boundary conditions for this function are depicted in figure \ref{FigA7S4}a. At large values of 
$r=\sqrt{x^2+y^2}$ this function behaves as 
\bea\label{HrmA7S4As}
\Phi\sim 8L^3\log r+c+O(r^{-1}),\quad r\rightarrow \infty
\eea
and any solution which asymptotes to $AdS_7\times S^4$ should obey this boundary condition. 
In particular this implies that for the geometries on the $AdS_7\times S^4$ branch the dark regions 
on the $x$ axis should be bounded, and a typical coloring of $y=0$ is shown in figure \ref{FigA7S4}b. 
One can easily write the harmonic function corresponding to this picture: 
\bea\label{HarmFinSup}
\Phi&=&\sum\left[-2(x-\xi)+2y\arctan\frac{x-\xi}{y}+(x-\xi)
\log[(x-\xi)^2+y^2]
\right]_{x_{2m-1}}^{x_{2m}}\nonumber\\
\d_y\Phi&=&2\sum
\left(\arctan\frac{x - x_{2m}}{y} -\arctan\frac{x - x_{2m-1}}{y}\right)\\
\d_x\Phi&=&\sum \log\frac{(x-x_{2m})^2+y^2}{(x-x_{2m-1})^2+y^2}
\nonumber
\eea
In the next subsection we will use a perturbation theory to show that any such function 
$\Phi$ leads to a unique regular solution.

Notice that since $\d_x\Phi$ diverges at the transition points $x=x_p$, the analysis of section 
\ref{SubsTopolA4S7} implies that the geometry does not have non--contractible seven--manifolds. This agrees with a general statement of section \ref{SubsM2BndrCnd} that such manifolds could exist only 
for $q=-1/2$ (and we are working with $q=1$).

To summarize, we showed that if the geometry belongs to the branch defined by (\ref{c12A7S4}) and at least one of the fluxes $f_1$, $f_3$ vanishes, then the solution is $AdS_7\times S^4$. We also extracted the harmonic function (\ref{HarmA7S4}) corresponding to this geometry and a relation to the more standard coordinates (\ref{A7S4Coord}). 

\subsection{Perturbation theory}

Let us discuss the excitations of $AdS_7\times S^4$. If we assume that the boundary conditions for the harmonic function $\Phi$ are such that $\d_y\Phi(x,y)|_{y=0}=0$ for sufficiently large $|x|$, then the solution asymptotes to $AdS_7\times S^4$ and one can construct it using a perturbation theory around that space. The small parameter controlling the series is
$L/r$ and one expect a convergence for the large values of $r=\sqrt{x^2+y^2}$. On the physical grounds it appears that the series should converge everywhere, but we will not attempt to prove this rigorously. As in section \ref{SubsA4S7Pert} we will just outline the construction of perturbation series and show that for 
any harmonic function $\Phi$ the $n$--th term in the series is uniquely defined. Our goal would be to demonstrate that $\Phi$ completely specifies the solution rather than to find the explicit geometries. 

As in section \ref{SubsA4S7Pert} we introduce a perturbation parameter $\eps$ and write the harmonic 
function $\Phi$ as 
\bea
\Phi=\Phi_0+\eps\Phi_1
\eea
but now $\Phi_0$ corresponds to $AdS_7\times S^4$ with appropriate radius (which should be chosen by requiring that $\Phi_1$ goes to zero at large $r=\sqrt{x^2+y^2}$). Next we introduce the expansions for the warp factors:
\bea\label{Tmp22}
A=\log(2L\cosh\rho)+\sum_{n=1}^\infty\eps^n A^{(n)},\quad 
B=\log(L\sin\theta)+\sum_{n=1}^\infty\eps^n B^{(n)}
\eea
$$
C=\log(2L\sinh\rho)+\sum_{n=1}^\infty\eps^n C^{(n)}
$$
and for the function $f_0$:
\bea\label{Tmp22a}
f_0=-(2L)^3\cosh^2\rho\sinh\rho+\sum_{n=1}^\infty \eps^n f_0^{(n)}
\eea
The zeroth order in perturbation theory is already written down in (\ref{Tmp22}), (\ref{Tmp22a}), 
to perform the induction we assume that all orders up to $n-1$--th are known. Then equation 
(\ref{Eqnf0}) allows one to find {\it algebraic} expressions for $G^{(n)}$ and $H^{(n)}$ in terms of 
$f_0^{(n)}$ and contributions from the previous orders. Plugging the result into (\ref{Eqnf1}), 
we find 
the expression for $df_1^{(n)}$ in terms of $f_0^{(n)}$ and its derivatives, then (\ref{Eqnf2*}) and 
(\ref{Eqnf3*}) give $*df_2^{(n)}$ and $*df_3^{(n)}$. At this point we know the left hand side of 
(\ref{DefPsi1}) in terms of $f_0^{(n)}$ and contributions from the lower orders, it is important that the resulting expression is purely algebraic in $f_0^{(n)}$ and its derivatives. Now one needs to treat (\ref{DefPsi1}) as a differential equation for $\Psi^{(n)}_1$, $\Psi^{(n)}_2$ and to solve it we first act on both sides by the exterior derivative to produce a Poisson equation for $\Psi^{(n)}_1$. This equation has a unique solution satisfying the boundary conditions (\ref{DefPsi2}), then (\ref{DefPsi1}) can be solved for $\Psi^{(n)}_2$\footnote{Notice that $\Psi^{(0)}_2$ goes to zero at large $r$, this implies that the same is true for the derivatives of $\Psi^{(n)}_2$. Then we can choose an integration constant by requiring $\Psi^{(n)}_2|_{r\rightarrow\infty}\rightarrow 0$. This determines the solution uniquely.}. At this point we have one unknown function $f_0^{(n)}$ and everything else is uniquely expressed in terms of it either algebraically of by solving Poisson equation:
\bea
f_0^{(n)}\quad\rightarrow\quad
\begin{array}l
A^{(n)},B^{(n)},C^{(n)},df_1^{(n)},df_2^{(n)},df_3^{(n)}\ \mbox{(algebraic expressions)}\\
\Psi^{(n)}_1,\ \Psi^{(n)}_2\mbox{(integral expressions)}
\end{array}\nonumber
\eea
To determine the function $f_0^{(n)}$ we should use the equations (\ref{SolnThrPhi}). At each order one gets linear integro--differential equations and the solution is unique. The different solutions of the entire system are parameterized by different "seeds" $\Phi_1$, and in turn they
are specified by the boundary conditions.

The perturbation theory which was described above can be applied to a solution with arbitrary asymptotics. To carry out the outlined procedure one has to start with $\Phi^{(0)}$ which corresponds to a {\it know nonlinear solution} and look at harmonic functions $\Phi$ which approach $\Phi^{(0)}$ at large values of $r=\sqrt{x^2+y^2}$. Then writing 
\bea
\Phi_\eps=\Phi^{(0)}+\eps(\Phi-\Phi^{(0)}):\qquad \Phi_1=\Phi
\eea
we can perform a perturbative expansion and it will converge for large values of $r$ even if 
$\eps=1$. In particular the construction described above would work for the geometries with $AdS_4\times S^7$ asymptotics, but in this case the alternative approach discussed in section 
\ref{SubsA4S7Pert} was somewhat simpler, moreover that expansion was a direct analog of 
perturbations around $AdS_5\times S^5$ which was constructed in \cite{myWils}. 

To summarize, we have shown that if the harmonic function satisfies the boundary conditions 
(\ref{y0Regul}), (\ref{HrmA7S4As}) with a compact dark region (see figure \ref{FigA7S4}b), 
then this function is given by (\ref{HarmFinSup}), the corresponding solution can be constructed 
as a perturbation theory around 
$AdS_7\times S^4$ and this procedure yields a unique regular geometry. In the next subsection we will discuss the interpretation of such solution in terms of branes.

\subsection{Relation to the brane probe analysis.}

Let us consider a harmonic function $\Phi$ which leads to a solution with $AdS_7\times S^4$ asymptotics. A typical boundary condition for such function is depicted in figure \ref{FigA7S4}b. Although the 
supergravity solutions have no sources, we expect that a good effective description of a 
small light or dark region is given by probe branes, and here we will identify these objects.

As in section \ref{SubsTopolA4S7} we begin with analyzing the fluxes for a generic solution from 
$AdS_7\times S^4$ branch. Such solution is described by a harmonic function (\ref{HarmFinSup}) which has transition 
points\footnote{We recall that transition points were introduced in section \ref{SubsTopolA4S7} as 
places on $y=0$ line where the boundary conditions changes.} at $x=x_k$ and one can see that 
$\d_x\Phi$ diverges at these points. This implies that geometry does not have non--contractible 
seven--manifolds which can support the electric flux (see section \ref{SubsTopolA4S7} for details), 
and the topology is completely specified by the set of the three--spheres which can be extracted from 
the coloring of the line. A typical perturbation of $AdS_7\times S^4$ is depicted in figure \ref{FigA7S4}c, 
it has light defects corresponding to the M5 branes discussed in section \ref{SubsA7M5S} and dark 
defects describing 
the branes from section \ref{SubsA7M5A}. The map between coordinates can be easily read off from 
(\ref{A7S4Coord}):
\bea
x_{light}=2L^3\cos\theta,\qquad x_{dark}=2L^3\cosh 2\rho,
\eea
One can also insert dark defects at negative values of $x$, they correspond to the counterparts of the branes studied in section \ref{SubsA7M5A} which are placed at the south pole of the sphere 
(i.e. they have 
$\theta=\pi$ rather than $\theta=0$). 

\subsection{Comments on more general solutions}
\label{SubsCmnts}

In this section we discussed the excitations of $AdS_7\times S^4$: the asymptotic behavior fixed the values of $c_1,c_2$ (\ref{c12A7S4}) as well as the scaling of the harmonic function (\ref{HrmA7S4As}). In section 
\ref{SectSugraA4S7} we saw that on $AdS_4\times S^7$ branch the values of $c_1,c_2$ were also 
fixed, but they were different from the ones discussed here.
In this subsection we will make some comments about general values of $c_1,c_2$.

\begin{figure}
\begin{center}
\epsfxsize=5in \epsffile{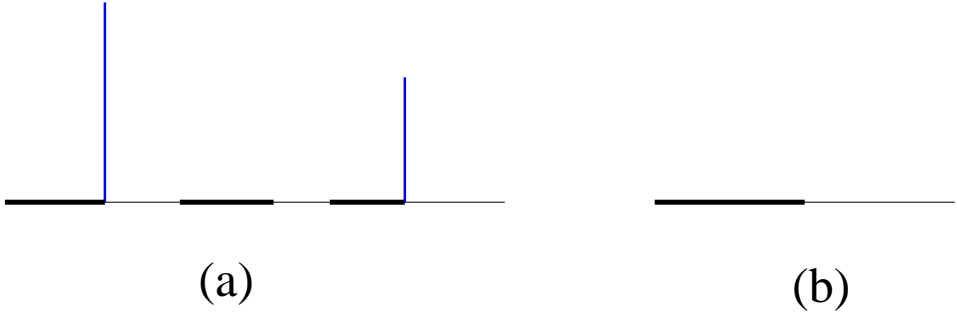}
\end{center}
\caption{
A bird's eye view of boundary conditions on $AdS_4\times S^7$ branch:
all finite--size effects (a) disappear and the distribution of dark and light regions becomes universal (b).
} \label{FigAverA4S7}
\end{figure}

The parameters $c_1,c_2$ should be determined by the behavior of the solution at large values of 
$r=\sqrt{x^2+y^2}$ where one sees the "average" coloring of the $y=0$ line and all finite size effects are washed away. For example, it we start with $AdS_7\times S^4$ solution and go to large distances, the harmonic function would be well--approximated by $\Phi=0$, i.e. in the leading order the boundary conditions are "light everywhere". The same fact is true for any solution on the $AdS_7\times S^4$ 
branch.
Similarly if one starts from a generic boundary condition on the $AdS_4\times S^7$ branch (such as one depicted in figure \ref{FigAverA4S7}a) and looks at large values or $r$, then boundary conditions for the 
harmonic function would be well--approximated by figure \ref{FigAverA4S7}b. In general after an averaging one effectively gets "grey" boundary conditions where $0<|\d_y\Phi|<2\pi$. Of course, such coloring would lead to singular solutions, but the averaging breaks down as we approach $y=0$ line, so when talking about "grey" boundary conditions at infinity, we always imply that one takes some regular solutions and averages them out in $x$. 


Generically we expect to have different shades of grey at positive and negative values of $x$, so the average boundary conditions can be modeled in a following way:
\bea\label{FarBoundCond}
\d_y\Phi|_{y=0,x<0}=\mu+\frac{\pi\la}{2},\quad \d_y\Phi|_{y=0,x>0}=\mu-\frac{\pi\la}{2},\quad
|\la|\le 2
\eea
Here $\mu$ and $\la$ are two constants parameterizing the solution.  
Notice that even if the branch cuts were present in the original solution they are washed away 
by averaging, so the last equation gives a complete boundary condition for the upper half plane. 

One can construct $\Phi$ satisfying the boundary condition (\ref{FarBoundCond}) using a 
Green's function, we will only need the expressions for the derivatives:
\bea
\d_y\Phi=\mu-\lambda\arctan\frac{x}{y},\quad \d_x\Phi=-\frac{\la}{2}\log(x^2+y^2)+\sigma
\nonumber
\eea
Here $\sigma$ is an integration constant which from now on will be set to zero. Taking into account the boundary condition (\ref{DefPsi2}), we can simplify the relations (\ref{SolnThrPhi}) in the "far region" which we are considering now:
\bea\label{GenBndrE1}
\log\frac{e^A-e^F}{e^A+e^F}+\Psi_2=-\frac{\la}{2}\log(x^2+y^2),\quad
4\arctan\frac{c_2e^B}{c_1e^C}=\mu-\lambda\arctan\frac{x}{y}
\eea
As one goes to large values of $y$, the warp factor in front of AdS space should go to infinity (due to the relation $y=e^{A+B+C}$) and some of the sphere warp factors might diverge as well. We will first 
assume that in such "decompactification limits" a contribution of $\Psi_2$ does not cancel the right hand side in 
(\ref{GenBndrE1})\footnote{This assumption is correct for solutions with $AdS_4\times S^7$ or 
$AdS_7\times S^4$ asymptotics.}. Let us analyze (\ref{GenBndrE1}) for different values of $\la$.

{\bf General case: $2\ge\la> 0$.} The second equation in (\ref{GenBndrE1}) implies that $e^B\sim e^C$, while from the first equation we conclude that
\bea\label{AsmptnWrng}
\frac{c_1^2 e^{2C}+c_2^2 e^{2B}}{4e^{2A}}=\frac{e^{-\Psi_2}}{(x^2+y^2)^{\la/2}}
\eea
Combining this with relation $y=e^{A+B+C}$, we arrive at the scaling:
\bea
e^A=\alpha yr^{(\la-2)/3}e^{\Psi_2/3},\quad e^B=\frac{\beta}{c_2} r^{(2-\la)/6}e^{-\Psi_2/6},\quad 
e^C=\frac{\gamma}{c_1} r^{(2-\la)/6}e^{-\Psi_2/6}
\eea
Here $\alpha$, $\beta$ and $\gamma$ are functions of $x/r$ and $y/r$. We have three equations for these functions:
\bea
\alpha\beta\gamma=c_1c_2,\quad \beta^2+\gamma^2=\frac{4\alpha^2y^2}{r^2},\quad
4\arctan\frac{\beta}{\gamma}=\mu-\lambda\arctan\frac{x}{y}
\eea
and they can be easily solved in terms of polar coordinates in $(x,y)$ plane ($x+iy=re^{i\zeta}$):
\bea
\beta=\frac{2\alpha y}{r}\sin\frac{\mu-\la\zeta}{4},\quad 
\gamma=\frac{2\alpha y}{r}\cos\frac{\mu-\la\zeta}{4},\quad
\alpha^3=\frac{c_1c_2 r^2}{2y^2\sin \frac{\mu-\la\zeta}{2}}.
\eea
Notice that since the warp factors on the spheres must be positive, the sign of $\frac{\beta}{\gamma}$ cannot jump, this leads to inequality
\bea\label{MuInequv}
\mu^2\ge \frac{\pi^2\la^2}{4}
\eea
Let us substitute this data into the equation (\ref{Eqnf0}):
\bea
df_0&=&-\frac{2\alpha^2}{c_2^2\beta^{-2}+c_1^2\gamma^{-2}}\left[-\frac{\beta\gamma}{c_1c_2}dy+
(\frac{\gamma^2}{c_1}-\frac{\beta^2}{c_2})dx\right]\nonumber\\
&=&-\frac{2c_1c_2}{(c_2^2c^2+c_1^2s^2)}\left[-sc~dy+(c_2 c^2-c_1s^2)dx\right]\nonumber\\
&&s\equiv \sin\frac{\mu-\la\zeta}{4},\quad c\equiv \cos\frac{\mu-\la\zeta}{4}
\eea
Integrability condition for this equation require $c_1=c_2=-\frac{1}{2}$, $\la=2$, $\mu=\pi$, this brings us 
to $AdS_4\times S^7$ asymptotics. We showed that no other values of $\la$ is allowed if one makes an assumption (\ref{AsmptnWrng}), and relaxation of this assumption requires a cancellation in 
(\ref{GenBndrE1}):
\bea\label{HadSlvBrnch}
\Psi_1=0,\quad \Psi_2=-\frac{\la}{2}\log(x^2+y^2),\quad
4\arctan\frac{c_2e^B}{c_1e^C}=\mu-\lambda\arctan\frac{x}{y}
\eea
We conjecture that for any $(\mu,\la)$ satisfying inequality (\ref{MuInequv}), one can start with these relations and solve (\ref{EqnMetr})--(\ref{DefPsi1}) for a unique value of $q(\mu,\la)$. We will not 
attempt to prove this statement.

{\bf Degenerate solutions: $\la=0$, $0<|\mu|< 2\pi$}. The second equation in (\ref{GenBndrE1}) can be easily solved:
\bea\label{EbThrH}
e^B=\frac{e^H}{c_2}\sin\frac{\mu}{4},\quad e^C=\frac{e^H}{c_1}\cos\frac{\mu}{4},\quad
e^F=\sqrt{e^{2A}-e^{2H}}
\eea
To find $\Psi_2$ it is convenient to use (\ref{Eqnf2*}), (\ref{Eqnf3*}) and eliminate $f_2$, $f_3$ from 
(\ref{DefPsi1}):
\bea
-(d\Psi_2+*d\Psi_1)&=&\frac{e^{B+C}}{e^{2A}-e^{2F}}\left[-e^{-3A-B-C}(c^2_2e^{2B}+c_1^2 e^{2C})df_1-
(c_1+c_2)e^{-A-B-C}df_0\right.\nonumber\\
&&\left.+2c_1c_2*d(B-C)\right]\nonumber\\
&=&\frac{e^{-A}}{e^{2A}-e^{2F}}\left[-(1-e^{2F-2A})df_1+df_0+2c_1c_2e^A *d(B-C)\right]\nonumber\\
&=&\frac{e^{-A}}{e^{2A}-e^{2F}}\left[e^{2F-2A}df_1-de^{2A+F}+2c_1c_2e^A *d(B-C)\right]
\eea
To perform the above transformations we used (\ref{EqnMetr}), (\ref{UnivABC}), (\ref{Eqnf1}). The last 
term disappears due to relations (\ref{EbThrH}), and if one assumes that $e^F$ remains
finite as $e^A$ goes to infinity, then both $\Psi_1$ and $\Psi_2$ vanish. Notice that this assumption is consistent, since for finite $e^F$, the first equation in (\ref{GenBndrE1}) leads to $\Psi_2=0$. 

We conclude that for $\la=0$ one can consistently set $\Psi_1=\Psi_2=0$ in the asymptotic region, this implies that $e^H=e^A$. Notice that for equation (\ref{Eqnf0}) to be integrable, both terms in the square bracket should be of the same order, this implies a vanishing of a leading contribution to
\bea
c_1 e^{2C}-c_2e^{2B}=e^{2H}\left[\frac{\cos^2\frac{\mu}{4}}{c_1}-\frac{\sin^2\frac{\mu}{4}}{c_2}\right]
\nonumber
\eea
This gives a simple relation between $q$ appearing in (\ref{UnivABC}) and a parameter $\mu$:
\bea\label{QthrMu}
q=-\cos^2\frac{\mu}{4}
\eea
Notice that to arrive at this conclusion we have assumed that $e^B$ and $e^C$ scale in the same way, and this assumption breaks down for $\mu=0,2\pi$.

{\bf Special points: $\la=\mu=0$ and $\la=0,\mu=-2\pi$.} Let us look at the case $\la=\mu=0$. The second equation in (\ref{GenBndrE1}) implies that either $c_2=0$ (then we are dealing with a limit of (\ref{QthrMu})) or $e^B\ll e^C$. One can show that the latter case leads to a scaling
\bea
e^A\sim e^C\gg e^F\sim e^B\sim 1,\quad \Psi_2=0\nonumber
\eea
The leading contribution to the equation (\ref{Eqnf0}) becomes very simple:
\bea
df_0=-2e^{2A}y^{-2}e^{2C}(c_1 e^{2C}dx)=-2c_1 dx
\eea
and its integrability condition does not lead to restrictions on $q$. 

Since function $\Phi$ is harmonic and it satisfies the boundary conditions
\bea
\d_y\Phi|_{r\rightarrow\infty}\rightarrow 0,\quad \d_x\Phi|_{r\rightarrow\infty}\rightarrow 0,
\eea
we can write its expansion:
\bea
\Phi=Q\log r+O(r^{-1}),\quad \d_y\Phi=\frac{Qy}{r^2}+O(r^{-2}),\quad 
\d_x\Phi=\frac{Qx}{r^2}+O(r^{-2})
\eea
and generically we expect to have a nonzero value of $Q$. Comparing this with (\ref{HrmA7S4As}) we conclude that geometry asymptotes to $AdS_7\times S^4$ with $L=Q^{1/3}/2$, then the analysis of section \ref{SectA7S4Brc} implies that $q=1$. 

Similarly for solutions with $\la=0$, $\mu=-2\pi$ we find that in addition to the solution (\ref{QthrMu}), there exists an $AdS_7\times S^4$ branch corresponding to $q=-2$: it can be found by interchanging the spheres in the previous paragraph.

Let us summarize the results of this subsection. We showed that an asymptotic behavior of a generic solution can be modeled by the boundary conditions (\ref{FarBoundCond}) imposed at large values of 
$|x|$. In the case of nonzero $\la$ one finds two different behaviors: for $\la=2$ the equations can be formulated entirely in terms of the warp factors and they lead to $AdS_4\times S^7$ asymptotics, while 
for $\la<2$ one needs to solve the system (\ref{Eqnf1})--(\ref{Eqnf0}), (\ref{DefPsi1}) along with equations
(\ref{HadSlvBrnch}). On the physical grounds we expect such solution to exist for some value of 
$q=q(\mu,\la)$ although we have not demonstrated this fact. For $\la=0$ we found a simple relation 
(\ref{QthrMu}) between $\mu$ and $q$ and in the special cases $\mu=0,-2\pi$ we also saw an existence of special solutions with $AdS_7\times S^4$ asymptotics. 

The goal of this subsection was to illustrate how an asymptotic behavior of function $\Phi$ fixes the values 
of $c_1$ and $c_2$: we expect that the boundary conditions for $\Phi$ determine the solution completely 
and in particular they should lead to the unique value of $q$. We did not prove this fact rigorously, but the 
discussion presented here gave some evidence for such proposal. It would be nice to study this problem further, in particular it would be interesting to find some explicit 1/2--BPS solutions which do not 
asymptote to $AdS_m\times S^n$.

\section{Decompactification limits}
\renewcommand{\theequation}{7.\arabic{equation}}
\setcounter{equation}{0}

\label{SectDecomp}

In the last three section we discussed various branches of the geometries with 
$SO(2,2)\times SO(4)\times SO(4)$ symmetries. While we were not able to write the explicit solutions, we showed that the metrics were uniquely parameterized by a harmonic function with well--defined boundary conditions. The difficulty in solving the differential equations stems from the 
fact that a spinor has a nontrivial dependence on the sphere and AdS coordinates, and one may hope that if these manifolds were replaced by flat space, then the equations would simplify. Such simplification indeed happens, moreover in this limit one recovers some geometries which were constructed before. We explore such relations in this section.

If a geometry has a sphere (or AdS) factor, one may consider a limit when the radius of such sphere goes to infinity. To keep the resulting metric finite, the coordinates on the sphere should be rescaled, this leads to a replacement of $S^n$ by $R^n$ in the metric. Such limit leads to simplifications in the equations for the Killing spinor: for example the derivative along $AdS_3$ space is given by (\ref{SpinDer1}), (\ref{SpinDer2}):
\bea
{\nabla}^H_m\eta=\frac{1}{2}e^{-A}\Gamma_H\gamma_m\eta-
\frac{1}{2}{\gamma^\mu}_m\d_\mu A\eta,
\eea
and rescaling $A$ by an infinite constant factor, we find the derivative on $R^{1,2}$:
\bea
{\nabla}^H_m\eta\rightarrow-
\frac{1}{2}{\gamma^\mu}_m\d_\mu A\eta,
\eea
Since the metric (\ref{InAns}) has three warp factors ($e^A$, $e^B$, $e^C$), it seems that one can 
rescale them independently. However for function $e^F$ defined by (\ref{EqnMetr}) to remain real, a decompactification of any of the spheres should be accompanied by sending $e^A$ to infinity. Thus there are three different ways to perform decompactifications, they produce geometries with 
$ISO(2,1)\times SO(4)^2$, $ISO(2,1)\times ISO(3)\times SO(4)$ or $ISO(2,1)\times ISO(3)^2$ isometries. We consider these cases separately.

\subsection{M2 brane with mass deformation.}

\label{SubsMsDfrmM2}

We begin with sending $e^A$ to infinity while keeping the other warp factor fixed. To be more precise, we consider a rescaling
\bea
e^A\rightarrow \Lambda e^A,\quad e^F\rightarrow \Lambda e^F,\quad 
x+iy\rightarrow \Lambda(x+iy),\quad f_1\rightarrow\Lambda^3 f_1,\quad
q\rightarrow \Lambda q,\quad f_0\rightarrow \Lambda^2 f_0
\nonumber
\eea
and take a large $\Lambda$ limit. Notice that both $q$ appearing in
(\ref{UnivABC}) and $f_0$ should be rescaled to yield a nontrivial solution. Sending $\Lambda$ to infinity, we find a simplified version of 
the system (\ref{EqnMetr})--(\ref{Eqnf0}):
\bea
\label{Lim1Metr}
&&ds^2=e^{2A}dw_{1,2}^2+e^{2B}ds_S^2+e^{2C}ds_{\tilde S}^2+
g_y^2(dx^2+dy^2)\\
&&F_4=df_1\wedge d^3 w+df_2\wedge d\Omega_3+
df_3\wedge d{\tilde\Omega}_3\nonumber
\eea
\bea
\label{Lim1Start}
&&y=e^{A+B+C},\quad 
g_y^{-1}=y\sqrt{e^{-2B}+e^{-2C}},\quad 
e^F\equiv \sqrt{e^{2A}-q^2(e^{2C}+e^{2B})}\\
\label{Lim1f1}
&&df_1=-2q^2ye^{2A-2F}*d(B-C)+de^{4A-F},\quad 
f_1=e^{2A+F}\\
\label{Lim1f0}
&&df_0=-2e^{2A}g_y^2q(e^{2C}+e^{2B})dx\\
\label{Lim1Cite}
&&e^{A-3B}df_2=2e^{-A-B}df_0+2qe^C*d(2B+C)-qe^{B-3A}df_1\\
&&e^{A-3C}df_3=2e^{-A-C}df_0+2qe^B*d(2C+B)+qe^{C-3A}df_1\nonumber\\
&&e^{F-3B}*df_2=qe^{C-3A}df_1+e^{-A-C}df_0+2qe^B*d(2B+C)\nonumber\\
&&e^{F-3C}*df_3=qe^{B-3A}df_1-e^{-A-B}df_0-2qe^C*d(B+2C)\nonumber\\
\label{Lim1Last}
&&d\log\frac{e^A-e^F}{e^A+e^F}-4*d\arctan e^{B-C}=
-\frac{qe^{F+B+C}}{e^{2A}-e^{2F}}\left[e^{-4C}*df_3+e^{-4B}*df_2\right]
\eea
Here $dw_{1,2}^2$ is a metric on $R^{2,1}$ and $d^3w$ is a volume factor on the same space. Notice that parameter $q$ can be eliminated by an appropriate rescaling of $w_i,e^A,e^F,f_1,f_0$, so without loss of generality we can set $q=1$. 

It turns out that the solutions of the form (\ref{Lim1Metr}) were studied in the past \cite{BenaWarn,LLM} and it might be useful to relate the system (\ref{Lim1Metr})--(\ref{Lim1Last}) with description presented in 
\cite{LLM}. To make the comparison we parameterize the warp factors 
of the spheres in terms of $e^A,y,e^G$ and introduce a useful function $h$:
\bea\label{Lim1Save1}
e^{2B}=ye^{-A+G},\quad e^{2C}=ye^{-A-G},\quad h^{-2}=y(e^G+e^{-G})
\eea
It is also convenient to parameterize $e^F$ by a function $V$ and rewrite the relations 
(\ref{Lim1Start}) in terms of new variables:
\bea\label{Lim1Save3}
V\equiv -h^2e^{F-A}:\qquad h^{-2}V^2=h^2-e^{-3A},\quad g_y^{-1}=e^{A/2}h^{-1}
\eea
The equations (\ref{Lim1f1}) take a very simple form:
\bea\label{Lim1ExprF1}
&&f_1=-h^{-2}e^{3A}V,\quad 
dV^{-1}=-2yh^4 V^{-2}*dG,
\eea
and the last relation can be rewritten in terms of the complete differentials:
\bea
ydV=-\frac{1}{2}*d\frac{1}{e^{2G}+1}
\eea
The integrability conditions for this equation imply an existence of a function $z$ which satisfies a linear differential equation, and the warp factors can be recovered from it:
\bea\label{Lim1Save2}
z=\frac{1}{2}\tanh G:\quad d(y^{-1}*dz)=0,\quad ydV=\frac{1}{2}*dz,\quad 
e^{-3A}=h^2-h^{-2}V^2
\eea
These relations as well as the expression (\ref{Lim1ExprF1}) for $f_1$ are in a perfect agreement with geometry presented in \cite{LLM}. To recover the fluxes on the spheres one should solve (\ref{Lim1f0}) ($f_0=-2x$) and treat five equations (\ref{Lim1Cite})--(\ref{Lim1Last}) as an overdefined system for $f_2,f_3$. We do not present the details here, but straightforward
computations allow one to recover the flux found in \cite{LLM}:
\bea
F_4=-d(h^{-2}e^{3A}V)\wedge d^3 w-\frac{e^{-3A}}{4}\left[
e^{-3G}*d(y^2e^{2G})\wedge d{\tilde\Omega}_3+
e^{3G}*d(y^2e^{-2G})\wedge d{\Omega}_3\right]\nonumber
\eea
To summarize, we showed that by sending the volume of AdS space to infinity, the system 
(\ref{EqnMetr})--(\ref{Eqnf0}) leads to the metrics produced by the mass--deformed theory on M2 
brane \cite{BenaWarn,LLM}. We see that the geometry is again parameterized in terms of 
a harmonic function $z$, but in this degenerate case one can write explicit expressions for the warp factors in terms of $z$. Let us now discuss other decompactifications.

\subsection{Relation to Russo--Tseytlin solution}

In this subsection we look at the geometries with $ISO(2,1)\times ISO(3)\times SO(4)$ isometries. Starting with general solution (\ref{InAns})--(\ref{MasterEqn16}) one should make a rescaling:
\bea
(e^A,e^B,e^F)\rightarrow \Lambda (e^A,e^B,e^F),\ (x+iy,f_0)\rightarrow\Lambda^2 (x+iy,f_0),\
(f_1,f_2)\rightarrow\Lambda^3 (f_1,f_2)
\eea
and take a limit $\Lambda\rightarrow\infty$. Then AdS space an one of the spheres are replaced by flat three dimensional spaces and we arrive at the system:
\bea\label{Lim2Metr}
&&ds^2=e^{2A}dw_{1,2}^2+e^{2B}dz_3^2+e^{2C}d{\tilde\Omega}_3^2+
g_y^2(dx^2+dy^2)\\
&&F_4=df_1\wedge d^3 w+df_2\wedge d^3 z+df_3\wedge d{\tilde\Omega}_3\nonumber
\eea
\bea\label{Lim2Start}
&&y=e^{A+B+C},\quad g_y^{-1}=ye^{-C},\quad 
e^F= \sqrt{e^{2A}-c_2^2 e^{2B}}\\
\label{Lim2f1}
&&df_1=de^{4A-F},\quad f_1=e^{2A+F}\\
\label{Lim2f0}
&&df_0=2c_2e^{2A+2B}g_y^2dx\\
\label{Lim2f2}
&&e^{A-3B}df_2=c_2e^{B-3A}df_1\\
\label{Lim2Rest}
&&e^{A-3C}df_3=2e^{-A-C}df_0-2c_2e^B*d(2C+B)\\
\label{Lim2Rest1}
&&e^{F-3B}*df_2=e^{-A-C}df_0-2c_2e^B*d(2B+C)\\
\label{Lim2f3}
&&e^{F-3C}*df_3=-c_2e^{B-3A}df_1\\
\label{Lim2Last}
&&d\log\frac{e^A-e^F}{e^A+e^F}=
\frac{e^{F+B+C}}{e^{2A}-e^{2F}}c_2e^{-4C}*df_3
\eea
We begin with analyzing a degenerate case of $c_2=0$. Then conditions for supersymmetry written above
lead to the following solution:
\bea
&&ds^2=e^{2A}d{\bf w}_{1,2}^2+e^{-A}\left[y e^Gd{\bf z}_3^2+y^{-1} e^{-G}(y^2d{\tilde\Omega}_3^2+
dx^2+dy^2)\right]\\
&&F_4=de^{3A}\wedge d^3 w\nonumber
\eea
The functions $e^A$ and $e^G$ are still undetermined, and to find them one has to look at the equations of motion. They lead to the relation $e^G=y$ and to the requirement that $e^{-3A}$ is a harmonic function on a five--dimensional space with metric
\bea
ds_5^2=y^2d{\tilde\Omega}_3^2+dx^2+dy^2
\eea
Thus we arrive at a solution describing M2 branes smeared in three transverse directions $z_i$. 

If $c_2$ is not equal to zero we can set $c_2=1$ by rescaling fluxes, warp factors and coordinates. 
It is convenient to parameterize the warp factors by two functions $f$ and $k$:
\bea
e^{F+2A}=f^{-1},\quad e^{2A}=k^{1/3}f^{-2/3}
\eea
Then the integrability condition for equations (\ref{Lim2f1}) gives a relation between these functions and an arbitrary constant c:
\bea
c=e^{4A-F}-e^{2A+F}=kf^{-2}f-f^{-1}:\qquad k=1+cf
\eea
We also evaluate the remaining warp factors:
\bea
e^{2B}=e^{2A}-e^{2F}=e^{2A}(1-f^{-2}e^{-6A})=ck^{-2/3}f^{1/3},\quad
e^{2C}=c^{-1}y^2(kf)^{1/3}
\eea
Equation (\ref{Lim2f2}) leads to the expression for $f_2$:
\bea
df_2=\frac{c^2f^2}{k^2}df^{-1}:\qquad f_2=\frac{c}{1+cf}=\frac{c}{k}
\eea
Further we can simplify the equation (\ref{Lim2f3}):
\bea
*df_3=-e^{3C-F+B-3A}df_1=-c^{-1}y^3f^2 df^{-1}
\eea
The integrability condition leads to the Laplace equation for $f$:
\bea
d(y^3*df)=0
\eea
and we can recover the complete solution in terms of this function:
\bea\label{RussTsetl1}
&&ds^2=(kf)^{1/3}\left[f^{-1}dx_{1,2}^2+k^{-1}dx_3^2\right]+\frac{1}{c(kf)^{1/3}}
\left[y^2d{\tilde\Omega}_3^2+dx^2+dy^2\right]\\
&&F_4=df^{-1}\wedge d^3 w+d\frac{c}{k}\wedge d^3 z-c^{-1}y^3*df\wedge d{\tilde\Omega}_3,\qquad
k=1+cf\nonumber
\eea
This solution was originally derived in \cite{RussoTseytl} using T dualities and rotations. We did not look at equations (\ref{Lim2Rest}), (\ref{Lim2Rest1}), (\ref{Lim2Last}), but one can easily see that they are satisfied.

\subsection{Complete decompactification.}

Finally let us send the radii of AdS and both spheres to infinity and look at solutions with 
$ISO(2,1)\times ISO(3)^2$ symmetry. The rescaling of the warp factors and fluxes is obvious:
\bea
(e^A,e^B,e^C,e^F)\rightarrow \Lambda (e^A,e^B,e^C,e^F),\qquad
(f_1,f_2,f_3)\rightarrow \Lambda^3(f_1,f_2,f_3),
\eea
then from (\ref{EqnMetr}) one finds that $y\rightarrow \Lambda^3 y$, $g_y\rightarrow \Lambda^{-2}g_y$. This implies that to arrive at nontrivial regular metric the differentials $dx$ and $dy$ should scale as 
$\Lambda^2$. A different scaling of $y$ and $dy$ implies that in the leading approximation $y$ is a constant\footnote{The fact that $y_0=e^{A+B+C}$ becomes a constant in this limit can also be seen directly from the equation (\ref{GeomProj}) once we set $a=b=c=0$.}:
\bea
x+iy\rightarrow i\Lambda^3y_0+\Lambda^2(x+iy),\quad d(x+iy)\rightarrow \Lambda^2d(x+iy)
\eea
This implies that $f_0$ scales as $\Lambda^2$ while $c_1,c_2$ do not depend on $\Lambda$. 
We also notice that in the large $\Lambda$ limit the relation (\ref{UnivABC}) between these two numbers disappears and they become independent (this can be seen from the relation (\ref{11Dabc}) with $a=b=c=0$). Making the rescalings described above and taking the limit $\Lambda\rightarrow \infty$ in (\ref{InAns})--(\ref{MasterEqn16}), we find
\bea\label{Lim3Metr}
&&ds^2=e^{2A}d{\bf w}_{1,2}^2+e^{2B}d{\bf u}_3^2+e^{2C}d{\bf v}_3^2+
g_y^2(dx^2+dy^2)\\
\label{Lim3Func}
&&e^{A+B+C}=y_0,\quad 
e^F\equiv \sqrt{e^{2A}-c_1^2 e^{2C}-c_2^2 e^{2B}}\\
&&df_0=-2e^Ag_y^2\left[-e^{F+B+C}dy+e^A(c_1e^{2C}-c_2e^{2B})dx\right]\nonumber\\
\label{Lim3f1}
&&df_1=2c_1c_2e^{2A-2F}*d(B-C)+de^{4A-F},\quad 
f_1=e^{2A+F}\\
\label{Lim3f2}
&&e^{A-3B}df_2=2c_1e^C*d(2B+C)+c_2e^{B-3A}df_1\\
&&e^{A-3C}df_3=-2c_2e^B*d(2C+B)+c_1e^{C-3A}df_1\nonumber\\
&&e^{F-3B}*df_2=c_1e^{C-3A}df_1-2c_2e^B*d(2B+C)\nonumber\\
&&e^{F-3C}*df_3=-c_2e^{B-3A}df_1-2c_1e^C*d(B+2C)\nonumber
\eea
\bea
\label{Lim3Last}
d\log\frac{e^A-e^F}{e^A+e^F}+4*d\arctan\frac{c_2 e^{B}}{c_1e^C}=
\frac{e^{F+B+C}}{e^{2A}-e^{2F}}\left[c_2e^{-4C}*df_3-c_1e^{-4B}*df_2\right]
\eea
We begin with degenerate case $c_1=0$. Then the equations (\ref{Lim3f1})--(\ref{Lim3Last}) reduce to 
a modification of the system (\ref{Lim2f1}), (\ref{Lim2f2})--(\ref{Lim2Last}) with $f_0=0$ and 
$e^{A+B+C}=1$. Repeating the derivation of (\ref{RussTsetl1}), we conclude that in the present case one finds another version of the Russo--Tseytlin solution\footnote{As before, the "doubly degenerate" case of $c_1=c_2=0$ corresponds to a metric of M2 brane. We also notice that to determine $g_y$ in 
(\ref{Lim3Metr}), (\ref{RussTsetl2}) one needs to look at the equations of motion.}:
\bea\label{RussTsetl2}
&&ds^2=(kf)^{1/3}\left[f^{-1}d{\bf w}_{1,2}^2+k^{-1}d{\bf u}_3^2\right]+\frac{1}{c(kf)^{1/3}}
\left[d{\bf v}_3^2+dx^2+dy^2\right]\\
&&F_4=df^{-1}\wedge d^3 w+d\frac{c}{k}\wedge d^3 u-c^{-1}*df\wedge d^3v,\quad
k=1+cf,\quad d*df=0\nonumber
\eea

Assuming that neither $c_1$ nor $c_2$ is equal to zero, we can rescale the warp factors and the fluxes to set $c_1=c_2=y_0=1$ in (\ref{Lim3Func}) and (\ref{Lim3f2})--(\ref{Lim3Last}), then the solutions are labeled by one parameter $c$ which still appears in (\ref{Lim3f1}):
\bea
\label{Lim3f1p}
&&df_1=2ce^{2A-2F}*d(B-C)+de^{4A-F},\quad 
f_1=e^{2A+F}
\eea
One can use (\ref{Lim3Func}) to introduce a convenient parameterization of the warp factors:
\bea
e^{2B}=e^{-A+G},\quad e^{2C}=e^{-A-G},\quad e^{2F}=e^{2A}-2e^{-A}\cosh G\equiv e^{2A}f^{-2}
\eea
Substituting this into (\ref{Lim3f1p}), we find a relation
\bea
d[f(e^G+e^{-G})]=-2c~f^2*dG:\qquad d\left[\frac{1}{f(e^G+e^{-G})}\right]=\frac{c}{2}*d\tanh G
\eea
We conclude that the solution is parameterized by a harmonic function $z$ and a field $V$ which is dual to it:
\bea\label{Lim3Sv1}
z=\frac{1}{2}\tanh G,\quad d*dz=0,\quad dV=-c*dz,\quad V^{-1}=-f(e^G+e^{-G})
\eea
To compute the warp factors in terms of $z$ and $V$, we introduce a function $h^{-2}\equiv e^G+e^{-G}$ 
which can be easily recovered from $z$. Then using the definition of $f$ we find:
\bea\label{Lim3Sv2}
e^{-3A}=h^2-h^{-2}V^2,\quad e^{2B}=e^{-A+G},\quad e^{2C}=e^{-A-G},\quad e^F=-Vh^{-2}e^A
\eea
This solution can be viewed as a limit of the geometry discussed in section \ref{SubsMsDfrmM2}. 
To be more precise, we start with relations (\ref{Lim1Save1}), (\ref{Lim1Save3}), (\ref{Lim1Save2}) 
and make a rescaling
\bea
(e^B,e^C)\rightarrow \Lambda (e^B,e^C),\quad x+iy\rightarrow i\Lambda^2 y_0+\Lambda(x+iy),\quad
(h^{2},V,e^{-3A})\rightarrow \Lambda^{-2}(h^2,V,e^{-3A})\nonumber
\eea
After sending $\Lambda$ to infinity, we can rewrite (\ref{Lim1Save1}), (\ref{Lim1Save3}), 
(\ref{Lim1Save2}) as
\bea
&&e^{2B}=y_0e^{-A+G},\quad e^{2C}=y_0e^{-A-G},\quad h^{-2}=y_0(e^G+e^{-G}),\quad
e^F=-Vh^{-2}e^A\nonumber\\
&&z=\frac{1}{2}\tanh G:\quad d*dz=0,\quad y_0dV=\frac{1}{2}*dz,\quad 
e^{-3A}=h^2-h^{-2}V^2
\eea
Additional finite rescaling brings this system to the form (\ref{Lim3Sv1}), (\ref{Lim3Sv2}). One can 
show that the fluxes determined by (\ref{Lim3f2})--(\ref{Lim3Last}) are matched by the decompactifying the solution discussed in section \ref{SubsMsDfrmM2} as well. For completeness we write down the 
resulting geometry:
\bea
ds^2&=&e^{2A}d{\bf w}_{1,2}^2+e^{-A}\left[e^G d{\bf u}_3^2+e^{-G}d{\bf v}_3^2+h^2(dx^2+dy^2)\right]\\
F_4&=&-d(e^{3A}h^{-2}V)\wedge d^3w-\frac{e^{-3A}}{4}\left[e^{-3G}*de^{2G}\wedge d^3v+
e^{3G}*de^{-2G}\wedge d^3u\right]\nonumber\\
&&h^{-2}= e^G+e^{-G},\quad z=\frac{1}{2}\tanh G, \quad dV=\frac{1}{2}*dz,\quad 
e^{-3A}=h^2-h^{-2}V^2\nonumber
\eea 
Notice that the harmonic function $z$ is defined on the entire $(x,y)$ plane, so to have nontrivial solutions 
this function should have sources in this plane. This implies that $e^G$ diverges at some points and the solution cannot be regular since a warp factor in front of $R^3$ cannot go to zero without creating a singularity. 

\bigskip

One can also look at ten dimensional geometries which are dual to Wilson lines in ${\cal N}=4$ SYM. 
The relevant solutions with $SO(2,1)\times SO(3)\times SO(5)$ isometries were constructed in 
\cite{myWils}, and it might be interesting to analyze their behavior when various warp factors are sent 
to infinity.
These metrics and their decompactification limits are discussed in the Appendix B, where it is shown that three possible geometries can be recovered: they describe either fundamental strings, D5 branes, or D3 branes with fluxes. The latter configuration was introduced in \cite{HashItz,MaldRuss} 
as a dual description of a non--commutative gauge theory.  

\bigskip

This concludes the discussion of geometries with $SO(2,2)\times SO(4)\times SO(4)$ isometries and now we will turn to the solution describing M2 branes in $AdS_4\times S^7$.

\section{Solutions with $SO(2,1)\times SO(6)$ isometries}
\renewcommand{\theequation}{8.\arabic{equation}}
\setcounter{equation}{0}

\label{SecAnalLLM}

In section \ref{SubsM2Intrs} we considered one dimensional defects in the field theory dual to 
$AdS_4\times S^7$. We saw that the bulk description of such defects was given in terms of 
M2 branes which preserved $SO(2,1)\times SO(6)$ symmetry. As number of M2 
branes becomes large, they are expected to produce changes in the geometry and in this section we describe the relevant metrics. Fortunately one can easily find the local geometry by making an analytic continuation of the metrics which are already known. We begin with recalling eleven dimensional supersymmetric solutions with 
$SO(3)\times SO(6)$ isometries which were constructed in 
\cite{LLM}:
\bea\label{LLMsol}
ds_{11}^2&=&-4e^{2\la}(1+y^2 e^{-6\la})(dt+V_i dx^i)^2+\frac{e^{-4\la}}{1+y^2 e^{-6\la}}
\left[dy^2+e^D(dx_1^2+dx_2^2)\right]\nonumber\\
&&+4e^{2\la}d\Omega_5^2+y^2 e^{-4\la}d{\tilde\Omega}_2^2\nonumber\\
F_4&=&F\wedge d^2{\tilde\Omega},\qquad e^{-6\la}=\frac{\d_y D}{y(1-y\d_y D)},\quad
V_i=\frac{1}{2}\eps_{ij}\d_j D
\eea
The solution is governed by one scalar function $D$ which satisfied a continual Toda equation
\bea\label{TodaEqn}
(\d_1^2+\d_2^2)D+\d_y^2 e^D=0
\eea
and the two--form $F$ is expressed in terms of derivatives of $D$ (we refer to \cite{LLM} for the details). To construct the geometries with $SO(2,1)\times SO(6)$ isometries one can perform a following analytic continuation:
\bea
d{\tilde\Omega}_2^2\equiv d{\tilde\alpha}^2+\cos^2{\tilde\alpha} d{\phi}^2\rightarrow 
-(d\alpha^2-\cosh^2\alpha d\phi^2)=-dH_2^2,\qquad
y\rightarrow iy,\quad x_i\rightarrow ix_i\nonumber
\eea
The new solution becomes\footnote{We also made a replacement $t\rightarrow \chi$ to stress the 
fact that now $\chi$ is a spacelike coordinate.}:
\bea\label{LLMCont}
ds_{11}^2&=&4e^{2\la}(y^2 e^{-6\la}-1)(d\chi+V_i dx^i)^2+\frac{e^{-4\la}}{y^2 e^{-6\la}-1}
\left[dy^2+e^D(dx_1^2+dx_2^2)\right]\nonumber\\
&&+4e^{2\la}d\Omega_5^2+y^2 e^{-4\la}dH_2^2\\
F_4&=&{\tilde F}\wedge d^2H,\qquad e^{-6\la}=\frac{\d_y D}{y(y\d_y D-1)},\quad
V_i=\frac{1}{2}\eps_{ij}\d_j D\nonumber\\
{\tilde F}&=&d\left[-4y^3 e^{-6\la}(d\chi+V)\right]+2~^*_3\left[e^{-D}y^2(\d_y \frac{1}{y}\d_y e^D)dy+
y\d_i\d_y Ddx^i\right]\nonumber
\eea
Notice that the Toda equation (\ref{TodaEqn}) is not affected by the continuation, one can also see that 
the two--form ${\tilde F}$ which appears in the new solution is real. It is interesting to note that the 
expression for the metric does not change if $y$ is replaced by $z=-y$: such reparameterization simply changes the sign of the flux.
Locally the 
relations (\ref{LLMCont}) give a unique solution for any function $D$ which satisfies a Toda equation (\ref{TodaEqn}). However to avoid singularities, one should also add some global constraints on function $D$.

We begin with observing that a coordinate $y$ is related to the radii of the sphere and AdS in a very simple way: $y=\frac{1}{2}R_2R_5^2$. Since AdS space cannot contract in a regular way, we conclude that $y$ can be equal to zero at a certain point if and only if the five dimensional sphere goes to zero size at that point. In particular this implies that the space "ends" there, i.e. one should
exclude negative values of $y$ (alternatively, one may consider $y$ which 
never becomes positive). In order to avoid singularities at such points, the AdS warp factor 
(we call it $R_2$) should remain finite, then $e^{2\la}=y/R_2$, $e^D\sim y$ and 
\bea
\frac{e^{-4\la}}{y^2 e^{-6\la}-1}dy^2+4e^{2\la}d\Omega_5^2\sim \frac{1}{R_2}\left[
\frac{dy^2}{y}+4yd\Omega_5^2\right]
\eea
The expression in the square bracket gives a metric on a patch of the flat $R^6$, this demonstrates  regularity of the metric in $(y,S^5)$ subspace. It is easy to see that the remaining part of the metric also 
remains regular as $y$ goes to zero. To summarize, we see that to preserve regularity at $y=0$, the five sphere should shrink to zero size and $e^D$ should scale like $y$. 
The same condition was derived for the original solution (\ref{LLMsol}) in \cite{LLM}, however 
in that case there was also a possibility for $S^2$ to contract as $y$ approached zero and now we don't have this option. However, in the present case $g_{\psi\psi}$ can go to zero at certain points away from $y=0$, and additional regularity conditions should be imposed at such points. 

Let us consider a vicinity of a point where $y=y_0>0$ and $y^2e^{-6\la}=1$. In this region it is convenient to rewrite the first line in the metric (\ref{LLMCont}) in terms of $D$:
\bea
ds_4^2&=&e^{-4\la}\left[\frac{4e^{6\la}}{y\d_y D-1}(d\chi+V_i dx^i)^2+(y\d_y D-1)
\left(dy^2+e^D(dx_1^2+dx_2^2)\right)\right]\nonumber\\
&&V_i=\frac{1}{2}\eps_{ij}\d_j D,\qquad y^2e^{-6\la}=1+\frac{1}{y\d_y D-1}\nonumber
\eea
We are interested in the vicinity of the point where $\d_y D$ goes to infinity, then the leading contribution to the above metric can be written as
\bea\label{LimitHyper}
ds_4^2\sim y_0^{-1/3}\left[\frac{4}{\d_y D}(d\chi+V_i dx^i)^2+\d_y D
\left(dy^2+e^D(dx_1^2+dx_2^2)\right)\right],
\eea
while the Toda equation (\ref{TodaEqn}) and relation between $D$ and $V_i$ (\ref{LLMCont}) still hold. We observe that the metric in the square brackets describes a four--dimensional hyper--Kahler manifold with a rotational Killing vector \cite{Bachas}. 

In it interesting to note that there is an alternative way of recovering hyper--Kahler manifold from the solution (\ref{LLMCont})\footnote{This construction was introduced in \cite{LLM} for the solutions with $SO(2,4)\times SO(3)$ isometries.}: one begins with making a shift in $D$
\bea\label{HyperScale}
D(x_1,x_2,y)={\tilde D}(x_1,x_2,{\tilde y})-2\log C,\quad {\tilde y}\equiv Cy
\eea
and then takes a limit where $y$ goes to infinity while $Cy^{1/3}$ is kept fixed. This leads to the four dimensional hyper--Kahler metric 
\bea\label{AlternHyper}
ds_4^2\sim \frac{1}{Cy^{1/3}}\left[\frac{4}{\d_{\tilde y} {\tilde D}}(d\chi+V_i dx^i)^2+
\d_{\tilde y} {\tilde D}
\left(d{\tilde y}^2+e^{\tilde D}(dx_1^2+dx_2^2)\right)\right]
\eea
While this expression is similar to (\ref{LimitHyper}), the effects of the two limits on the remaining part of the metric are different: the scaling (\ref{HyperScale}) leads to decompactification of $AdS_2$ and $S^5$ at large $y$, while (\ref{LimitHyper}) corresponds to finite radii. Another difference between the two ways of recovering a hyper--Kahler space is hidden in the nature of the limits: the derivative $\d_y D$ should go to infinity in 
(\ref{LimitHyper}), while $\d_{\tilde y} {\tilde D}$ remains finite in (\ref{AlternHyper}). 

\subsection{Boundary conditions}

Let us go back to the hyper--Kahler metric (\ref{LimitHyper}) and analyze the constraints imposed by regularity. The space should locally reduce to $R^4$, so it might be useful to start from a flat metric:
\bea\label{Flat4D}
ds_{flat}^2=dr^2+r^2 d\theta^2+r^2\sin^2\theta d\phi^2+r^2\cos^2\theta d\psi^2
\eea
and rewrite it in a form similar to (\ref{LimitHyper}). There are two non--equivalent ways of doing this, and they lead to two types of boundary conditions. We begin with observing that a Killing  spinor on (\ref{Flat4D}) has a nontrivial angular dependence\footnote{In fact there are also spinors 
which have different combinations of U(1) charges: $(\pm\frac{1}{2},\pm\frac{1}{2})$, but the same arguments apply to them as well.}: 
$\eps\sim \exp\left[\frac{i}{2}(\phi+\psi)\right]$. The Killing spinor on (\ref{LimitHyper}) depends on 
$\chi$ as $\eta\sim e^{i\chi/2}$, so we can make two different identifications:
\bea
I:&&\phi=\chi,\quad \psi=\alpha\nonumber\\
II:&&\phi=2\chi+\alpha,\quad \psi=-\chi\nonumber
\eea
One can also look at more general linear relations with $(\d_\chi\phi,\d_\chi\psi)=(n+1,-n)$, but they 
lead to conical defects if $n>1$. Let us consider the consequences of two identifications. 

{\bf Type I boundary condition.} Rewriting the flat metric in a form similar to (\ref{LimitHyper}), 
we find:
\bea
ds^2_{flat}=4Yd\chi^2+\frac{1}{Y}\left[dY^2+Y(dx^2+x^2 d\alpha^2)\right]
\eea
Here we defined $Y= \frac{1}{4}r^2\sin^2\theta$, $x=r\cos\theta$. Comparing with the 
expression in the square brackets of (\ref{LimitHyper}), we can make identifications:
\bea
x_1+ix_2=x e^{i\alpha},\quad e^D=Y=y-y_0
\eea
If the four dimensional geometry combines into a patch of flat space using this mechanism, then 
the $\chi$ circle should shrink along the plane $y=y_0$. Then the three--dimensional space 
$(x_1,x_2,y)$ terminates at this plane and the geometry is described by the Toda equation 
(\ref{TodaEqn}) at $y>y_0$ along with boundary conditions
\bea\label{TdBcY0}
D\sim \log (y-y_0),\quad y\rightarrow y_0
\eea
Such boundary conditions lead to different types of the solutions depending on the sign of $y_0$. 

\begin{figure}
\begin{center}
\epsfxsize=4.7in \epsffile{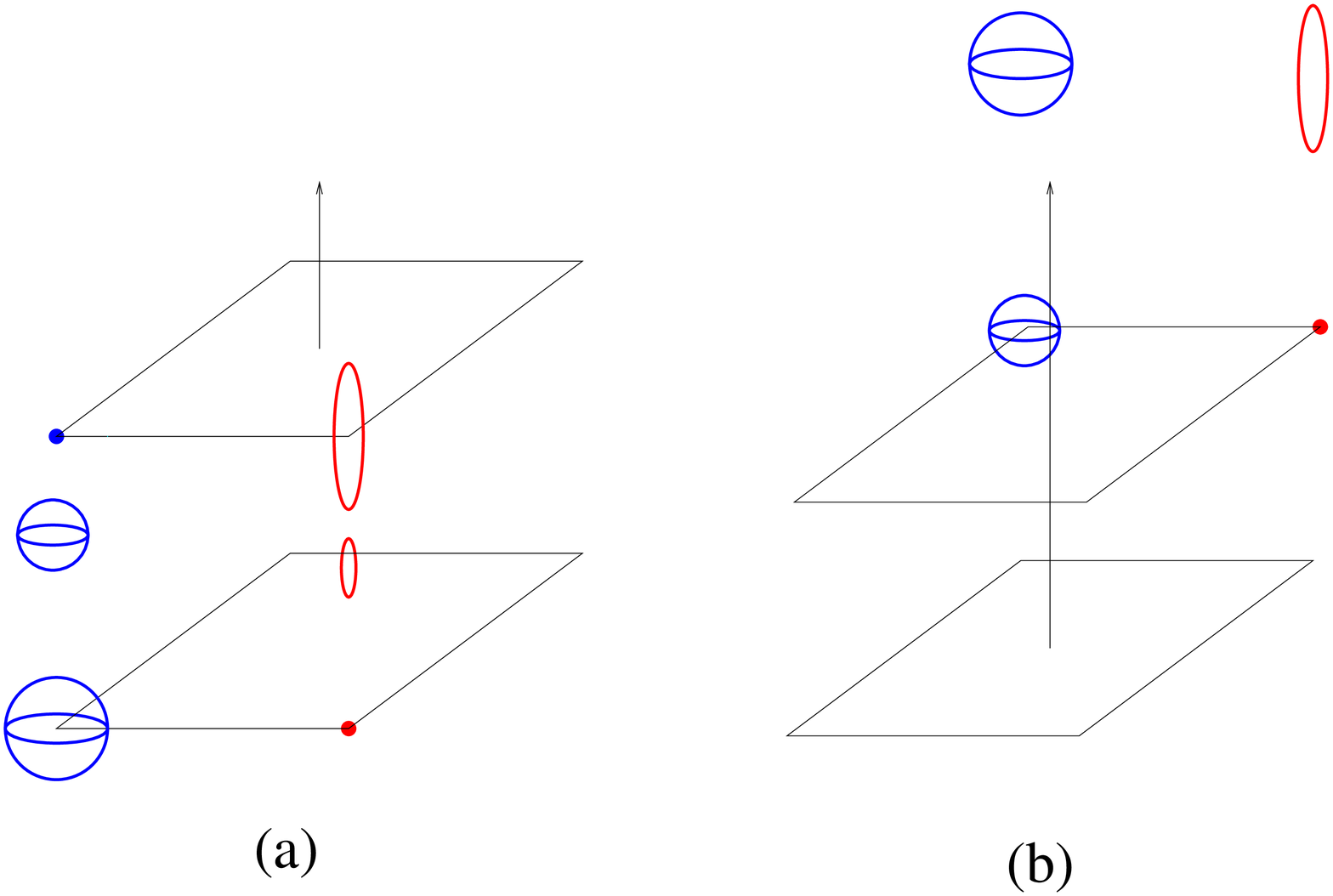}
\end{center}
\caption{
Type I boundary conditions: (a) negative value of $y_0$ leads to solutions with finite range of $y$; 
(b) for positive $y_0$ one should be able to find regular geometries with $AdS_4\times S^7$ asymptotics.
} \label{FigChiColPln}
\end{figure}

If $y_0<0$, then coordinate $y$ varies over a finite range: $y_0\le y\le 0$ and Toda equation 
(\ref{TodaEqn}) should be supplemented by two boundary conditions:
\bea
e^D\sim (y-y_0),\quad y\rightarrow y_0;\qquad
e^D\sim -y,\quad y\rightarrow 0
\eea
This situation is depicted in figure \ref{FigChiColPln}a: a five dimensional sphere contracts on the upper plane and the $\chi$ circle shrinks on the lower plane. One can see that the geometry is regular everywhere as long as condition $\d_y D>0$ is satisfied between the planes. While this inequality holds on the boundaries, a better understanding of Toda equation is required to prove it in the bulk. It would be very interesting to 
study this class of smooth geometries in more details, but since such solutions do not asymptote to 
$AdS_4\times S^7$, we will not discuss them further. 


A positive value of $y_0$ in the boundary condition (\ref{TdBcY0}) leads to another class of geometries: 
the five--dimensional sphere never collapses to zero size, while the $\chi$ circle should shrink along the plane $y=y_0$ (see figure \ref{FigChiColPln}b). The allowed range of parameters is
$y>y_0>0$ and one needs one more boundary condition at large values of $y$. This condition can 
come from the $AdS_4\times S^7$ asymptotics (see (\ref{TodaBCA4S7})):
\bea
e^D=\frac{y}{L^3}+O(r^{-1})
\eea
It would be very interesting to construct solutions with these boundary conditions and find a clear physical 
interpretation of the parameter $y_0$. We will not attempt to do this here.

{\bf Type II boundary condition.} Let us now introduce a reparameterization of type II and rewrite 
the flat metric (\ref{Flat4D}) in terms of $\chi$ and $\alpha$:
\bea
&&ds^2_{flat}=4(X^2+Y^2)\left[d\chi+\frac{X^2 d\alpha}{2(X^2+Y^2)}\right]^2\nonumber\\
&&\qquad+\frac{1}{X^2+Y^2}\left[
X^2Y^2d\alpha^2+(X^2+Y^2)\left(dX^2+4dY^2\right)
\right]\nonumber\\
&&X=r\sin\theta,\quad Y=\frac{1}{2}r\cos\theta\nonumber
\eea
Using some guesswork, we can find the canonical coordinates and function $D$:
\bea
x=XY^2,\quad y=\frac{X^2}{2}-Y^2+y_0,\quad x_1+ix_2=xe^{i\alpha},\quad e^D=\frac{1}{Y^2}.
\eea
One can check that this choice leads to the correct relations:
\bea
g_{\chi\chi}=\frac{4}{\d_y D},\quad V_\alpha=-\frac{1}{2}x\d_x D\nonumber
\eea
The $\chi$ circle shrinks to zero size when both $x$ and $y-y_0$ are equal to zero, this happens at a 
point in $(x_1,x_2,y)$ space. However we see that $e^D$ goes to infinity not only at that point, but also along the line $x=0$, so one might suspect that a geometry becomes singular on this line unless some additional restriction on $D$ is imposed. In the case of the flat space the singularity is avoided due to 
a boundary condition on a half--line:
\bea\label{HalfBC}
D\sim -\log x,\quad x\rightarrow 0, \ y>y_0.
\eea
One also notices that $\chi$ circle collapses at the end point of this line. 

It appears that for a general geometry, the relation (\ref{HalfBC}) 
is sufficient to guarantee regularity and one does not need to impose an extra condition at 
$y=y_0$. It is interesting to compare this situation with regularity condition for the geometries 
with $SO(2,2)\times SO(4)\times U(1)$ isometry. In \cite{linMald} such geometries were constructed in terms of harmonic function (rather than a function satisfying Toda equation which 
we have here) and it was also shown that $U(1)$ circle collapses at a point in $(x_1,x_2,y)$ space.
The regularity required the harmonic function to have a pole with specific residue at this point (see 
\cite{linMald} for details). In the present case it seems that we do not need any special restriction at a point, but rather we need a boundary condition (\ref{HalfBC}) on a line approaching the point. 

So far we assumed that the geometry combines into a patch of flat space near a point 
$(x_1,x_2,y)=(0,0,y_0)$ by mean of type II identification, this led to the conclusion that $e^D$ diverged on a vertical line (at least in the vicinity of $(0,0,y_0)$ where the flat approximation was valid) and regularity led to the requirement (\ref{HalfBC}). In particular we conclude that for any regular solution which has a collapsing $\chi$ cycle, there always exists a line where $e^D$ diverges. Let us show that this line must have a form $(x_1,x_2)=(x^{(0)}_1,x^{(0)}_2)$ and that 
a condition (\ref{HalfBC}) should hold in its vicinity. To demonstrate this we take an arbitrary 
{\it internal} point $(x_1,x_2,y)=(x^{(0)}_1,x^{(0)}_2,y^{(0)})$ on this line where $e^{2\la}$ and 
$g_{\chi\chi}$ must be finite and non--zero. In particular this implies that $\d_yD$ is bounded, so one can decompose $D$ into divergent $y$--independent piece and finite contribution:
\bea
D=D_\infty(x_1,x_2)+D_f(x_1,x_2,y)
\eea
It is convenient to use polar coordinates $(x,\alpha)$ with an origin at $(x^{(0)}_1,x^{(0)}_2)$
since $D_\infty$ should only depend on $x$. Let us look at the Toda equation (\ref{TodaEqn}):
\bea
\Delta_x D_\infty+\Delta_x D_f+e^{D_\infty}\d_y^2 e^{D_f}=0\nonumber
\eea
Assuming\footnote{This is not a crucial assumption, and we are making it only to simplify the 
argument.} that $\d_y^3 e^{D_f}\ne 0$, we find that the first term in the last equation should vanish,
this implies that $D_\infty=c\log x$ and we arrive at an equation for $D_f$:
\bea
\left[\frac{1}{x}\d_x(x\d_x)+\frac{1}{x^2}\d_\alpha^2\right] D_f+x^c\d_y^2 e^{D_f}=0\nonumber
\eea
At a generic point on a "singular line" we expect to have a good expansion for $D_f$:
\bea
D_f=D_0(y)+xD_1(y)+\dots\nonumber
\eea
which implies that $c=-1$. In other words, we showed that near any internal point on a "singular line" one has an expansion
\bea
D=-\log |{\bf x}-{\bf x}^{(0)}|+D_0(y)+\dots
\eea
This implies that $e^D$ vanishes along a straight line $(x_1,x_2)=(x^{(0)}_1,x^{(0)}_2)$ and the last equation is equivalent to the boundary condition (\ref{HalfBC}). One still needs to show that near this line the space remains regular, but this will become obvious from the coordinate transformation which will be introduced below (it will make the value of $e^D$ finite and non--zero in the internal points of the rod).

\begin{figure}
\begin{center}
\epsfxsize=5in \epsffile{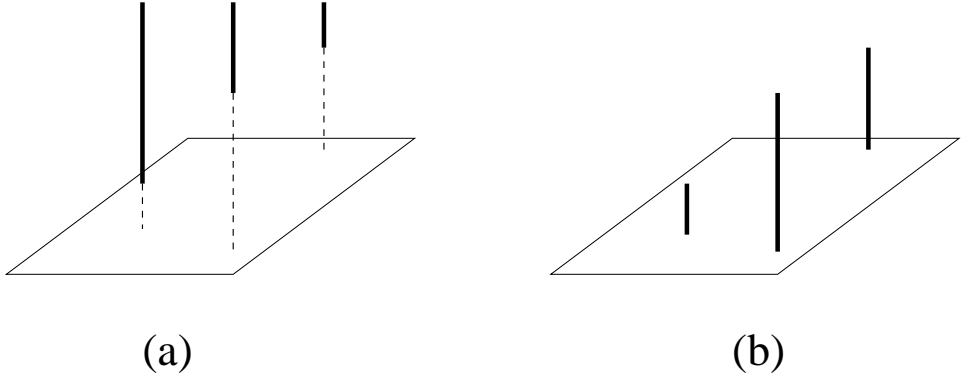}
\end{center}
\caption{
Two formulations of Type II boundary conditions: (a) description (\ref{TypeIIBound}) is convenient for proving regularity, but it leads to infinite rods; (b) complementary formulation (\ref{AltTypeIIBound}) has 
a natural interpretation in terms of the probe branes.
} \label{FigTypeII}
\end{figure}

To summarize, we found that the type II conditions should be imposed along vertical lines which terminate at some positive values of $y$, so the regular solutions in this sector are 
parameterized by the set $({\bf x}^{(i)},y^{(i)})$:
\bea\label{TypeIIBound}
&&D\sim \log y,\quad y\rightarrow 0\nonumber\\
&&D\sim -\log |{\bf x}-{\bf x}^{(i)}|,\quad {\bf x}\rightarrow {\bf x}^{(i)},\ y>y^{(i)}>0,\quad i=1,\dots n
\eea
The pictorial representation of these boundary conditions is given in figure \ref{FigTypeII}a. While the conditions (\ref{TypeIIBound}) came out in a natural way in the analysis of regularity, it turns out that there is an alternative way of writing the solutions which makes a comparison with brane probe a little easier. We will now discuss the transformation which leads to such "complementary" description.


{\bf Type II boundary condition: complementary formulation.} The boundary conditions
(\ref{TypeIIBound}) are imposed in $y=0$ plane and on the semi--infinite rods which extend from some value of $y$ to infinity. It is reasonable to assume that each rod corresponds to a stack of M2 branes 
(this assumption will be confirmed by a more detailed analysis in section \ref{SubsTodaProbe}), then 
it is natural to look 
for a description of a probe M2 brane. It seems that an addition of an extra brane changes the boundary conditions (\ref{TypeIIBound}) in a very radical way: one needs to introduce a rod which goes all the way to infinity. While the objects with very low co--dimension are expected to cause such drastic 
changes (for example, this is the case for a domain wall in 
SYM$_{3+1}$ \cite{myWils,gomRom}), probe M2 branes should not modify the $AdS_4\times S^7$ asymptotics. 
This implies that the rods which extend to infinity should not introduce significant changes to the geometry 
at large values of $y$. It turns out that one can rewrite the solution in a different form, which makes it clear that the modifications of the metric are localized.

We begin with recalling the action of conformal transformations in $(x_1,x_2)$ plane on the solutions of Toda equation. Let us consider two coordinate systems which are related by a holomorphic map: $x_1+ix_2=f(x'_1+ix'_2)$. Then starting with three dimensional metric in 
$(x_1,x_2,y)$ space, one can rewrite it in $(x'_1,x'_2,y)$ coordinates:
\bea
ds_3^2\equiv dy^2+e^D(dx^2_1+dx_2^2)=dy^2+e^{D'}((dx'_1)^2+(dx'_2)^2),\qquad
D'=D+\log |{\d f}|^2\nonumber
\eea
An important property of Toda equation is that if $D$ satisfies (\ref{TodaEqn}) in variables $(x_1,x_2,y)$,
then $D'$ satisfies the same equation in variables $(x'_1,x'_2,y)$. In particular we conclude that 
two solutions of the Toda equation which differ by a logarithm of any harmonic function  
$f(x_1,x_2)$ lead to the same two dimensional metric. Moreover, since the conformal transformation written above does not change $y$ or $\d_y D$, the warp factors and $g_{\chi\chi}$ appearing in 
(\ref{LLMCont}) are not affected by it, while $V_i$ is shifted in a very simple way. 

We can now use this "gauge transformation" to find an alternative set of boundary conditions. 
Let us start with function $D$ obeying (\ref{TypeIIBound}) and introduce $H(x_1,x_2)$ as a solution of an equation 
\bea
(\d_1^2+\d_2^2)H=-\sum_{i=1}^n \delta({\bf x}-{\bf x}^{(i)}),\qquad 
H|_{x_1^2+x_2^2\rightarrow \infty}=1
\eea
Notice that due to the maximum principle, this function never vanishes. Then one can find a new
solution of Toda equation ${\tilde D}$: $e^{\tilde D}=H^{-1}e^D$ and $e^{\tilde D}$ remains finite everywhere. This function still vanishes at $y=0$, but in addition it also vanishes on the rods which are complimentary to (\ref{TypeIIBound}): ${\bf x}={\bf x}^{(i)}$, $y<y^{(i)}$. Since ${\tilde D}$ describes the same solution as the original $D$, we conclude that there is an alternative set of boundary conditions which lead to regular geometry (we drop the tilde from ${\tilde D}$):
\bea\label{AltTypeIIBound}
&&D\sim \log y,\quad y\rightarrow 0\nonumber\\
&&D\sim \log |{\bf x}-{\bf x}^{(i)}|,\quad {\bf x}\rightarrow {\bf x}^{(i)},\ y<y^{(i)},\quad i=1,\dots n
\eea
One more condition should be added at at infinity, and to have $AdS_4\times S^7$ asymptotics we
require
\bea
e^D\sim y,\quad \sqrt{x^2+y^2}\rightarrow \infty
\eea
These boundary conditions are depicted in figure \ref{FigTypeII}b and in section \ref{SubsTodaProbe} 
we will use them to make a connection with probe M2 branes discussed in subsection \ref{SubsM2Intrs}.

\subsection{Example: $AdS_4\times S^7$}

As an example of the construction described above, we will rewrite the $AdS_4\times S^7$ metric 
in terms of parameterization (\ref{LLMCont}). Starting from the metric (\ref{AdSMetrU1}), and looking at the warp factors of $AdS_2$ and $S^5$, we find the expressions for $y$ and $e^{2\la}$. This allows one to write $g_{\chi\chi}$ in terms of $(\rho,\theta)$ and extract the relation between $\phi,\psi$ and 
$\chi$: 
\bea\label{A4S7Chi}
\phi=\chi,\quad \psi=2\chi+\alpha
\eea
Notice that at this stage there is a certain ambiguity in defining $\alpha$ (we can shift $\chi$ or rescale $\alpha$) which corresponds to the gauge freedom in $V_i$. This freedom was fixed in a particular way in the solution (\ref{LLMCont}) and it turns out that coordinates (\ref{A4S7Chi}) lead to the same gauge choice as (\ref{LLMCont}) (see (\ref{CheckGauge})).

At this point we know the explicit expressions for three coordinates 
$(y,\chi,\alpha)$, so one can define the last coordinate $x$ to be orthogonal to them. Rewriting the 
metric (\ref{AdSMetrU1}) in terms of $(x,y,\chi,\alpha)$, we can read off all ingredients appearing in 
(\ref{LLMCont}): 
\bea\label{A4S7TdSln}
&&x=L^3\sinh\rho \cos^2\theta,\quad y=L^3\cosh\rho\sin^2\theta,\quad 
x_1+ix_2=xe^{i\alpha}
\nonumber\\
&&e^{2\la}=L^2\sin^2\theta,\quad
V=-\frac{1}{2}\frac{\sinh^2\rho~d\alpha}{\sinh^2\rho+\cos^2\theta},
\quad e^D=\tan^2\theta\\
&&g_{\chi\chi}=4L^2(\sinh^2\rho+\cos^2\theta)\nonumber
\eea
One can check that $D$ satisfies the Toda equation (\ref{TodaEqn}) and the relations (\ref{LLMCont}) between $e^{-6\la}$, $V$ and $D$ hold. For example, one can check that 
(\ref{A4S7TdSln}) and (\ref{LLMCont}) are written in the same gauge:
\bea\label{CheckGauge}
\d_x D=-\frac{\sin\theta\sinh\rho}{2L^3\cos\theta(\sinh^2\rho+\cos^2\theta)}\d_\theta D=
-\frac{\sinh^2\rho}{x(\sinh^2\rho+\cos^2\theta)}=\frac{2}{x}V_\alpha
\eea
It is clear that this solution satisfies the boundary conditions (\ref{TypeIIBound}) with $n=1$:
\bea\label{BoundAAdS}
&&D\sim \log y,\quad y\rightarrow 0\nonumber\\
&&D\sim -\log x,\quad {\bf x}\rightarrow 0,\ y>L^3,
\eea
which corresponds to one stack of M2 branes. We also extract the behavior at infinity of 
$(x_1,x_2,y)$ space:
\bea
e^D=\frac{y}{x}+O(r^{-1}),\quad r\equiv \sqrt{x^2+y^2}\rightarrow\infty
\eea
This condition should be imposed on any geometry which asymptotes to $AdS_4\times S^7$.

We see that in terms of the general description (\ref{TypeIIBound}) this geometry corresponds to one rod which extends from $y=L^3$ to infinity, one might have suspected that this would be the case since we have only one stack of M2 branes. One can conjecture that additional stacks of M2 branes lead to more rods extending to infinity. However this picture is not very intuitive in the present coordinates: since $N$ scales like $L^6$, it seems that to add an extra stack of $k$ 
branes\footnote{We assume that the $AdS_4\times S^7$ is created by $N$ branes and add 
$k\ll N$ extra M2's. Notice that one should not consider the limit of small $N$, since it leads to singular geometry. To recover the flat space from $AdS_4\times S^7$, one should take $N$ to infinity, which means that a single rod starts at a very large value of $y$ and goes to infinity.}
one should introduce a rod which goes from infinity to $y\sim \sqrt{k}$. In particular as $k$ goes 
to zero, one is left with an extra rod going to $y=0$ which was not present in the original description (\ref{BoundAAdS}) of $AdS_4\times S^7$. As explained in the previous subsection, even with this extra rod one gets $AdS_4\times S^7$ geometry, but it is written in a 
non--conventional gauge for $D$, to recover (\ref{BoundAAdS}) one should use a complimentary description at least for the probe branes. When the number of probes becomes large, it is impossible to make a distinction between the M2's which created $AdS_4\times S^7$ and 
"extra branes", so one should go to the complimentary boundary conditions (\ref{BoundAAdS}) for all stacks, in this description $AdS_4\times S^7$ has the following parameterization:
\bea\label{TypeIIAdSMap}
x_1+ix_2=\frac{L^3}{2}\sqrt{\sinh\rho} \cos\theta e^{i\alpha/2},\quad 
e^D=\sinh\rho\sin^2\theta,\quad {\tilde\chi}=\chi-\frac{\alpha}{2}
\eea
and the expressions for $y$ and various warp factors remain unchanged. At large values of $\rho$ 
we find an asymptotic behavior of $D$:
\bea\label{TodaBCA4S7}
e^D=\frac{y}{L^3}+O(r^{-1})
\eea
Pictorially the boundary conditions for $AdS_7\times S^4$ solution (\ref{TypeIIAdSMap}) are represented by one finite rod.

\subsection{Relation to the brane probes.}

\label{SubsTodaProbe}

Let us connect the gravity solutions described here with brane probe analysis of section 
\ref{SubsM2Intrs}.
We begin with $AdS_4\times S^7$ (which corresponds to one rod at ${\bf x}=0$, $y<L^3$ in 
(\ref{AltTypeIIBound})) and introduce a small change in the boundary condition:
\bea
&&D\sim \log y, \quad y\rightarrow 0\nonumber\\
&&D\sim \log |{\bf x}|,\ {\bf x}\rightarrow 0,\ y<L^3;\quad
D\sim \log |{\bf x}-{\bf x}^{(1)}|,\ {\bf x}\rightarrow {\bf x}^{(1)},\ y<y_1\ll L^3\nonumber
\eea
The extra rod is located at ${\bf x}^{(1)}=(R\cos \frac{\alpha_1}{2},R\sin \frac{\alpha_1}{2})$ and 
for very small value 
of $y_1$ it degenerates into a point 
$(x_1,x_2,y)=(R\cos \frac{\alpha_1}{2},R\sin \frac{\alpha_1}{2},0)$. Using the map (\ref{A4S7Chi}), 
(\ref{TypeIIAdSMap}), we can find this location in the more standard coordinates on $AdS_4\times S^7$:
\bea
\theta=0,\quad \sinh\rho=\frac{4R^2}{L^6},\quad \phi={\tilde\chi}+\frac{\alpha_1}{2},\quad 
\psi=2{\tilde\chi}+2\alpha_1
\eea
In other words, we find that $\psi=2\phi+\alpha_1$, so one recovers a location of a probe
 M2 brane (\ref{PsiAndPhi}). If we put many such branes on top of each other, the rod at 
 ${\bf x}={\bf x}^{(1)}$ 
starts growing and it is interesting to find the relation between its length and the number of branes. 


\begin{figure}
\begin{center}
\epsfxsize=5in \epsffile{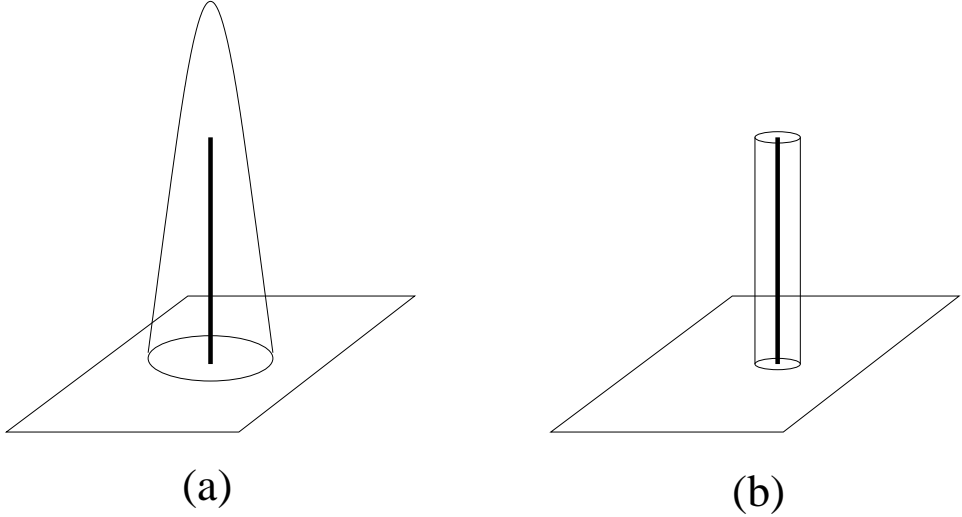}
\end{center}
\caption{
(a) To construct a non--contractible seven--manifold one should fiber $S^5$ over two--dimensional cap covering a rod.\newline
(b) To compute a flux over such manifold it it convenient to deform a cap into a thin cylinder, this leads to 
(\ref{rodFlux}).
} \label{FigTypeIICap}
\end{figure}


To compute the number of M2 branes one should evaluate a value of a flux over some closed seven--manifold. 
It is easy to identify such manifolds for a solution with generic boundary conditions (\ref{AltTypeIIBound}): one can take any rod and cover it with a cap terminating at 
$y=0$ (see figure \ref{FigTypeIICap}a). This gives a two dimensional surface and one can attach an 
$S^5$ fiber 
over each point of this surface. Then restricting the metric (\ref{LLMCont}) to this seven 
dimensional space, one finds that the resulting manifold is smooth and compact (since $S^5$ degenerates on $y=0$ plane). One can smoothly deform such manifolds into each other, but to avoid singularities, the two dimensional surface cannot touch the rods. This gives a topological classification of the 7--manifolds in the geometry: there is one non--contractible $S^7$ 
for each rod. Once the seven manifolds are classified, one can compute the flux of $*F_4$ through 
them. In particular we can deform a single cap into a small cylinder surrounding the rod (see figure 
\ref{FigTypeIICap}b), then the expression for the flux becomes especially simple:
\bea\label{rodFlux}
\int_{V_7} *F_4&=&\Omega_5 \int_{\Sigma_2}2^7 e^{12\la-D}(y^2e^{-6\la}-1)(d\chi+V)\wedge 
\d_y(y^{-1}\d_y e^D)dy\nonumber\\
&=&2^7\pi \Omega_5 \int_0^{y^{(i)}} 
\left.\frac{e^{12\la-D}dy}{y\d_y D-1}\d_y(y^{-1}\d_y e^D)\right|_{{\bf x}-{\bf x}^{(i)}}
\eea
We used polar coordinates $x_1+ix_2=re^{i\alpha/2}$ in the vicinity of the rod as well as 
the leading term in $V=\frac{1}{2}d\alpha$.

To summarize, we showed that the boundary conditions (\ref{AltTypeIIBound}) have a nice interpretation
in terms of stacks of M2 branes and the number of branes in each stack can be expressed as an integral along the corresponding rod (\ref{rodFlux}). We also reproduced the description of the probe M2 branes in $AdS_4\times S^7$ and found that such branes are placed at points 
$(x_1,x_2,y)=\frac{L^3}{2}\sqrt{\sinh\rho}(\cos\frac{\psi-2\phi}{2},\sin\frac{\psi-2\phi}{2},0)$ in supergravity description.

\section{Discussion}
\renewcommand{\theequation}{9.\arabic{equation}}
\setcounter{equation}{0}

In this article we considered various brane configurations which preserve $16$ supercharges in eleven dimensions. We began with analysis in the probe approximation and showed that multiple membranes
on $AdS_m\times S^n$ background expand into M5 branes with fluxes. This phenomenon have been encountered before for gravitons 
\cite{giantGrav,dualGiant} and fundamental strings \cite{CalMald,PawRey}. 
Just as giant gravitons, the M5 branes with fluxes appear in two varieties: the value of the membrane charge is bounded for one class of M5s and it can be arbitrarily large for another. We also saw that supersymmetric branes corresponding to the defects in field theory preserve either 
$SO(2,2)\times SO(4)^2$ or $SO(2,1)\times SO(6)$ symmetries and we constructed supergravity solutions for both cases. 

We showed that the metrics with $SO(2,2)\times SO(4)^2$ isometries are completely specified by a 
harmonic function of two variables and we classified the boundary conditions which lead to regular solutions. It turns out that the geometries have a very rich topological structure which can be read off from the boundary conditions and the brane probes are also encoded in these conditions in a very natural way. 
Our discussion focused on solutions with $AdS_4\times S^7$ and $AdS_7\times S^4$ asymptotics (in this case the amount of preserved supersymmetry is enhanced at infinity), and it would be very interesting to study more general asymptotic behavior. 

The geometries with $SO(2,1)\times SO(6)$ symmetry are constructed by analytic continuation of the 
solutions derived in \cite{LLM} and they are specified by one function satisfying a Toda equation. While 
such continuation gives a good local solution, the regularity conditions found in this paper are different from the ones given in \cite{LLM} since the topological structure of the solution is altered. By requiring regularity, we arrived at two classes of boundary conditions (called type I and type II) and one of them recovers all known brane probes. The other condition changes geometry in a more drastic way and it would be very nice to find its interpretation. 

Our results can be extended in various directions. From the point of view of AdS/CFT the most natural 
generalization involves fixing of $AdS_m\times S^n$ asymptotics and looking for states preserving $8$ or $4$ supercharges\footnote{Some examples of relevant eleven--dimensional geometries have been constructed in \cite{GauntJuly}.}. 
This problem is notoriously 
hard and only partial progress has been reported in 
$AdS_3\times S^3$ \cite{Qrt33} and $AdS_5\times S^5$ \cite{Qrt55} cases. Perhaps one 
should get a 
better understanding of those solutions before looking at the M theory setup. Another interesting extension involves the same brane configurations that were discussed in this paper and going beyond the near horizon limit. While such exercise is not useful for AdS/CFT correspondence, it might lead to a better understanding of intersecting branes in flat space.  Finally one can try to look for the backreaction of 
non--normalizable modes in $AdS_m\times S^n$, they parameterize the supersymmetric 
deformations of the dual field theories, so the gravity solutions might teach us something about their
dynamics.

In this paper we only discussed a bulk side of the duality and it would be very nice to find the 
corresponding nonlocal operators on the field theory side and compare the results. Unfortunately neither 
(2,0) theory in six dimensions nor three--dimensional superconformal theory with $32$ supercharges are 
understood well enough to perform such comparison. 

\section*{Acknowledgments}

I want to thank David Kutasov and Savdeep Sethi for useful discussions.
This work was supported by DOE grant DE-FG02-90ER40560.

\appendix

\section{Derivation of 1/2--BPS solutions in M theory}

\label{SectAppA}

\renewcommand{\theequation}{A.\arabic{equation}}
\setcounter{equation}{0}

In sections \ref{SectA7S4}, \ref{SectA4S7} we discussed various branes preserving $16$ supercharges. 
In particular we analyzed three classes of branes which preserve 
$SO(2,1)\times SO(3)\times SO(3)$ symmetry (see subsections \ref{SubsA7M5S}, \ref{SubsA7M5A} 
and \ref{SubsA4M5S}) and in this appendix we construct the supergravity solutions produced by such 
branes. Logically the analysis of this appendix is similar to the derivation of $1/2$ BPS geometries in type IIB supergravity presented in \cite{myWils}, but the resulting geometries are not related to the ones constructed there.

\subsection{Ansatz and equations for Killing spinors}

\label{SubsAppAns}

We begin with recalling the equation for Killing spinors in 11 dimensional supergravity:
\bea\label{Orig11DVer}
\nabla_m\eta+\frac{1}{288}\left[{\gamma_m}^{npqr}-
8\delta^n_m\gamma^{pqr}\right]G_{npqr}\eta=0
\eea
Using analogy with geometries describing Wilson lines, we consider an ansatz with 
$SL(2)\times SO(4)\times SO(4)$ symmetry \cite{yama}:
\bea
ds^2&=&e^{2A}ds_H^2+e^{2B}ds_S^2+e^{2C}ds_{\tilde S}^2+
h_{ij}dx^i dx^j\nonumber\\
F_4&=&df_1\wedge dH_3+df_2\wedge d\Omega_3+
df_3\wedge d{\tilde\Omega}_3
\eea
Here $ds^2_S$ and $ds^2_{\tilde S}$ represent metrics on unit spheres $S^3$, ${\tilde S}^3$, and 
$ds_H^2$ is a metric on $AdS_3$ with unit radius. We also have an undetermined metric in two dimensions $h_{ij}dx^i dx^i$, and all scalars are functions of $x_1,x_2$. It is convenient to choose a specific basis of gamma matrices:
\bea
x_1,x_2:&&\Gamma_{1,2}=\sigma_{1,2}\otimes 1_{16},\nonumber\\
AdS:&&\Gamma_m=\sigma_3\otimes \sigma_1\otimes \sigma_m\otimes 1_{4},\nonumber\\
S:&&\Gamma_m=\sigma_3\otimes \sigma_2\otimes 1_2\otimes \sigma_m\otimes 1_{2}\\
{\tilde S}:&&\Gamma_m=\sigma_3\otimes \sigma_2\otimes 1_4\otimes \sigma_m\nonumber
\eea
Next we impose the symmetries on the Killing spinors. If we introduce three matrices\footnote{Notice that these matrices satisfy a relation 
$\Gamma_H\Gamma_S\Gamma_\Omega\Gamma_1\Gamma_2=-1$}\label{GamFoot}
\bea
\Gamma_H=\Gamma^0\Gamma^3\Gamma^4,\quad 
\Gamma_S=-i\Gamma^{567},\quad
\Gamma_{\Omega}=-i\Gamma^{89(10)}
\eea
then the invariant spinors are defined by
\bea\label{SpinDer1}
{\tilde\nabla}^H_m\eps=\frac{a}{2}e^{-A}\Gamma_H\gamma_m\eps,\quad
{\tilde\nabla}^S_m\eps=\frac{ib}{2}e^{-B}\Gamma_S\gamma_m\eps,\quad
{\tilde\nabla}^\Omega_m\eps=\frac{ic}{2}e^{-C}\Gamma_\Omega\gamma_m\eps
\eea
Here ${\tilde\nabla}^H_m$ is a derivative on a unit AdS, which is related to a total derivative as
\bea\label{SpinDer2}
{\nabla}^H_m={\tilde\nabla}^H_m-\frac{1}{2}{\gamma^\mu}_m\d_\mu A.
\eea
Similar relations hold for the spheres as well. We also introduced three independent signs 
$(a,b,c)$: $a^2=b^2=c^2=1$.

To analyze the equation (\ref{Orig11DVer}) it is convenient to introduce a matrix ${\not G}$:
\bea
\frac{1}{288}{\not G}=\frac{1}{12}\left[e^{-3A}{\not\d}f_1\Gamma_H+
ie^{-3B}{\not\d}f_2\Gamma_S+ie^{-3C}{\not\d}f_3\Gamma_\Omega
\right]\equiv G_H+G_S+G_\Omega
\eea
and rewrite (\ref{Orig11DVer}) in terms of it
\bea\label{Mod11DVer}
&&\nabla_m\eta+\frac{1}{288}
\left[-\frac{1}{2}\gamma_m {\not G}+\frac{3}{2}{\not G}\gamma_m\right]\eta=0
\eea
Taking various components of this equation, we find\footnote{Here and below all indices take two values corresponding to $x_1$ and $x_2$.} 
\bea\label{11DEa}
&&\left[\frac{a}{2}e^{-A}\Gamma_H+\frac{1}{2}{\not\d}A+(G_S+G_\Omega-2G_H)\right]\eta=0\\
\label{11DEb}
&&\left[\frac{ib}{2}e^{-B}\Gamma_S+\frac{1}{2}{\not\d}B+(-2G_S+G_\Omega+G_H)\right]\eta=0\\
\label{11DEc}
&&\left[\frac{ic}{2}e^{-C}\Gamma_\Omega+\frac{1}{2}{\not\d}C+
(G_S-2G_\Omega+G_H)\right]
\eta=0\\
\label{11DEqn}
&&\nabla_\mu\eta+
\left[-\frac{1}{2}\gamma_\mu (G_H+G_S+G_\Omega)+\frac{3}{2}(G_H+G_S+G_\Omega)
\gamma_\mu\right]\eta=0
\eea
Combining the first three relations, we can construct one projector that does not contain fluxes:
\bea\label{GeomProj}
\left[ae^{-A}\Gamma_H+ibe^{-B}\Gamma_S+ice^{-C}\Gamma_\Omega+{\not\d}(A+B+C)\right]\eta=0
\eea
This suggests a convenient choice of a coordinate: $y=e^{A+B+C}$. Let us also choose the last coordinate $x$ to be orthogonal 
to $y$, then taking $g_{yy}=g_y^2$, we can rewrite the last equation as
\bea
\left[ae^{-A}\Gamma_H+ibe^{-B}\Gamma_S+ice^{-C}\Gamma_\Omega+\frac{1}{yg_y}\Gamma_y\right]\eta=0
\eea
The square of this projector relates $g_y$ with various warp factors:
\bea\label{gYY}
g^{-1}_y=y\sqrt{-e^{-2A}+e^{-2B}+e^{-2C}}
\eea
To restrict the metric further it is convenient to look at various bilinears. 


\subsection{Scalar bilinears}


In this subsection we consider the spinor bilinears which do not carry any indices. The simplest such bilinear is $\eta^\dagger\eta$, but we can also sandwich any combination of matrices 
$\Gamma_\Omega,\Gamma_S,\Gamma_H$ between two spinors. This leads to $1+3+3+1=8$ independent expressions. Notice that due to the relation 
$\Gamma_H\Gamma_S\Gamma_\Omega=-\Gamma_{12}$, one can use 
$\eta^\dagger\Gamma_{12}\eta$ instead of a bilinear containing three matrices.

We begin with fixing normalization of $\eta$. To this end we compute
\bea\label{11DEqnDag}
&&\nabla_\mu \eta^\dagger +\eta^\dagger
\left[-\frac{1}{2}(-G_H+G_S+G_\Omega)\gamma_\mu+\frac{3}{2}\gamma_\mu(-G_H+G_S+G_\Omega)
\right]=0\\
&&\nabla_\mu(\eta^\dagger\eta)+\eta^\dagger
\left[\gamma_\mu (-2G_H+G_S+G_\Omega)+(2G_H+G_S+G_\Omega)
\gamma_\mu\right]\eta=0\nonumber\\
&&\nabla_\mu(\eta^\dagger\eta)+\eta^\dagger
\left\{\gamma_\mu, -\frac{a}{2}e^{-A}\Gamma_H-\frac{1}{2}{\not\d}A\right\}\eta=0\nonumber
\eea
This implies a normalization $\eta^\dagger\eta=e^A$. Notice that equation (\ref{11DEqnDag}) will be useful later on.

Next we look at (\ref{GeomProj}) and consider various bilinears. The conjugate equation is
\bea\label{GeomProjCnj}
&&\eta^\dagger\left[ae^{-A}\Gamma_H-ibe^{-B}\Gamma_S-ice^{-C}\Gamma_\Omega+{\not\d}\log y\right]=0
\eea
Multiplying this equation by $\gamma_\mu dx^\mu\Gamma_S\Gamma_\Omega\eta$ and combining the results with (\ref{GeomProj}), we find
\bea
dy~\eta^\dagger\Gamma_S\Gamma_\Omega\eta=0:\quad \eta^\dagger\Gamma_S\Gamma_\Omega\eta=0
\eea
Similarly, multiplications by $M\eta$ with various $M$ lead to relations
\bea\label{OmSbil1}
\eta:&&\eta^\dagger\left[ibe^{-B}\Gamma_S+ice^{-C}\Gamma_\Omega\right]\eta=0\\
\label{GamHSndw}
\Gamma_H\eta:&&
\eta^\dagger\left[ae^{-A}-ibe^{-B}\Gamma_S\Gamma_H-ice^{-C}\Gamma_\Omega\Gamma_H\right]\eta=0\\
\label{Gam56BilA}
\Gamma_S\Gamma_{12}\eta:&&
ae^{-A}\eta^\dagger\Gamma_S\Gamma_H\Gamma_{12}\eta +ibe^{-B}\eta^\dagger\Gamma_{12}\eta=0\\
\label{Gam56BilB}
\Gamma_\Omega\Gamma_{12}:&&ae^{-A}\eta^\dagger\Gamma_\Omega\Gamma_H\Gamma_{12}\eta +ice^{-C}\eta^\dagger\Gamma_{12}\eta=0\\
\label{PreDefYForm}
\gamma_\mu dx^\mu\eta:&&\eta^\dagger\left[-ibe^{-B}\Gamma_S\gamma_\mu-ice^{-C}\Gamma_\Omega\gamma_\mu\right]\eta 
dx^\mu+e^A d\log y=0\\
\label{GamHy}
&&ae^{-A}\eta^\dagger\Gamma_H\gamma_\mu\eta~dx^\mu
-*d \log y~\eta^\dagger\Gamma_{12}\eta=0
\eea
Next  we subtract equation (\ref{11DEc}) from (\ref{11DEb}) and take a hermitean conjugate of the result:
\bea
&&\left[ibe^{-B}\Gamma_S-ice^{-C}\Gamma_\Omega+{\not\d}(B-C)+3(G_\Omega-G_S)\right]\eta=0
\nonumber\\
&&\eta^\dagger\left[ibe^{-B}\Gamma_S-ice^{-C}\Gamma_\Omega-{\not\d}(B-C)-3(G_\Omega-G_S)\right]=0
\nonumber
\eea 
These equations can be combined to find a relation between scalar bilinears similar to 
(\ref{OmSbil1}):
\bea
be^{-B}\eta^\dagger\Gamma_S\eta-ce^{-C}\eta^\dagger\Gamma_\Omega\eta=0
\eea
However due to the difference in the coefficients, we conclude that
\bea
\eta^\dagger\Gamma_S\eta=\eta^\dagger\Gamma_\Omega\eta=0
\eea
Then (\ref{Gam56BilA}) and (\ref{GamHy}) imply that
\bea\label{11DZeroVec}
\eta^\dagger\Gamma_{12}\eta=0,\qquad \eta^\dagger\Gamma_H\gamma_\mu\eta=0
\eea

At this point we have only four nontrivial scalar bilinears, one of them was computed 
($\eta^\dagger\eta=e^A$), and 
now we will derive the differential equations for three remaining scalars: 
$\eta^\dagger\Gamma_H\eta$,
$\eta^\dagger\Gamma_H\Gamma_S\eta$, $\eta^\dagger\Gamma_H\Gamma_\Omega\eta$.

We begin with computing a derivative of $\eta^\dagger\Gamma_H\eta$:
\bea
&&\nabla_\mu (\eta^\dagger\Gamma_H\eta)+\eta^\dagger\Gamma_H
\left[-\frac{1}{2}\gamma_\mu(G_H+G_S+G_\Omega)+\frac{3}{2}(G_H+G_S+G_\Omega)\gamma_\mu
\right]\eta\nonumber\\
&&\qquad+\eta^\dagger
\left[-\frac{1}{2}(-G_H+G_S+G_\Omega)\gamma_\mu+\frac{3}{2}\gamma_\mu(-G_H+G_S+G_\Omega)
\right]\Gamma_H\eta=0\nonumber
\eea
This expression can be simplified in two different ways: we can either write it in terms of scalar bilinears:
\bea
&&\nabla_\mu (\eta^\dagger\Gamma_H\eta)+2\eta^\dagger\Gamma_H
\left[G_H+G_S+G_\Omega,\gamma_\mu\right]\eta=0
\nonumber\\
\label{DerGamH1}
&&\nabla_\mu(\eta^\dagger\Gamma_H\eta)-\frac{1}{3}\eta^\dagger\Gamma_H
(e^{-3A}\d_\mu f_1\Gamma_H+ie^{-3B}\d_\mu f_2\Gamma_S+ie^{-3C}\d_\mu f_3\Gamma_\Omega)\eta=0,
\eea
or we can eliminate $f_2$ and $f_3$: 
\bea
&&\nabla_\mu(\eta^\dagger\Gamma_H\eta)+2\eta^\dagger
\left[-(-G_H+G_S+G_\Omega)\gamma_\mu+\gamma_\mu(-G_H+G_S+G_\Omega)\right]\Gamma_H\eta
=0\nonumber\\
&&\nabla_\mu(\eta^\dagger\Gamma_H\eta)+\eta^\dagger
\left[6G_H\gamma_\mu-6\gamma_\mu G_H\right]\Gamma_H\eta-
2\eta^\dagger\left[ae^{-A}\gamma_\mu-\d_\mu A\Gamma_H\right]\eta=0\nonumber\\
&&\nabla_\mu(\eta^\dagger\Gamma_H\eta)-e^{-3A}\d_\mu f_1\eta^\dagger\eta-
2ae^{-A}\eta^\dagger\gamma_\mu\eta+2\d_\mu A\eta^\dagger\Gamma_H\eta=0\nonumber
\eea
The last equation can be rewritten as an expression for one of the vector bilinears:
\bea\label{VectorExpr}
2ae^A \eta^\dagger\gamma_\mu\eta=\d_\mu(e^{2A}\eta^\dagger\Gamma_H\eta)-\d_\mu f_1
\eea
In the process of making simplifications we used two relations between bilinears which can be obtained by combining various projectors:
\bea
&&\eta^\dagger\gamma_\mu\Gamma_H
\left[\frac{a}{2}e^{-A}\Gamma_H+\frac{1}{2}{\not\d}A+(G_S+G_\Omega-2G_H)\right]\eta\nonumber\\
&&\qquad-
\eta^\dagger\left[\frac{a}{2}e^{-A}\Gamma_H+\frac{1}{2}{\not\d}A+(G_S+G_\Omega+2G_H)\right]
\gamma_\mu\Gamma_H\eta=0\nonumber\\
&&\eta^\dagger\left[\gamma_\mu(G_S+G_\Omega+2G_H)-(G_S+G_\Omega+2G_H)\gamma_\mu\right]
\Gamma_H\eta\nonumber\\
&&\qquad+
\eta^\dagger\left[ae^{-A}\gamma_\mu-\d_\mu A\Gamma_H\right]\eta=0\nonumber
\eea

Let us now look at derivatives of the two remaining scalars. Taking an appropriate bilinear of the differential equation
and using various projectors, we find
\bea
&&\nabla_\mu (\eta^\dagger\Gamma_\Omega\Gamma_H\eta) +\eta^\dagger
\left[-\frac{1}{2}(-G_H+G_S+G_\Omega)\gamma_\mu+\frac{3}{2}\gamma_\mu(-G_H+G_S+G_\Omega)
\right]\Gamma_\Omega\Gamma_H\eta\nonumber\\
&&\qquad+\eta^\dagger\Gamma_\Omega\Gamma_H
\left[-\frac{1}{2}\gamma_\mu(G_H+G_S+G_\Omega)+\frac{3}{2}(G_H+G_S+G_\Omega)\gamma_\mu
\right]\eta=0\nonumber\\
&&\nabla_\mu (\eta^\dagger\Gamma_\Omega\Gamma_H\eta) +\eta^\dagger
\left[(-G_H-2G_\Omega+G_S)\gamma_\mu\Gamma_\Omega\Gamma_H+
\Gamma_\Omega\Gamma_H\gamma_\mu(G_H+G_S-2G_\Omega)
\right]\eta=0\nonumber\\
&&2\nabla_\mu (\eta^\dagger\Gamma_\Omega\Gamma_H\eta) +\eta^\dagger
\left[(ice^{-C}\Gamma_\Omega-{\not\d}C)\gamma_\mu\Gamma_\Omega\Gamma_H+
\Gamma_\Omega\Gamma_H\gamma_\mu(-ice^{-C}\Gamma_\Omega-{\not\d}C)
\right]\eta=0\nonumber
\eea
This implies a very simple relation
\bea
&&\d_\mu(e^{-C}\eta^\dagger\Gamma_\Omega\Gamma_H\eta)=0:\quad 
\eta^\dagger\Gamma_\Omega\Gamma_H\eta=ic_1 e^C
\eea
and similar manipulations lead to
\bea
&&\d_\mu(e^{-B}\eta^\dagger\Gamma_S\Gamma_H\eta)=0:\quad 
\eta^\dagger\Gamma_S\Gamma_H\eta=ic_2 e^B
\eea
The constants $c_1$ and $c_2$ are not determined, but one relation between them can be found 
using (\ref{GamHSndw}):
\bea\label{11Dabc}
bc_2+cc_1=-a
\eea
It appears that there is no further restrictions which allows one to fix $c_1$ and $c_2$ completely. In fact, 
there are two known solutions with $AdS_3\times S^3\times S^3$ symmetries ($AdS_7\times S^4$ and 
$AdS_4\times S^7$) and they have different values of $c_1$ and $c_2$. 

Let us summarize the content of this subsection. We found that four out of eight scalar bilinears vanish and we also determined the other three:
\bea
&&\eta^\dagger\Gamma_S\eta=\eta^\dagger\Gamma_\Omega\eta=
\eta^\dagger\Gamma_S\Gamma_\Omega\eta=\eta^\dagger\Gamma_{12}\eta=0
\nonumber\\
&&\eta^\dagger\eta=e^A,\quad \eta^\dagger\Gamma_\Omega\Gamma_H\eta=ic_1 e^C,\quad
\eta^\dagger\Gamma_S\Gamma_H\eta=ic_2 e^B
\eea
The last scalar bilinear $\eta^\dagger\Gamma_H\eta$ is still undetermined, but its derivative can be expressed in terms of fluxes using  (\ref{DerGamH1}):
\bea\label{DerGamH}
d(\eta^\dagger\Gamma_H\eta)=\frac{1}{3}\left[
e^{-2A}d f_1+c_2e^{-2B}d f_2+c_1e^{-2C}d f_3\right]
\eea
To find an algebraic expression for this last scalar we use Fierz identities. If we define four dimensional vectors as
\bea
&&K_m=(\eta^\dagger \gamma_\mu\eta,\eta^\dagger \Gamma_S\eta,\eta^\dagger \Gamma_\Omega\eta),
\qquad
L_m=(\eta^\dagger \Gamma_H\gamma_\mu\eta,\eta^\dagger \Gamma_H\Gamma_S\eta,
          \eta^\dagger \Gamma_H\Gamma_\Omega\eta)\nonumber
\eea
then Fierz identities lead to the following relations:
\bea
&&K_m L^m=0,\quad K^2=-L^2=(\eta^\dagger\eta)^2-(\eta^\dagger\Gamma_H\eta)^2\nonumber
\eea
Writing this more explicitly and using the fact that $\eta^\dagger\Gamma_H\gamma_\mu\eta=0$, 
we find a relation
\bea\label{11DFierz}
(\eta^\dagger \gamma_\mu\eta)(\eta^\dagger \gamma^\mu\eta)=
-(\eta^\dagger \Gamma_H\Gamma_S\eta)^2-(\eta^\dagger \Gamma_H\Gamma_\Omega\eta)^2=
(\eta^\dagger\eta)^2-(\eta^\dagger\Gamma_H\eta)^2
\eea
which in particular leads to
\bea\label{11DLastScal}
\eta^\dagger\Gamma_H\eta=e_1\sqrt{e^{2A}-c_1^2 e^{2C}-c_2^2 e^{2B}}\equiv e_1 e^F,\quad e_1^2=1
\eea

As a byproduct of this analysis of scalars, we also found three relations (\ref{PreDefYForm}), 
(\ref{11DZeroVec}), (\ref{VectorExpr}) containing vector bilinears
\bea\label{11DAccVect}
&&\eta^\dagger\Gamma_H\gamma_\mu\eta=0\nonumber\\
&&2ae^A \eta^\dagger\gamma_\mu\eta=\d_\mu(e^{2A}\eta^\dagger\Gamma_H\eta)-\d_\mu f_1\\
&&\left[ibe^{C}\eta^\dagger\Gamma_S\gamma_\mu\eta+
ice^{B}\eta^\dagger\Gamma_\Omega\gamma_\mu\eta\right]dx^\mu=dy\nonumber
\eea 
In the next subsection we will analyze the differential equations for the three nontrivial vectors which
appear in these equations.


\subsection{Vector bilinears}


Let us now write differential equations for the vector bilinears. Due to the restriction on the product of gamma matrices (see footnote on page \pageref{GamFoot}), only the following 
vectors are independent:
\bea
\eta^\dagger\gamma_\mu\eta,\quad \eta^\dagger\gamma_\mu\Gamma_H\eta,\quad 
\eta^\dagger\gamma_\mu\Gamma_S\eta,\quad \eta^\dagger\gamma_\mu\Gamma_\Omega\eta.
\eea
We already know that $ \eta^\dagger\gamma_\mu\Gamma_H\eta=0$, and now we look at the equations for the 
three remaining vectors. We begin with
\bea
&&\nabla_\mu (\eta^\dagger\gamma_\nu\eta)+\eta^\dagger\gamma_\nu
\left[-\frac{1}{2}\gamma_\mu(G_H+G_S+G_\Omega)+\frac{3}{2}(G_H+G_S+G_\Omega)\gamma_\mu
\right]\eta
\nonumber\\
&&\qquad+\eta^\dagger
\left[-\frac{1}{2}(-G_H+G_S+G_\Omega)\gamma_\mu+\frac{3}{2}\gamma_\mu(-G_H+G_S+G_\Omega)
\right]\gamma_\nu\eta=0\nonumber
\eea
Antisymmetric part of this equation can be simplified further
\bea\label{Vec2Apr}
&&\nabla_{[\mu} (\eta^\dagger\gamma_{\nu]}\eta)+\eta^\dagger
\left[-\frac{1}{2}\gamma_{\nu\mu}(G_H+G_S+G_\Omega)+3\gamma_{[\nu}G_H\gamma_{\mu]}
\right]\eta
\nonumber\\
&&\qquad-\eta^\dagger\frac{1}{2}(-G_H+G_S+G_\Omega)\gamma_{\mu\nu}\eta=0
\nonumber\\
&&\nabla_{[\mu} (\eta^\dagger\gamma_{\nu]}\eta)-
\eta^\dagger\gamma_{\nu\mu}(G_S+G_\Omega)\eta=0
\eea
To eliminate fluxes from this equation, we multiply (\ref{11DEa}) by $\eta^\dagger\Gamma_{12}$ and 
subtract the conjugate expression:
\bea
&&\eta^\dagger\Gamma_{12}\left[{\not\d}A+2(G_S+G_\Omega)\right]\eta=0
\eea
This allows us to simplify the equation (\ref{Vec2Apr}) for the vector:
\bea
&&\nabla_{[\mu} (\eta^\dagger\gamma_{\nu]}\eta)+
\frac{1}{2}\eps_{\nu\mu}\eta^\dagger\Gamma_{56}{\not\d}A\eta=0\nonumber\\
&&\nabla_{[\mu} (\eta^\dagger\gamma_{\nu]}\eta)+
\frac{1}{2}\eps_{\nu\mu}\d_\alpha A(-\eps^{\alpha\beta})\eta^\dagger \gamma_\beta\eta=0\nonumber
\eea
Rewriting this relation in terms of forms:
\bea
d(\eta^\dagger\gamma_{\mu}\eta~dx^\mu)+dA\wedge \eta^\dagger\gamma_{\mu}\eta~dx^\mu=0:\qquad 
d(e^A \eta^\dagger\gamma_{\mu}\eta~dx^\mu)=0,
\eea
we observe that the resulting equation can be viewed as integrability condition of (\ref{VectorExpr}) and does not lead to new information.
The trace part of the differential equation gives
\bea\label{VectLapl1}
&&\nabla_\mu (\eta^\dagger\gamma^\mu\eta)+\eta^\dagger
\left[-(G_H+G_S+G_\Omega)-(-G_H+G_S+G_\Omega)\right]\eta=0
\nonumber\\
&&\nabla_\mu (\eta^\dagger\gamma^\mu\eta)-2\eta^\dagger(G_S+G_\Omega)\eta=0\nonumber\\
&&\nabla_\mu (\eta^\dagger\gamma^\mu\eta)+\eta^\dagger(ae^{-A}\Gamma_H+{\not\d}A)\eta=0
\nonumber\\
&&\nabla_\mu (e^A\eta^\dagger\gamma^\mu\eta)+a\eta^\dagger\Gamma_H\eta=0
\eea
Next we look at a derivative of the vector $\eta^\dagger\Gamma_\Omega\gamma_\nu\eta$:
\bea
&&\nabla_\mu (\eta^\dagger\Gamma_\Omega\gamma_\nu\eta) +\eta^\dagger
\left[-\frac{1}{2}(-G_H+G_S+G_\Omega)\gamma_\mu+\frac{3}{2}\gamma_\mu(-G_H+G_S+G_\Omega)
\right]\Gamma_\Omega\gamma_\nu\eta\nonumber\\
&&\qquad+\eta^\dagger\Gamma_\Omega\gamma_\nu
\left[-\frac{1}{2}\gamma_\mu(G_H+G_S+G_\Omega)+\frac{3}{2}(G_H+G_S+G_\Omega)\gamma_\mu
\right]\eta=0\nonumber
\eea
Antisymmetric part of this expression gives
\bea
&&\nabla_{[\mu} (\eta^\dagger\Gamma_\Omega\gamma_{\nu]}\eta) +\eta^\dagger
\left[\frac{1}{2}(-G_H+G_S+G_\Omega)\gamma_{\mu\nu}\Gamma_\Omega\right]\eta\nonumber\\
&&\qquad+\frac{1}{2}\eta^\dagger\Gamma_\Omega\gamma_{\mu\nu}(G_H+G_S+G_\Omega)\eta=0
\nonumber\\
&&\nabla_{[\mu} (\eta^\dagger\Gamma_\Omega\gamma_{\nu]}\eta) +\frac{1}{2}\eta^\dagger
\left[\frac{ib}{2}e^{-B}\Gamma_S-\frac{1}{2}{\not\d}B+3G_S\right]
\gamma_{\mu\nu}\Gamma_\Omega\eta\nonumber\\
&&\qquad+\frac{1}{2}\eta^\dagger\Gamma_\Omega\gamma_{\mu\nu}
(-\frac{ib}{2}e^{-B}\Gamma_S-\frac{1}{2}{\not\d}B+3G_S)\eta=0
\nonumber\\
&&\nabla_{[\mu} (\eta^\dagger\Gamma_\Omega\gamma_{\nu]}\eta) +\eta^\dagger
\left[\frac{ib}{2}e^{-B}\Gamma_S-\frac{1}{2}{\not\d}B\right]
\gamma_{\mu\nu}\Gamma_\Omega\eta=0\nonumber
\eea
Using the relation $\Gamma_S\Gamma_\Omega\Gamma_{12}=-\Gamma_H$, we can simplify this equation:
\bea\label{11DVecDer1}
\nabla_{[\mu} (e^B\eta^\dagger\Gamma_\Omega\gamma_{\nu]}\eta) -
\frac{ib}{2}\eps_{\mu\nu}\eta^\dagger\Gamma_H\eta=0
\eea
Similarly, for the last vector bilinear we find
\bea\label{11DVecDer2}
\nabla_{[\mu} (e^C\eta^\dagger\Gamma_S\gamma_{\nu]}\eta) +
\frac{ic}{2}\eps_{\mu\nu}\eta^\dagger\Gamma_H\eta=0
\eea
The trace part of the equation for $\eta^\dagger\Gamma_\Omega\gamma_\nu\eta$ gives
\bea
&&\nabla_\mu (\eta^\dagger\Gamma_\Omega\gamma^\mu\eta) +\eta^\dagger
\left[(-G_H+G_S+G_\Omega)\Gamma_\Omega-\Gamma_\Omega(G_H+G_S+G_\Omega)\right]\eta=0\nonumber\\
&&\nabla_\mu (\eta^\dagger\Gamma_\Omega\gamma^\mu\eta) +\eta^\dagger
\left[(\frac{ib}{2}e^{-B}\Gamma_S-\frac{1}{2}{\not\d}B)\Gamma_\Omega-\Gamma_\Omega
(-\frac{ib}{2}e^{-B}\Gamma_S-\frac{1}{2}{\not\d}B)\right]\eta=0\nonumber
\eea
This equation along with its counterpart for $\eta^\dagger\Gamma_S\gamma^\mu\eta$ leads to "conserved currents":
\bea\label{VectLapl2}
&&\nabla_\mu (\eta^\dagger\Gamma_\Omega\gamma^\mu\eta) -
\eta^\dagger{\not\d}B\Gamma_\Omega\eta=0:\quad 
\nabla_\mu (e^{B}\eta^\dagger\Gamma_\Omega\gamma^\mu\eta)=0\\
\label{VectLapl3}
&&\nabla_\mu (\eta^\dagger\Gamma_S\gamma^\mu\eta) -
\eta^\dagger{\not\d}C\Gamma_S\eta=0:\quad 
\nabla_\mu (e^{C}\eta^\dagger\Gamma_S\gamma^\mu\eta)=0
\eea
At this point it is useful to recall the last equation in (\ref{11DAccVect}):
\bea
dy=ibe^{C}\eta^\dagger\Gamma_S\gamma_\mu\eta dx^\mu+ice^{B}\eta^\dagger\Gamma_\Omega\gamma_\mu\eta  dx^\mu
\eea
In particular, the relations (\ref{VectLapl2}), (\ref{VectLapl3}) imply that $d*dy=0$, then we can define a coordinate $x$ such that 
$dx=*dy$. With this choice the metric becomes
\bea\label{XYmetr}
ds^2=e^{2A}ds_H^2+e^{2B}ds_S^2+e^{2C}ds_{\tilde S}^2+g_y^2(dx^2+dy^2)
\eea
with $g_y$ given by (\ref{gYY}).

Finally let us evaluate various components of the vector $\eps^\dagger \gamma_\mu\eps$. 
Multiplying (\ref{GeomProj}) by $\eta^\dagger$ and adding the conjugate relation, we find:
\bea
\frac{1}{yg_y}\eta^\dagger\Gamma_y\eta=-ae^{-A}\eta^\dagger\Gamma_H\eta,\nonumber
\eea
and similar multiplication by $\eta^\dagger\Gamma_{xy}$ gives
\bea
\frac{1}{yg_y}\eta^\dagger\Gamma_x\eta&=&-ibe^{-B}\eta^\dagger\Gamma_S\Gamma_{xy}\eta-
ice^{-C}\eta^\dagger\Gamma_\Omega\Gamma_{xy}\eta=
-ibe^{-B}\eta^\dagger\Gamma_\Omega\Gamma_H\eta+
ice^{-C}\eta^\dagger\Gamma_S\Gamma_H\eta\nonumber\\
&=&bc_1e^{C-B}-cc_2e^{B-C}\nonumber
\eea
This leads to a simple expression for the one--form:
\bea\label{11DBasicVec}
\eta^\dagger\gamma_\mu \eta dx^\mu=g_y^2\left[-ae_1 e^{F+B+C}dy+e^A(bc_1e^{2C}-cc_2e^{2B})dx\right]
\eea
To summarize, in this subsection we have completely determined the form of the metric and we derived six equations for the vector bilinears (\ref{11DVecDer1}), (\ref{11DVecDer2}), 
(\ref{VectLapl1}), (\ref{VectLapl2}),  (\ref{VectLapl3}), (\ref{11DBasicVec}). This concludes the analysis 
of the differential equation (\ref{11DEqn}) and in the next subsection we will use the projectors 
(\ref{11DEa})--(\ref{11DEc}) to extract information about the fluxes.


\subsection{Equations for the fluxes.}


In this subsection we will use various projectors to find the expressions for the fluxes $f_1,f_2,f_3$. 
Analysis of subsection \ref{SubsAppAns} shows that there are three projectors annihilating the spinor 
(see (\ref{11DEa})--(\ref{11DEc})), 
however since we are looking for a solution which preserves $16$ supercharges, only one projector is
expected to be independent. In particular, it is convenient to impose (\ref{GeomProj}) then 
(\ref{11DEa})--(\ref{11DEc}) would follow from it. 

Let us consider two combinations of (\ref{11DEa})--(\ref{11DEc}) which apriori seem complementary to (\ref{GeomProj}):
\bea\label{11DExtraP1}
&&\left[\frac{a}{2}e^{-A}\Gamma_H+\frac{1}{2}{\not\d}A+(G_S+G_\Omega-2G_H)\right]\eps=0\\
\label{11DExtraP2}
&&\left[ibe^{-B}\Gamma_S-ice^{-C}\Gamma_\Omega+
{\not\d}(B-C)-\frac{i}{2}(e^{-3B}{\not\d}f_2\Gamma_S-
e^{-3C}{\not\d}f_3\Gamma_\Omega)\right]\eta=0
\eea
Let us take the conjugate of the last equation:
\bea
&&\eta^\dagger\left[-ibe^{-B}\Gamma_S+ice^{-C}\Gamma_\Omega+
{\not\d}(B-C)-\frac{i}{2}(e^{-3B}{\not\d}f_2\Gamma_S-
e^{-3C}{\not\d}f_3\Gamma_\Omega)\right]=0\nonumber
\eea
Multiplying this relation by $\Gamma_S\gamma_\mu\eta$ and (\ref{11DExtraP2}) by 
$\eta^\dagger\gamma_\mu\Gamma_S$,
we find a relation between bilinears:
\bea
&&\eta^\dagger\left[-ibe^{-B}\gamma_\mu+\eps_{\mu\nu}
\Gamma_S\Gamma_{12}\d^\nu (B-C)-
\frac{i}{2}e^{-3B}{\d}_\mu f_2+
\frac{i}{2}\eps_{\mu\nu}\Gamma_S\Gamma_\Omega\Gamma_{12}
e^{-3C}{\d}^\nu f_3\right]\eta=0\nonumber
\eea
This equation and analogous relation obtained by interchanging $\Gamma_S$ and $\Gamma_\Omega$ lead to
the expressions for the fluxes\footnote{We also used the relations 
$\Gamma_S\Gamma_{12}\eta=\Gamma_\Omega\Gamma_H\eta$,
$\Gamma_S\Gamma_\Omega\Gamma_{12}\eta=-\Gamma_H\eta$.}:
\bea
\label{OldFluxes1}
&&e^{A-3B}df_2=abe^{-A-B}df_0+2c_1e^C*d(B-C)-
e_1e^{F-3C}*df_3\\
\label{OldFluxes2}
&&e^{A-3C}df_3=ace^{-A-C}df_0+2c_2e^B*d(B-C)+
e_1e^{F-3B}*df_2
\eea
Here we used (\ref{VectorExpr}) to introduce a new function $f_0$:
\bea\label{11DefF0}
f_0=f_1-e_1 e^{2A+F},\qquad 2\eta^\dagger\gamma_\mu\eta dx^\mu=-ae^{-A}df_0
\eea

To find additional relations we use a differential equation (\ref{11DEqn}) and its conjugate to evaluate the derivative of $\eta^\dagger\Gamma_{HS}\eta$\footnote{Since we already have an expression for this derivative, these manipulations exploit the projectors, rather than differential equation.}:
\bea
&&\nabla_\mu(\eta^\dagger\Gamma_{HS}\eta)+
\eta^\dagger\Gamma_{HS}
\left[\frac{1}{2}\gamma_\mu (2G_H-4G_S+2G_\Omega)+\frac{1}{2}(2G_H+4G_S+
2G_\Omega)
\gamma_\mu\right]\eta=0\nonumber\\
&&\d_\mu(-ic_2e^B)+
\frac{1}{12}\eta^\dagger\Gamma_{HS}
\left[e^{-3A}[\gamma_\mu,{\not\d}f_1]\Gamma_H-4ie^{-3B}\d_\mu f_2\Gamma_S+
ie^{-3C}[\gamma_\mu,{\not\d}f_3]\Gamma_\Omega\right]\eta=0\nonumber\\
&&-12ic_2~de^B-4i e_1 e^{F-3B}df_2-2e^{-3A}(*df_1)\eta^\dagger\Gamma_S\Gamma_{56}\eta+
2ie^{-3C}(*df_3)
\eta^\dagger\Gamma_{HS\Omega}\Gamma_{56}\eta=0\nonumber
\eea
Substituting the expressions for the bilinears, we arrive at a relation between fluxes:
\bea\label{NewFluxes1}
&&-6c_2de^B-2e_1 e^{F-3B}df_2-c_1e^{C-3A}*df_1-e^{A-3C}*df_3=0
\eea
Similar manipulations for $\eta^\dagger\Gamma_H\Gamma_\Omega\eta$ give
\bea\label{NewFluxes2}
&&-6c_1de^C-2e_1 e^{F-3C}df_3+c_2e^{B-3A}*df_1+e^{A-3B}*df_2=0
\eea
Equations (\ref{OldFluxes1}), (\ref{OldFluxes2}), (\ref{NewFluxes1}), (\ref{NewFluxes2}) can be 
viewed as a linear system for $df_2$, $df_3$, $*df_2$, $*df_3$ and the solution is
\bea\label{Mast4e1}
e_1 e^{F-3C}*df_3&=&-c_2e^{B-3A}df_1-
abe^{-A-B}df_0-2c_1e^C*d(B+2C)\\
\label{Mast4e2}
e_1e^{F-3B}*df_2&=&c_1e^{C-3A}df_1+ace^{-A-C}df_0-2c_2e^B*d(2B+C)\\
\label{Mast4e3}
e^{A-3B}df_2&=&2abe^{-A-B}df_0+2c_1e^C*d(2B+C)+c_2e^{B-3A}df_1\\
\label{Mast4e4}
e^{A-3C}df_3&=&2ace^{-A-C}df_0-2c_2e^B*d(2C+B)+c_1e^{C-3A}df_1
\eea
These relations can be used to eliminate $f_2$, $f_3$ from (\ref{DerGamH}):
\bea
d(e_1 e^F)-\frac{1}{3}\left[-2e^{-2A}df_0+2c_1c_2 e^{B+C-A}*d(B-C)+e^{-4A}(2e^{2A}-e^{2F})df_1\right]=0\nonumber
\eea
Simplifying this equation, we express $f_1$ in terms of the warp factors:
\bea\label{Mast4e5}
df_1=2c_1c_2ye^{2A-2F}*d(B-C)+e_1de^{4A-F} 
\eea

To summarize, the relations (\ref{Mast4e1})--(\ref{Mast4e5}) (along with definition (\ref{11DefF0})) allow us to write all fluxes in terms of the warp factors. Then the equations for the components of the metric would come from the integrability conditions of (\ref{Mast4e1})--(\ref{Mast4e5}) combined with 
(\ref{11DBasicVec}), (\ref{11DefF0}). The situation is analogous to the system which was encountered in 
\cite{myWils} for the type IIB geometries.

Using an analogy with \cite{myWils}, it seems useful to combine (\ref{Mast4e1}) and 
(\ref{Mast4e2}):
\bea
&&e_1 e^{F}\left[c_2e^{-4C}*df_3-c_1e^{-4B}*df_2\right]\nonumber\\
&&\qquad=-e^{-3A}(c^2_2e^{B-C}+c^2_1e^{C-B})df_1+
e^{-A-B-C}df_0+2c_1c_2*d(B-C)\nonumber\\
&&\qquad=e^{-3A-B-C+2F}df_1-
e_1e^{-A-B-C}de^{2A+F}+2c_1c_2*d(B-C)\nonumber\\
&&\qquad=2e_1e^{2F-B-C}de^{A-F}+4c_1c_2*d(B-C)\nonumber
\eea
Dividing by $e^{2A}-e^{2F}$, we find
\bea
\frac{e_1 e^{F+B+C}}{e^{2A}-e^{2F}}\left[c_2e^{-4C}*df_3-c_1e^{-4B}*df_2\right]=
\frac{2e_1e^{2F}de^{A-F}}{e^{2A}-e^{2F}}+\frac{4c_1c_2e^{B+C}*d(B-C)}{c^2_1e^{2C}+c^2_2e^{2B}}\nonumber
\eea
In other words, we found that one of the consequences of the equations for the fluxes is a differential relation 
\bea\label{11DMaster}
e_1d\log\frac{e^A-e^F}{e^A+e^F}+4*d\arctan\frac{c_2 e^{B-C}}{c_1}=
\frac{e_1 e^{F+B+C}}{e^{2A}-e^{2F}}\left[c_2e^{-4C}*df_3-c_1e^{-4B}*df_2\right]
\eea
It is interesting to construct an alternative form of this relation which does not contain $f_2$. 
To find it we begin 
with extracting $*df_2$ from (\ref{OldFluxes2}):
\bea
&&\frac{e_1 c_1e^{F+C-3B}}{e^{2A}-e^{2F}}*df_2\nonumber\\
&&\quad=\frac{c_1}{e^{2A}-e^{2F}}\left[(e^{A-2C}df_3-ace^{-A}df_1)+
ace_1e^{-A}de^{2A+F}-2c_2 e^{B+C}*d(B-C)\right]\nonumber
\eea
The above expression would be much simpler if we had $d(A-F)$ instead of $d(2A+F)$. The simplest 
way to go from one differential to another is to use a consequence of (\ref{Mast4e5}):
\bea
e_1e^{A+F}d(4A-F)=e^{2F-3A}df_1-2c_1c_2e^{B+C}*d(B-C)\nonumber
\eea
This leads to the relation
\bea
&&\frac{e_1 c_1e^{F+C-3B}}{e^{2A}-e^{2F}}*df_2\nonumber\\
&&\quad=\frac{c_1}{e^{2A}-e^{2F}}\left[(e^{A-2C}df_3-ace^{-A}df_1)
-2c_2 e^{B+C}*d(B-C)\right]\nonumber\\
&&\qquad+\frac{acc_1}{e^{2A}-e^{2F}}\left[2e_1e^{A+F}d(F-A)+e^{2F-3A}df_1-
2c_1c_2e^{B+C}*d(B-C)\right]\nonumber\\
&&\quad=\frac{c_1e^{A-2C}}{e^{2A}-e^{2F}}df_3-acc_1e^{-3A}df_1+
acc_1e_1\log\frac{e^A+e^F}{e^A-e^F}+\frac{2abc_1 c^2_2 e^{B+C}}{e^{2A}-e^{2F}}*d(B-C)
\nonumber
\eea
Noticing that 
\bea
\frac{c_1 c^2_2 e^{B+C}}{e^{2A}-e^{2F}}*d(B-C)=
\frac{c_1 c^2_2 e^{B+C}}{c_1^2 e^{2C}+c_2^2 e^{2B}}*d(B-C)=c_2*d\arctan\frac{c_2 e^B}{c_1 e^C},
\eea
we can substitute the expression for $*df_2$ into (\ref{11DMaster}) and move the warp factors to the 
left hand side:
\bea\label{11DMasterA}
&&(1-acc_1)\left[e_1d\log\frac{e^A-e^F}{e^A+e^F}+2*d\arctan\frac{c_2 e^{B-C}}{c_1}\right]
\nonumber\\
&&\qquad=acc_1 e^{-3A}df_1+
\frac{e_1 e^{-3C}}{e^{2A}-e^{2F}}\left[c_2e^{F+B}*df_3-c_1e_1 e^{A+C}df_3\right]
\eea
Here we used a relation $4+2abc_2=2(1-acc_1)$. The last relation can be used to define a new harmonic function ${\tilde\Phi}$ in the same way as (\ref{MasterEqn16}) was used to define 
function $\Phi$ in (\ref{SolnThrPhi}). It appears that while $\Phi$ was natural for the solutions with 
$AdS_4\times S^7$ branch (the right hand side of (\ref{MasterEqn16}) vanished for the maximally symmetric solution), the function ${\tilde\Phi}$ might be appropriate for the $AdS_7\times S^4$ since the right hand side of (\ref{11DMasterA}) vanishes for the ground state in this sector. Unfortunately in this branch $c_1=ac$ and equation (\ref{11DMasterA}) does not contain the warp factors. Then it seems that there is no advantage in introducing the second harmonic function 
${\tilde\Phi}$, moreover it appears that while equation (\ref{MasterEqn16}) is well--suited for doing perturbation theory around $AdS_4\times S^7$ (and a similar equation for $AdS_5\times S^5$ asymptotics was derived in \cite{myWils}), there is no natural counterpart of this starting point in the $AdS_7\times S^4$ sector. This makes $AdS_7\times S^4$ sector somewhat special and it would be nice to understand the nature of this difference. 

\subsection{Evaluation of the Killing spinor}

Finally let us analyze the constrains imposed on Killing spinors. Naively it seems that we have three
projectors (\ref{GeomProj}), (\ref{11DExtraP1}), (\ref{11DExtraP2}), but to preserve a half of supersymmetries, only one of them should be independent. The projector (\ref{GeomProj}) and  
equations for bilinears can be used to derive (\ref{11DExtraP1}), (\ref{11DExtraP2}). In this subsection we use (\ref{GeomProj}) to find the restrictions imposed on the spinor.

We begin with recalling the expressions for the non--vanishing scalar bilinears:
\bea
\eta^\dagger\eta=e^A,\quad \eta^\dagger\Gamma_\Omega\Gamma_H\eta=ic_1 e^C,\quad
\eta^\dagger\Gamma_S\Gamma_H\eta=ic_2 e^B,\quad
\eta^\dagger\Gamma_H\eta=e_1 e^F
\eea
Due to reality of $e^F$ it is convenient to define two angles $\sigma$ and $\delta$:
\bea\label{11DefSD}
e^F=e^A\cos 2\sigma,\quad c_1 e^C=e^A\sin 2\sigma\cos 2\delta,\quad
c_2 e^B=e^A\sin 2\sigma\sin 2\delta
\eea
Let us define a rotated spinor $\eta_1$ as
\bea
\eta_1=e^{-\delta \Gamma_S\Gamma_\Omega}\eta
\eea
Then $\eta_1^\dagger\Gamma_S\Gamma_H\eta_1=0$ and there are only three 
non--trivial bilinears containing $\eta_1$:
\bea
\eta_1^\dagger\eta_1=e^A,\quad 
\eta_1^\dagger\Gamma_\Omega\Gamma_H\eta_1=ie^A\sin 2\sigma,
\quad
\eta_1^\dagger\Gamma_H\eta_1=e_1 e^A \cos 2\sigma
\eea
This suggests an additional rotation:
\bea
\eta_2=e^{ie_1\sigma\Gamma_\Omega}\eta_1
\eea
which leads to a spinor with only two non--vanishing bilinears:
\bea
\eta_2^\dagger\eta_2=e^A,\quad 
\eta_2^\dagger\Gamma_H\eta_2=e_1e^A 
\eea
This implies a simple projection on $\eta_2$. 

To summarize, we have the following expression 
for the original spinor:
\bea\label{11DSpinorA}
\eta=e^{\delta \Gamma_S\Gamma_\Omega}e^{-ie_1\sigma\Gamma_\Omega}\eta_2,\qquad
\Gamma_H\eta_2=e_1\eta_2
\eea
Notice that $e_1$ could be equal to either one or minus one and the same geometry could have 
spinors with {\it both} signs of $e_1$. So the last relation does not truncate the number of 
components of the Killing spinor, but rather it introduces two different branches.

Let us now look at the projector (\ref{GeomProj}).
We begin with rewriting it as
\bea
\left[ae^{-A}\Gamma_H+ibe^{-B}\Gamma_S+ice^{-C}\Gamma_\Omega+\frac{1}{yg_y}\Gamma_y\right]e^{\delta \Gamma_S\Gamma_\Omega}e^{-ie_1\sigma\Gamma_\Omega}\eta_2=0
\nonumber\\
\left[ae^{-A}\Gamma_H+i(be^{-B}c_{2\delta}-ce^{-C}s_{2\delta})\Gamma_S+
i(ce^{-C}c_{2\delta}+be^{-B}s_{2\delta})\Gamma_\Omega+\frac{1}{yg_y}\Gamma_y\right]
e^{-ie_1\sigma\Gamma_\Omega}\eta_2=0\nonumber
\nonumber
\eea
Simplifying further, we find
\bea
&&e_1\left[ae^{-A}+s_{2\sigma}(ce^{-C}c_{2\delta}+be^{-B}s_{2\delta})\right]\eta_2\nonumber\\
&&\quad+
\left[i(be^{-B}c_{2\delta}-ce^{-C}s_{2\delta})\Gamma_S+
ic_{2\sigma}(ce^{-C}c_{2\delta}+be^{-B}s_{2\delta})\Gamma_\Omega+\frac{1}{yg_y}\Gamma_y\right]
\eta_2=0\nonumber
\eea
The first line in this expression is proportional to $a+bc_2+cc_1$, so it vanishes due to (\ref{11Dabc}).
Thus the projector (\ref{GeomProj}) is equivalent to
\bea
&&\left[i(bc_1e^{C-B}-cc_2e^{B-C})\Gamma_S+
ic_{2\sigma}(cc_1+bc_2)\Gamma_\Omega+\frac{e^As_{2\sigma}}{yg_y}\Gamma_y\right]
\eta_2=0\nonumber\\
&&\left[i(bc_1e^{C-B}-cc_2e^{B-C})\Gamma_S-
iae^{F-A}\Gamma_\Omega+s_{2\sigma}\sqrt{-1+e^{2A-2B}+e^{2A-2C}}\Gamma_y\right]
\eta_2=0\nonumber
\eea
Notice that this is a consistent projector since
\bea
e^{2G-2A}&\equiv&(bc_1e^{C-B}-cc_2e^{B-C})^2+e^{2F-2A}\nonumber\\
&=&c_1^2 e^{2C}(e^{-2B}-e^{-2A})+
c_2^2 e^{2B}(e^{-2C}-e^{-2A})-2bcc_1c_2+1\nonumber\\
&=&(c_1^2 e^{2C}+c_2^2 e^{2B})(e^{-2B}+e^{-2C}-e^{-2A})\nonumber
\eea
Introducing one more angle $\tau$ by requiring
\bea\label{DefAngTau}
\sin 2\tau=e^{A-G}(bc_1e^{C-B}-cc_2e^{B-C}),\quad 
\cos 2\tau=e^{F-G},
\eea
we can rewrite the equation for $\eta_2$ as
\bea
\left[i\sin 2\tau\Gamma_S-ia\cos 2\tau\Gamma_\Omega+\Gamma_y\right]\eta_2=0
\eea
This implies an existence of a spinor with very simple projection:
\bea
\eta_0=e^{a\tau\Gamma_S\Gamma_\Omega}\eta_2,\quad 
\Gamma_\Omega\Gamma_y\eta_0=ia\eta_0
\eea
Combining this with (\ref{11DSpinorA}), we find 
the following relations for the spinor $\eta$:
\bea\label{11DSpinor}
\eta=e^{\delta \Gamma_S\Gamma_\Omega}e^{-ie_1\sigma\Gamma_\Omega}
e^{-a\tau\Gamma_S\Gamma_\Omega}\eta_0,\qquad
\Gamma_H\eta_0=e_1\eta_0,\quad \Gamma_\Omega\Gamma_y\eta_0=ia\eta_0
\eea
As we already mentioned, the same geometry has Killing spinors with both positive and negative values of $e_1$, so the last relation leads to only one nontrivial projection and thus it corresponds to a geometry preserving $16$ supercharges. 

\subsection{Summary of the equations}

In this appendix we analyzed the $1/2$--BPS geometries and we found a set of relations which should be satisfied by them. In this subsection we collect all relevant equations.

We began with an assumption of $SO(2,2)\times SO(4)\times SO(4)$ symmetry, this led to the expression for metric and fluxes (\ref{gYY}), (\ref{XYmetr}):
\bea
ds^2&=&e^{2A}ds_H^2+e^{2B}ds_S^2+e^{2C}ds_{\tilde S}^2+
g_y^2(dx^2+dy^2)\nonumber\\
F_4&=&df_1\wedge dH_3+df_2\wedge d\Omega_3+
df_3\wedge d{\tilde\Omega}_3\\
&&g^{-1}_y=y\sqrt{-e^{-2A}+e^{-2B}+e^{-2C}},\quad y=e^{A+B+C}\nonumber
\eea
The relations (\ref{11DLastScal}), (\ref{11DefF0}), (\ref{Mast4e1})--(\ref{Mast4e5}) allow one to write all fluxes in terms of warp factors:
\bea\label{AppBOsEqn1}
e^F&=&\sqrt{e^{2A}-c_1^2 e^{2C}-c_2^2 e^{2B}},\quad f_0=f_1-e_1 e^{2A+F}\nonumber\\
df_1&=&2c_1c_2ye^{2A-2F}*d(B-C)+e_1de^{4A-F}\nonumber\\
e_1 e^{F-3C}*df_3&=&-c_2e^{B-3A}df_1-
abe^{-A-B}df_0-2c_1e^C*d(B+2C)\\
e_1e^{F-3B}*df_2&=&c_1e^{C-3A}df_1+ace^{-A-C}df_0-2c_2e^B*d(2B+C)\nonumber\\
e^{A-3B}df_2&=&2abe^{-A-B}df_0+2c_1e^C*d(2B+C)+c_2e^{B-3A}df_1\nonumber\\
e^{A-3C}df_3&=&2ace^{-A-C}df_0-2c_2e^B*d(2C+B)+c_1e^{C-3A}df_1\nonumber
\eea
The constants $c_1$, $c_2$ cannot be fixed by the local analysis, but there is a constraint 
(\ref{11Dabc}) imposed on them:
\bea\label{SumABC}
bc_2+cc_1=-a
\eea
One can also arrive at a simple differential relation for $f_0$ by combining 
(\ref{11DefF0}) and (\ref{11DBasicVec})
\bea\label{AppBOsEqn2}
df_0=-2ae^Ag_y^2\left[-ae_1 e^{F+B+C}dy+e^A(bc_1e^{2C}-cc_2e^{2B})dx\right]
\eea
Once the constants $c_1$, $c_2$ and the boundary conditions are specified, the equations written above are sufficient for finding a unique geometry (we discuss this construction in the main part 
of the paper). To do the computations in practice, it is convenient to combine the equations for the fluxes into a relation (\ref{11DMaster}):
\bea
e_1d\log\frac{e^A-e^F}{e^A+e^F}+4*d\arctan\frac{c_2 e^{B-C}}{c_1}=
\frac{e_1 e^{F+B+C}}{e^{2A}-e^{2F}}\left[c_2e^{-4C}*df_3-c_1e^{-4B}*df_2\right]
\eea
Finally we recall that the Killing spinor has a very simple form (\ref{11DSpinor}):
\bea\label{Ap11DSpnr}
\eta=e^{\delta \Gamma_S\Gamma_\Omega}e^{-ie_1\sigma\Gamma_\Omega}
e^{-a\tau\Gamma_S\Gamma_\Omega}\eta_0,\qquad
\Gamma_\Omega\Gamma_y\eta_0=ia\eta_0,\quad
\Gamma_H\eta_0=e_1\eta_0
\eea
Notice that one of these projections is a matter of convenience (for example if one solves a SUSY 
variations in $R^2$ using polar coordinates, it is convenient to consider the spinors $\eta_\pm$ satisfying $\Gamma_{r\phi}\eps=\pm i\eps$ separately), so it does not affect the number of preserved supersymmetries. The second projector does break the maximal SUSY to 16 
conserved supercharges\footnote{For a more detailed explanation see the end of this 
subsection.}. 
The angles $\sigma,\delta,\tau$ were defined in (\ref{11DefSD}), (\ref{DefAngTau}):
\bea
&&e^F=e^A\cos 2\sigma,\quad c_1 e^C=e^A\sin 2\sigma\cos 2\delta,\quad
c_2 e^B=e^A\sin 2\sigma\sin 2\delta\nonumber\\
&&e^{2i\tau}=e^{F-G}+ie^{A-G}(bc_1e^{C-B}-cc_2e^{B-C})\nonumber
\eea
While we expect to find geometries for any combinations of signs $(a,b,c,e_1)$, it is sufficient to solve the problem for $(a,b,c,e_1)=(+,+,+,+)$ and the results for the remaining branches can be obtained using various symmetries. To see this we first solve (\ref{SumABC}) by parameterizing  $c_1$, $c_2$ in terms of a real number $q$:
\bea\label{SumABCq}
c_1=acq,\qquad c_2=-ab(q+1)
\eea
Then we observe that equations (\ref{AppBOsEqn1}), (\ref{AppBOsEqn2}) remain invariant if 
we flip the signs of any of the following four sets:
\bea\label{3Z3sym}
(a,e_1,x,f_0,f_1),\qquad (a,f_2,f_3),\qquad (b,x,f_2),\qquad (c,x,f_3)
\eea
while keeping the remaining functions as well as $q$ fixed. Notice that coordinate $x$ was defined to be orthogonal to $y$ and we also fixed the normalization of $dx^2$, but not the sign of $x$. 
The sign reversals in (\ref{3Z3sym}) can be used to map a solution from any branch into the one
with $(a,b,c,e_1)=(+,+,+,+)$. Thus without a loss of generality we can set
\bea\label{StandBranch}
a=b=c=e_1=1
\eea
and these conventions are used in the main part of the paper.

While the symmetries (\ref{3Z3sym}) can be used to map any solution into the standard branch (\ref{StandBranch}), the functional form of the metric may change under such map. For example, if we flip a sign of $x$ and function $e^{2A}$ is not even under such change, then one gets a new solution. The bosonic parts of  the maximally symmetric solutions ($AdS_4\times S^7$ and $AdS_7\times S^4$) are invariant under all four $Z_2$ symmetries defined in (\ref{3Z3sym}), then starting with a spinor in (\ref{StandBranch}) one can reconstruct all 32 supercharges. For a more general solution we should avoid the sign change in $x$, this leaves only three independent reflections in (\ref{3Z3sym}). It is easy to see that all of them leave a solution invariant, but one should also change orientations of the spheres:
\bea
(a,b,c),\qquad
(a,b,e_1,f_0,f_1,f_2,d^3H,d^3\Omega),\qquad (a,f_2,f_3,d^3\Omega,d^3{\tilde\Omega})
\eea
Using these reflections one can go to the branch $(a,b,c)=(+,+,+)$ while keeping the solution unchanged. This guarantees an existence of 16 supercharges for any geometry constructed in 
this appendix, and the solutions preserving different halves have opposite sign of $e_1$. This clarifies the discussion after equation (\ref{Ap11DSpnr}). 

\section{Decompactification limits of 10--dimensional geometries}

\renewcommand{\theequation}{B.\arabic{equation}}
\setcounter{equation}{0}

In section \ref{SectDecomp} we showed that by taking various decompactification limits of the new 
geometries one recovers known solutions. Here we discuss similar limits for the type IIB metrics constructed in 
\cite{myWils} and again we will see that some interesting solutions are recovered. 

We begin with recalling the geometries constructed in \cite{myWils}:
\bea\label{WilsEqn1}
&&ds^2=e^{2A}dH_2^2+e^{2B}d\Omega_2^2+e^{2C}d\Omega_4^2+
\frac{e^{-\phi}}{e^{2B}+e^{2C}}
(dx^2+dy^2)\\
&&F_5=df_3\wedge d\Omega_4+*_{10}(df_3\wedge d\Omega_4),\quad
H_3=df_1\wedge dH_2,\quad F_3=df_2\wedge d\Omega_2\nonumber\\
&&e^{2A}=ye^{H-\phi/2},\quad e^{2B}=ye^{G-\phi/2},\quad e^{2C}=ye^{-G-\phi/2},\quad 
F=\sqrt{e^{2A}-e^{2B}-e^{2C}}\nonumber
\eea
The warp factors and fluxes obey a system of differential equations:
\bea
&&df_1=-\frac{2e^{2A+\phi/2}}{e^{2A}-e^{2B}}\left[e^A Fd\phi-e^{B+C}*d\phi
\right],\nonumber\\
&&df_2=\frac{2e^{2B-\phi/2}}{e^{2A}-e^{2B}}\left[e^B Fd\phi-e^{A+C}
*d\phi\right]\nonumber\\
\label{WilsEqn2}
&&e^B e^{-4C}*df_3=e^A d(A-\frac{\phi}{4})+\frac{1}{4}F e^{-\phi/2-2A}df_1\\
&&d(H-G-2\phi)=-\frac{2}{y(e^{2B}+e^{2C})}(e^{2C}dy+ F e^{B+C-A}dx)\nonumber\\
&&*d\arctan e^G+\frac{1}{2}d\log\frac{e^A-F}{e^A+F}-
\frac{1}{2}e^{-\phi/2-2A}df_1=0\nonumber\\
&&e^A e^{-4C}*df_3=e^B d(B+\frac{\phi}{4})+\frac{1}{4}F e^{\phi/2-2B}df_2\nonumber
\eea
The boundary conditions for these differential equations were studied in \cite{myWils}, here we will proceed with local analysis and consider the limits where various warp factors go to infinity.

\subsection{Metric produced by a string}

We begin with decompactification of AdS space. To accomplish this goal we rescale various quantities as
\bea
(e^A,F)\rightarrow \Lambda (e^A,F),\quad f_1\rightarrow \Lambda^2 f_1
\eea
and send $\Lambda$ to infinity. Then the geometry takes the form
\bea
&&ds^2=e^{2A}dw_{1,1}^2+e^{2B}d\Omega_2^2+e^{2C}d\Omega_4^2+
\frac{e^{-\phi}}{e^{2B}+e^{2C}}
(dx^2+dy^2)\\
&&F_5=df_3\wedge d\Omega_4+*_{10}(df_3\wedge d\Omega_4),\quad
H_3=df_1\wedge d^2w,\quad F_3=df_2\wedge d\Omega_2\nonumber\\
&&e^{2A}=ye^{H-\phi/2},\quad e^{2B}=ye^{G-\phi/2},\quad e^{2C}=ye^{-G-\phi/2},\quad 
F=e^{A}\nonumber
\eea
and the system of differential equations can be rewritten as
\bea\label{Lim1ApE1}
&&df_1=-2e^{2A+\phi/2}d\phi,\qquad e^A d(A-\frac{\phi}{4})+\frac{1}{4}F e^{-\phi/2-2A}df_1=0\\
&&df_2=0,\quad df_3=0,\nonumber\\
\label{Lim1ApE2}
&&d(H-G-2\phi)=-\frac{2}{y(e^{2B}+e^{2C})}(e^{2C}dy+e^{B+C}dx)\\
\label{Lim1ApE3}
&&*d\arctan e^G+\frac{1}{2}d\log\frac{e^{2B}+e^{2C}}{4e^{2A}}-
\frac{1}{2}e^{-\phi/2-2A}df_1=0\nonumber
\eea
Eliminating $f_1$ from the second equation in (\ref{Lim1ApE1}), 
we find a relation between $A$ and the dilaton, then we can also evaluate the flux:
\bea
d(A-\frac{\phi}{4})-\frac{1}{2}d\phi=0:\qquad A=\frac{3\phi}{4},\quad f_1=-e^{2\phi}
\eea
Let us now simplify equation (\ref{Lim1ApE2}):
\bea
d(G+\log y)=\frac{2}{y(e^{G}+e^{-G})}(e^{-G}dy+dx):\quad
ydG=\frac{dx}{\cosh G}-\tanh G dy
\eea
This leads to the unique solution for $G$: 
\bea
\sinh G=\frac{x}{y},\quad e^G=\frac{x}{y}+\sqrt{(\frac{x}{y})^2+1},\quad
e^{2B}=e^{-\phi/2}(x+r),\quad e^{2C}=e^{-\phi/2}(r-x)\nonumber
\eea
and and to the complete solution in terms of the dilaton: 
\bea\label{LimIIFundMetr}
&&ds^2=e^{-\phi/2}\left[e^{2\phi}ds_{1,1}^2+(r+x)d\Omega_2^2+(r-x)d\Omega_4^2+
\frac{1}{2r}(dx^2+dy^2)\right]\nonumber\\
&&H_3=-de^{2\phi}\wedge d^3 w,\qquad r\equiv \sqrt{x^2+y^2}
\eea
Rewriting the metric in terms of polar coordinates in $(x,y)$ plane, one recognizes a geometry produced by fundamental string in flat space. One can check that the last equation (\ref{Lim1ApE3}) becomes an identity and it does not lead to an extra restriction on the dilaton. This agrees with a well--known fact that a geometry produced by fundamental string cannot be completely fixed by looking at supersymmetry variations: to determine a dilaton one has to look at the equations of motion which require $e^{-2\phi}$ to be harmonic.

To summarize, we see that by sending the AdS warp factor to infinity in the solutions of \cite{myWils} one recovers a geometry produced by a fundamental string.


\subsection{Relation to non--commutative theories}


Let us now decompatify AdS and at least one of the spheres. 
We begin with sending $e^A$ and $e^B$ to infinity while keeping $e^C$ fixed. To produce a well--defined geometry one should make the following rescalings:
\bea
(e^A,e^B,F,x,y)\rightarrow \Lambda (e^A,e^B,F,x,y),\quad (f_1,f_2)\rightarrow\Lambda^2 (f_1,f_2)
\eea
and take $\Lambda$ to infinity. Then the system (\ref{WilsEqn1})--(\ref{WilsEqn2}) becomes:
\bea
&&e^{\phi/2}ds^2=ye^Hd{\bf w}_{1,1}^2+ye^Gd{\bf u}_2^2+ye^{-G}d\Omega_4^2+
y^{-1}e^{-G}(dx^2+dy^2)\\
&&F_5=df_3\wedge d\Omega_4+*_{10}(df_3\wedge d\Omega_4),\quad
H_3=df_1\wedge d^2w,\quad F_3=df_2\wedge d^2 u\nonumber\\
\label{Wils1Flux1}
&&e^{2A}=ye^{H-\phi/2},\quad e^{2B}=ye^{G-\phi/2},\quad e^{2C}=ye^{-G-\phi/2},\quad 
F=\sqrt{e^{2A}-e^{2B}}\nonumber\\
&&df_1=-\frac{2e^{3A+\phi/2}}{\sqrt{e^{2A}-e^{2B}}}d\phi,\qquad
df_2=\frac{2e^{3B-\phi/2}}{\sqrt{e^{2A}-e^{2B}}}d\phi\\
&&e^B e^{-4C}*df_3=e^A d(A-\frac{\phi}{4})+\frac{1}{4}F e^{-\phi/2-2A}df_1\nonumber\\
&&e^A e^{-4C}*df_3=e^B d(B+\frac{\phi}{4})+\frac{1}{4}F e^{\phi/2-2B}df_2\nonumber\\
\label{Wils1FluxLst}
&&d(H-G-2\phi)=0,\qquad
\frac{1}{2}d\log\frac{e^A-F}{e^A+F}-
\frac{1}{2}e^{-\phi/2-2A}df_1=0
\eea
From the first equation in the last line we conclude that $e^B=e^Ae^{-\phi}$, then one can simplify the expressions for $f_1$, $f_2$ coming from (\ref{Wils1Flux1}) 
\bea\label{Wils1Fluxes}
-\frac{1}{2}e^{-\phi/2-2A}df_1&=&\frac{e^A}{\sqrt{e^{2A}-e^{2B}}}d\phi=
\frac{d\phi}{\sqrt{1-e^{-2\phi}}}=d\log[e^\phi+\sqrt{e^{2\phi}-1}]\\
\frac{1}{2}e^{\phi/2-2B}df_2&=&\frac{e^B}{\sqrt{e^{2A}-e^{2B}}}d\phi=\frac{d\phi}{\sqrt{e^{2\phi}-1}}
=d\arctan\sqrt{e^{2\phi}-1}\nonumber
\eea 
We observe that the second equation in (\ref{Wils1FluxLst}) becomes an identity. 
Let us simplify the equations for $f_3$:
\bea
~*df_3&=&e^{4C}\left[e^{\phi} d(A-\frac{\phi}{4})-\frac{1}{2}e^{\phi}d\phi\right]=
y^4e^{-4B-2\phi}e^{\phi}d(A-\frac{3\phi}{4})\nonumber\\
~*df_3&=&e^{4C}\left[e^{-\phi} d(B+\frac{\phi}{4})+\frac{1}{2}e^{-\phi}d\phi\right]=
y^4e^{-4B-2\phi}e^{-\phi}d(B+\frac{3\phi}{4})\nonumber
\eea
Integrability conditions for these relations imply an existence of two harmonic functions which differ by  
a constant:
\bea
e^{-4A+\phi}=f,\quad e^{-4A+3\phi}=e^{-4B-\phi}=f+c,\quad *df_3=-\frac{y^4}{4}df,\quad
d(y^4*df)=0
\eea
We can now express the warp factors and dilaton in terms of a harmonic function $f$ and constant $c$:
\bea
e^{2\phi}=1+cf^{-1},\quad e^{2A}=e^{\phi/2}f^{-1/2},\quad e^{2B}=e^{-3\phi/2}f^{-1/2},\quad
e^{2C}=y^2e^{-\phi}e^{-2B}
\eea
Finally we simplify the expressions (\ref{Wils1Fluxes}) for the fluxes:
\bea
df_1&=&
-\frac{2e^{\phi/2+2A}d\phi}{\sqrt{1-e^{-2\phi}}}=-\frac{1}{\sqrt{c}}de^{2\phi}=-\sqrt{c}df^{-1}
\nonumber\\
df_2&=&\frac{2e^{-\phi/2+2B}d\phi}{\sqrt{e^{2\phi}-1}}=-\frac{1}{\sqrt{c}}de^{-2\phi}=\sqrt{c}
d(f^{-1}e^{-2\phi})
\nonumber
\eea 
Thus we arrive at the following solution:
\bea\label{MaldRus}
ds_{E}^2&=&e^{\phi/2}\left[f^{-1/2}\left[-d{\bf w}_{1,1}^2+e^{-2\phi}d{\bf v}_2^2\right]+
f^{1/2}(dx^2+dy^2+y^2d\Omega_4^2)\right]\nonumber\\
&&e^{2\phi}=1+cf^{-1},\quad B=-\sqrt{c}f^{-1}d^2 w,\quad C^{(2)}=\sqrt{c}f^{-1}e^{-2\phi}d^2 v\\
&&F_5=df_3\wedge d\Omega_4+*_{10}(df_3\wedge d\Omega_4),\quad *df_3=-\frac{y^4}{4}df,\quad
d(y^4*df)=0\nonumber
\eea
This geometry was constructed in \cite{MaldRuss} as a gravity dual of non--commutative gauge theory 
(see also \cite{HashItz}). The derivation of \cite{MaldRuss,HashItz} involves T dualities and rotations, so it 
is much easier than the method used here, but the fact that we recover this metric as a special limit of solutions dual to Wilson lines serves as a nice cross--check of our computations. 


\subsection{Decompactifications of $S^4$.}

Let us now send $e^A$ and $e^C$ to infinity while keeping $e^B$ fixed. One way to accomplish this is 
to make following rescalings:
\bea
(e^A,e^C,F,x,y)\rightarrow \Lambda (e^A,e^C,F,x,y),\quad f_1\rightarrow\Lambda^2 f_1,\quad
f_3\rightarrow\Lambda^4 f_3
\eea
In the limit $\Lambda\rightarrow\infty$ the system (\ref{WilsEqn1})--(\ref{WilsEqn2}) becomes:
\bea\label{WilsLim2Eqn1}
&&ds^2=e^{2A}d{\bf w}_{1,1}^2+e^{2B}d\Omega_2^2+e^{2C}d{\bf v}_4^2+
e^{-\phi-2C}(dx^2+dy^2)\\
&&F_5=df_3\wedge d^4v+*_{10}(df_3\wedge d^4v),\quad
H_3=df_1\wedge d^2w,\quad F_3=df_2\wedge d\Omega_2\nonumber\\
&&e^{2A}=ye^{H-\phi/2},\quad e^{2B}=ye^{G-\phi/2},\quad e^{2C}=ye^{-G-\phi/2},\quad 
F=\sqrt{e^{2A}-e^{2C}}\nonumber\\
\label{WilsLim2Eqn3}
&&df_1=-2e^{A+\phi/2}Fd\phi,\quad
df_2=-2e^{2B+C-A-\phi/2}*d\phi\\
&&e^B e^{-4C}*df_3=e^A d(A-\frac{\phi}{4})+\frac{1}{4}F e^{-\phi/2-2A}df_1\nonumber\\
&&e^A e^{-4C}*df_3=\frac{1}{4}F e^{\phi/2-2B}df_2\nonumber\\
\label{WilsLim2Eqn2}
&&d(2A-2B-2\phi)=-2d\log y,\qquad
\frac{1}{2}d\log\frac{e^A-F}{e^A+F}-
\frac{1}{2}e^{-\phi/2-2A}df_1=0
\eea
The first equation in the last line allows us to express all warp factors in terms of $e^A$ and the dilaton:
\bea\label{WilsLim2Warp}
e^{B}=ye^{A-\phi},\quad e^C=e^{-A+\phi/2}
\eea
Let us combine the equation for $f_1$ with last relation in (\ref{WilsLim2Eqn2}):
\bea
\frac{1}{2}d\log\frac{e^A-F}{e^A+F}+Fe^{-A}d\phi=0
\eea
If we assume that $F\ne 0$, then this equation relates differentials of $A$ and $\phi$: $dA=S(\phi)d\phi$. 
We can now combine the equations for the fluxes to produce a relation between the warp factors:
\bea
-\frac{1}{2}e^{B-A}Fe^{\phi/2-2B}e^{2B+C-A-\phi/2}*d\phi=
e^A d(A-\frac{\phi}{4})-\frac{1}{2}F e^{-\phi/2-2A}e^{A+\phi/2}d\phi\nonumber
\eea
The left hand side of this equation is proportional to $*d\phi$, while the right hand side is proportional to 
$d\phi$, so each side should vanish. Since we want the warp factors and the dilaton to remain finite, this leads to the conclusion that $F=0$. This fact simplifies the system 
(\ref{WilsLim2Eqn3})--(\ref{WilsLim2Eqn2}):
\bea
f_1=f_3=0,\quad df_2=y^2*de^{-2\phi},\quad e^{4A}=e^\phi
\eea
Combining this with (\ref{WilsLim2Warp}), we arrive at the solution:
\bea
&&ds^2=e^{\phi/2}\left[d{\bf w}_{1,1}^2+d{\bf v}_4^2+e^{-2\phi}(dx^2+dy^2+y^2d\Omega_2^2)
\right]\\
&&F_3=y^2*de^{-2\phi}\wedge d\Omega_2,\quad d(y^2*de^{-2\phi})=0\nonumber
\eea
This is a geometry produced by smeared D5 branes. 

One can make a different rescaling which leads to the metric preserving
 $ISO(1,1)\times ISO(4)\times SO(3)$ isometry:
\bea
(e^{2A},F^2,f_1)\rightarrow \Lambda^4(e^{2A},F^2,f_1),\quad 
(e^{C},x+iy)\rightarrow \Lambda(e^{C},x+iy),\quad f_3\rightarrow \Lambda^3 f_3
\eea
In the limit $\Lambda\rightarrow\infty$ one finds $f_2=f_3=0$ and the remaining equations become
\bea
&&df_1=-2e^{2A+\phi/2}d\phi,\quad
d(A-\frac{\phi}{4})+\frac{1}{4}e^{-\phi/2-2A}df_1=0\nonumber\\
&&d(A-B-\phi)=-\frac{dy}{y},\quad
\frac{1}{2}d\log\frac{e^{2C}}{4e^{2A}}-
\frac{1}{2}e^{-\phi/2-2A}df_1=0\nonumber
\eea
Using these relations, one can parameterize the solution in terms of one unknown function $e^\phi$:
\bea
&&ds^2=e^{3\phi/2}d{\bf w}_{1,2}^2+e^{-\phi/2}\left[d{\bf v}_4^2+dx^2+dy^2+y^2d\Omega_2^2\right]\\
&&H_3=-de^{2\phi}\wedge d{\bf w}_2
\eea
This geometry is produced by smeared fundamental strings, and one needs to use the equations of motion to show that $e^{-2\phi}$ is harmonic. In fact this solution can be obtained by taking a decompactification limit of (\ref{LimIIFundMetr}). 

One can also send all three warp factors to infinity, there are various ways of doing this, but all solutions can be viewed as further limits of metrics produced by fundamental strings, D5 branes or D3 branes with fluxes (\ref{MaldRus}).


\begin{thebibliography}{99}

\bibitem{mald}
J.~M.~Maldacena,
  Adv.\ Theor.\ Math.\ Phys.\  {\bf 2}, 231 (1998)
  [Int.\ J.\ Theor.\ Phys.\  {\bf 38}, 1113 (1999)], hep-th/9711200.
%
\bibitem{gkpw}
S.~S.~Gubser, I.~R.~Klebanov and A.~M.~Polyakov,
  Phys.\ Lett.\  B {\bf 428}, 105 (1998), hep-th/9802109;\\
E.~Witten,
  Adv.\ Theor.\ Math.\ Phys.\  {\bf 2}, 253 (1998), hep-th/9802150.
  %
\bibitem{Freedm}
D.~Z.~Freedman, S.~D.~Mathur, A.~Matusis and L.~Rastelli,
  Nucl.\ Phys.\  B {\bf 546}, 96 (1999), hep-th/9804058;
  Phys.\ Lett.\  B {\bf 452}, 61 (1999), hep-th/9808006;\\
 S.~M.~Lee, S.~Minwalla, M.~Rangamani and N.~Seiberg,
  Adv.\ Theor.\ Math.\ Phys.\  {\bf 2}, 697 (1998), hep-th/9806074;\\
E.~D'Hoker, D.~Z.~Freedman, S.~D.~Mathur, A.~Matusis and L.~Rastelli,
  Nucl.\ Phys.\  B {\bf 562}, 353 (1999), hep-th/9903196.
%
\bibitem{ppWave}
D.~Berenstein, J.~M.~Maldacena and H.~Nastase,
  JHEP {\bf 0204}, 013 (2002), hep-th/0202021.
%
\bibitem{FrlTs}
S.~S.~Gubser, I.~R.~Klebanov and A.~M.~Polyakov,
  Nucl.\ Phys.\  B {\bf 636}, 99 (2002), hep-th/0204051;\\
S.~Frolov and A.~A.~Tseytlin,
  JHEP {\bf 0206}, 007 (2002), hep-th/0204226;
  Nucl.\ Phys.\  B {\bf 668}, 77 (2003), hep-th/0304255;
  JHEP {\bf 0307}, 016 (2003), hep-th/0306130.
\bibitem{IntgrBlt}
J.~A.~Minahan and K.~Zarembo,
  JHEP {\bf 0303}, 013 (2003), hep-th/0212208;\\
N.~Beisert, J.~A.~Minahan, M.~Staudacher and K.~Zarembo,
  JHEP {\bf 0309}, 010 (2003), hep-th/0306139;\\
N.~Beisert,
  Nucl.\ Phys.\  B {\bf 676}, 3 (2004), hep-th/0307015;\\
N.~Beisert and M.~Staudacher,
  Nucl.\ Phys.\  B {\bf 670}, 439 (2003), hep-th/0307042;\\
G.~Arutyunov, S.~Frolov, J.~Russo and A.~A.~Tseytlin,
  Nucl.\ Phys.\  B {\bf 671}, 3 (2003), hep-th/0307191;\\
N.~Beisert, S.~Frolov, M.~Staudacher and A.~A.~Tseytlin,
  JHEP {\bf 0310}, 037 (2003), hep-th/0308117;\\
V.~A.~Kazakov, A.~Marshakov, J.~A.~Minahan and K.~Zarembo,
  JHEP {\bf 0405}, 024 (2004), hep-th/0402207;\\
D.~M.~Hofman and J.~M.~Maldacena,
  J.\ Phys.\ A  {\bf 39}, 13095 (2006), hep-th/0604135.
%
\bibitem{Witt20}
E.~Witten,
  arXiv:hep-th/9507121.
\bibitem{Seiberg}
N.~Seiberg,
  Nucl.\ Phys.\ Proc.\ Suppl.\  {\bf 67}, 158 (1998), hep-th/9705117.
%
\bibitem{CFT20}
O.~Aharony, M.~Berkooz, S.~Kachru, N.~Seiberg and E.~Silverstein,
  Adv.\ Theor.\ Math.\ Phys.\  {\bf 1}, 148 (1998), hep-th/9707079;\\
O.~Aharony, M.~Berkooz and N.~Seiberg,
  Adv.\ Theor.\ Math.\ Phys.\  {\bf 2}, 119 (1998), hep-th/9712117;\\
R.~G.~Leigh and M.~Rozali,
  Phys.\ Lett.\  B {\bf 431}, 311 (1998), hep-th/9803068.
\bibitem{CFT3D}
O.~J.~Ganor and S.~Sethi,
  JHEP {\bf 9801}, 007 (1998), hep-th/9712071;\\
M.~Berkooz and A.~Kapustin,
  JHEP {\bf 9902}, 009 (1999), hep-th/9810257.
\bibitem{AdSCorrel}
L.~Castellani, R.~D'Auria, P.~Fre, K.~Pilch and P.~van Nieuwenhuizen,
  Class.\ Quant.\ Grav.\  {\bf 1}, 339 (1984);\\
P.~van Nieuwenhuizen,
  Class.\ Quant.\ Grav.\  {\bf 2} (1985) 1;\\
O.~Aharony, Y.~Oz and Z.~Yin,
  Phys.\ Lett.\  B {\bf 430}, 87 (1998), hep-th/9803051;\\
S.~Minwalla,
  JHEP {\bf 9810}, 002 (1998), hep-th/9803053;\\
E.~Halyo,
  JHEP {\bf 9804}, 011 (1998), hep-th/9803077.
%
\bibitem{lunMath}
O.~Lunin and S.~D.~Mathur,
  Nucl.\ Phys.\  B {\bf 623}, 342 (2002), hep-th/0109154;\\
O.~Lunin, J.~M.~Maldacena and L.~Maoz,
  arXiv:hep-th/0212210.
%
\bibitem{LLM}
H.~Lin, O.~Lunin and J.~M.~Maldacena,
  JHEP {\bf 0410}, 025 (2004), hep-th/0409174.
  %
\bibitem{yama}
S.~Yamaguchi,
  Int.\ J.\ Mod.\ Phys.\  A {\bf 22}, 1353 (2007), hep-th/0601089.
%
\bibitem{myWils}
O.~Lunin,
  JHEP {\bf 0606}, 026 (2006), hep-th/0604133.
\bibitem{gomRom}
J.~Gomis and C.~Romelsberger,
  JHEP {\bf 0608}, 050 (2006), hep-th/0604155;\\
J.~Gomis and S.~Matsuura,
  arXiv:0704.1657 [hep-th].
\bibitem{giantGrav}
J.~McGreevy, L.~Susskind and N.~Toumbas,
  JHEP {\bf 0006}, 008 (2000), hep-th/0003075.
\bibitem{dualGiant}
M.~T.~Grisaru, R.~C.~Myers and O.~Tafjord,
  JHEP {\bf 0008}, 040 (2000), hep-th/0008015.
\bibitem{GiantAki}
A.~Hashimoto, S.~Hirano and N.~Itzhaki,
  JHEP {\bf 0008}, 051 (2000), hep-th/0008016.
%
\bibitem{berens}
S.~Corley, A.~Jevicki and S.~Ramgoolam,
  Adv.\ Theor.\ Math.\ Phys.\  {\bf 5}, 809 (2002), hep-th/0111222;\\
D.~Berenstein,
  JHEP {\bf 0407}, 018 (2004), hep-th/0403110.
%
\bibitem{ReyYee}
S.~J.~Rey and J.~T.~Yee,
  Eur.\ Phys.\ J.\  C {\bf 22}, 379 (2001), hep-th/9803001.
%
\bibitem{PawRey}
J.~Pawelczyk and S.~J.~Rey,
  Phys.\ Lett.\  B {\bf 493}, 395 (2000), hep-th/0007154.
%
\bibitem{DrukFiol}
N.~Drukker and B.~Fiol,
  JHEP {\bf 0502}, 010 (2005), hep-th/0501109.
\bibitem{GomisYama}
S.~Yamaguchi,
  JHEP {\bf 0605}, 037 (2006), hep-th/0603208;\\
J.~Gomis and F.~Passerini,
  JHEP {\bf 0608}, 074 (2006), hep-th/0604007;
  JHEP {\bf 0701}, 097 (2007), hep-th/0612022.
\bibitem{maldWils}
J.~M.~Maldacena,
  Phys.\ Rev.\ Lett.\  {\bf 80}, 4859 (1998), hep-th/9803002.
\bibitem{SkendTayl}
K.~Skenderis and M.~Taylor,
  JHEP {\bf 0206}, 025 (2002), hep-th/0204054;\\
P.~Bain, K.~Peeters and M.~Zamaklar,
  Phys.\ Rev.\  D {\bf 67}, 066001 (2003), hep-th/0208038.
\bibitem{KimYee}
N.~Kim and J.~T.~Yee,
  Phys.\ Rev.\  D {\bf 67}, 046004 (2003), hep-th/0211029.
\bibitem{cardy}
J.~L.~Cardy,
  Nucl.\ Phys.\  B {\bf 240}, 514 (1984).
\bibitem{CalMald}
C.~G.~Callan and J.~M.~Maldacena,
  Nucl.\ Phys.\  B {\bf 513}, 198 (1998), hep-th/9708147.
\bibitem{AganSchw}
M.~Perry and J.~H.~Schwarz,
  Nucl.\ Phys.\  B {\bf 489}, 47 (1997), hep-th/9611065;\\
M.~Aganagic, J.~Park, C.~Popescu and J.~H.~Schwarz,
  Nucl.\ Phys.\  B {\bf 496}, 191 (1997), hep-th/9701166;\\
I.~A.~Bandos, K.~Lechner, A.~Nurmagambetov, P.~Pasti, D.~P.~Sorokin and M.~Tonin,
  Phys.\ Lett.\  B {\bf 408}, 135 (1997), hep-th/9703127.
\bibitem{PST}
P.~Pasti, D.~P.~Sorokin and M.~Tonin,
  Phys.\ Lett.\  B {\bf 398}, 41 (1997), hep-th/9701037;\\
I.~A.~Bandos, K.~Lechner, A.~Nurmagambetov, P.~Pasti, D.~P.~Sorokin 
and M.~Tonin,
  Phys.\ Rev.\ Lett.\  {\bf 78}, 4332 (1997), hep-th/9701149.
\bibitem{Douglas}
M.~R.~Douglas,
  arXiv:hep-th/9512077;\\
C.~Bachas, M.~R.~Douglas and C.~Schweigert,
  JHEP {\bf 0005}, 048 (2000), hep-th/0003037.
\bibitem{ramal1}J.~M.~Camino, A.~Paredes and A.~V.~Ramallo,
  JHEP {\bf 0105}, 011 (2001), hep-th/0104082.
\bibitem{ramal2}
D.~S.~Berman and P.~Sundell,
  Phys.\ Lett.\  B {\bf 529}, 171 (2002), hep-th/0105288;\\
D.~Arean, A.~V.~Ramallo and D.~Rodriguez-Gomez,
  JHEP {\bf 0705}, 044 (2007), hep-th/0703094.
\bibitem{strom}
C.~Beasley, D.~Gaiotto, M.~Guica, L.~Huang, A.~Strominger and X.~Yin,
  arXiv:hep-th/0608021.
%
\bibitem{zarembo}
K.~Zarembo,
  Nucl.\ Phys.\  B {\bf 643}, 157 (2002), hep-th/0205160.
  %
\bibitem{yamaDef}
S.~Yamaguchi,
  JHEP {\bf 0306}, 002 (2003), hep-th/0305007.
\bibitem{BenaWarn}
I.~Bena and N.~P.~Warner,
  JHEP {\bf 0412}, 021 (2004), hep-th/0406145.
\bibitem{RussoTseytl}
J.~G.~Russo and A.~A.~Tseytlin,
  Nucl.\ Phys.\  B {\bf 490}, 121 (1997), hep-th/9611047.
\bibitem{HashItz}
A.~Hashimoto and N.~Itzhaki,
  Phys.\ Lett.\  B {\bf 465}, 142 (1999), hep-th/9907166.
  %
\bibitem{MaldRuss}
J.~M.~Maldacena and J.~G.~Russo,
  JHEP {\bf 9909}, 025 (1999), hep-th/9908134.
\bibitem{Bachas}
I.~Bakas and K.~Sfetsos,
  Int.\ J.\ Mod.\ Phys.\  A {\bf 12}, 2585 (1997), hep-th/9604003.
\bibitem{linMald}
H.~Lin and J.~M.~Maldacena,
  Phys.\ Rev.\  D {\bf 74}, 084014 (2006), hep-th/0509235.
\bibitem{Qrt33}
O.~Lunin,
  JHEP {\bf 0404}, 054 (2004), hep-th/0404006;\\
S.~Giusto, S.~D.~Mathur and A.~Saxena,
  Nucl.\ Phys.\  B {\bf 701}, 357 (2004), hep-th/0405017;
  Nucl.\ Phys.\  B {\bf 710}, 425 (2005), hep-th/0406103;\\
S.~Giusto and S.~D.~Mathur,
  Nucl.\ Phys.\  B {\bf 729}, 203 (2005), hep-th/0409067;\\
I.~Bena and N.~P.~Warner,
  Phys.\ Rev.\  D {\bf 74}, 066001 (2006), hep-th/0505166;\\
P.~Berglund, E.~G.~Gimon and T.~S.~Levi,
  JHEP {\bf 0606}, 007 (2006), hep-th/0505167;
I.~Bena, C.~W.~Wang and N.~P.~Warner,
  hep-th/0604110;
  JHEP {\bf 0611}, 042 (2006), hep-th/0608217;\\
J.~Ford, S.~Giusto and A.~Saxena,
  hep-th/0612227.
%
\bibitem{Qrt55}
N.~Kim,
  JHEP {\bf 0601}, 094 (2006), hep-th/0511029;\\
A.~Donos,
  Phys.\ Rev.\  D {\bf 75}, 025010 (2007), hep-th/0606199;
  hep-th/0610259;\\
E.~Gava, G.~Milanesi, K.~S.~Narain and M.~O'Loughlin,
  hep-th/0611065;\\
B.~Chen {\it et al.},
  arXiv:0704.2233 [hep-th].
\bibitem{GauntJuly}
J.~P.~Gauntlett, O.~A.~P.~Mac Conamhna, T.~Mateos and D.~Waldram,
  JHEP {\bf 0611}, 053 (2006), hep-th/0605146;\\
O.~A.~P.~Mac Conamhna and E.~O Colgain,
  JHEP {\bf 0703}, 115 (2007), hep-th/0612196;\\
J.~P.~Gauntlett, N.~Kim and D.~Waldram,
  JHEP {\bf 0704}, 005 (2007), hep-th/0612253;\\

\end{thebibliography}
\end{document}